\documentclass[iop,letter]{emulateapj}

\newcommand{\kms}{\,km\,s$^{-1}$}   \newcommand{\sqcm}{\,cm$^{-2}$}  
\newcommand{\os}{\ion{O}{6}}       \newcommand{\cf}{\ion{C}{4}}
\newcommand{\sif}{\ion{Si}{4}}     \newcommand{\nf}{\ion{N}{5}}       
\newcommand{\sit}{\ion{Si}{3}}     \newcommand{\siw}{\ion{Si}{2}} 
      
\newcommand{\hi}{\ion{H}{1}}       \newcommand{\hw}{\ion{H}{2}}
\newcommand{\oi}{\ion{O}{1}}       \newcommand{\cw}{\ion{C}{2}}
\newcommand{\caw}{\ion{Ca}{2}}     \newcommand{\nao}{\ion{Na}{1}}
\newcommand{\sw}{\ion{S}{2}}       \newcommand{\hst}{\emph{HST}}
  
    \newcommand{\msun}{\,M$_\odot$}
\newcommand{\tm}{\tablenotemark}   \newcommand{\tn}{\tablenotetext}
\newcommand{\nsl}{69}              \newcommand{\ndet}{56} 
\newcommand{\ncloudy}{13}          \newcommand{\nwham}{nine}
\newcommand{\ha}{H$\alpha$}        \newcommand{\msper}{81\%}
\newcommand{\smy}{\,M$_\odot$\,yr$^{-1}$}   

\begin{document}
\shorttitle{Ionized Gas in the Magellanic Stream}
\shortauthors{Fox et al.}
\title{The COS/UVES Absorption Survey of the Magellanic Stream. III: 
Ionization, Total Mass, and Inflow Rate onto the Milky Way\footnotemark[1]}
%Magellanic Gas Covers One Quarter of the Sky\footnotemark[1]}
\footnotetext[1]{Based on observations taken under programs 
  11520, 11524, 11541, 11585, 11598, 11632, 11686, 11692, 12025, 12029,
  12038, 12172, 12204, 12212, 12248, 12263, 12264, 12275, 12533, 12536, 12569, 
  12593, and 12604 of the NASA/ESA Hubble Space Telescope, obtained at the 
  Space Telescope Science Institute, which is operated by the Association of 
  Universities for Research in Astronomy, Inc., under NASA contract 
  NAS 5-26555.} 
 \author{Andrew J. Fox$^{2}$, Bart P. Wakker$^3$, Kathleen A. Barger$^{4,5}$, 
  Audra K. Hernandez$^3$, Philipp Richter$^{6,7}$, Nicolas Lehner$^4$, 
  Joss Bland-Hawthorn$^8$, Jane C. Charlton$^9$, Tobias Westmeier$^{10}$, 
  Christopher Thom$^2$, Jason Tumlinson$^2$, Toru Misawa$^{11}$, 
  J. Christopher Howk$^4$,  L. Matthew Haffner$^{3,12}$, Justin Ely$^2$, 
  Paola Rodriguez-Hidalgo$^{13,14}$, \& Nimisha Kumari$^{2,15}$}  
\affil{$^2$ Space Telescope Science Institute, 3700 San Martin Drive,
  Baltimore, MD 21218\\
$^3$ Department of Astronomy, University of
  Wisconsin--Madison, 475 North Charter St., Madison, WI 53706\\
$^4$ Department of Physics, University of Notre Dame, 225 Nieuwland 
Science Hall, Notre Dame, IN 46556\\
$^5$ National Science Foundation Astronomy and Astrophysics 
Postdoctoral Fellow\\
$^6$ Institut f\"ur Physik und Astronomie, Universit\"at Potsdam, Haus
   28, Karl-Liebknecht-Str. 24/25, 14476, Potsdam, Germany\\
$^7$ Leibniz-Institut f\"ur Astrophysik Potsdam (AIP),
   An der Sternwarte 16, 14482 Potsdam, Germany\\
$^8$ Institute of Astronomy, School of Physics, University of Sydney, 
NSW 2006, Australia\\
$^9$ Department of Astronomy and Astrophysics, Pennsylvania State 
University, University Park, PA 16802\\
$^{10}$ ICRAR, The University of Western Australia, 35 Stirling Highway, 
Crawley WA 6009, Australia\\
$^{11}$ School of General Education, Shinshu University, 3-1-1 Asahi, Matsumoto, 
Nagano 390-8621, Japan\\
$^{12}$ Space Science Institute, 4750 Walnut Street, Suite 205, Boulder, 
CO 80301\\
$^{13}$ Department of Physics and Astronomy, York University, 4700 Keele St., 
Toronto, Ontario, M3J 1P3, Canada\\
$^{14}$ Department of Astronomy and Astrophysics, University of Toronto, 
50 St. George Street, Toronto, Ontario, M5S 3H4, Canada\\
$^{15}$ Ecole Polytechnique, Route de Saclay, 91128 Palaiseau, France}
\email{afox@stsci.edu}

\begin{abstract} 
Dynamic interactions between the two Magellanic Clouds
have flung large quantities of gas into the halo of the Milky Way.
The result is a spectacular arrangement of gaseous structures
including the Magellanic Stream, the Magellanic Bridge, and the Leading Arm
(collectively referred to as the Magellanic System).
In this third paper of a series studying the Magellanic gas in 
absorption, we analyze the gas ionization level 
using a sample of \nsl\ \emph{Hubble Space Telescope}/Cosmic Origins 
Spectrograph sightlines that pass through or within 30\degr\ of 
the 21\,cm-emitting regions.
%For nine of these sightlines we include optical \ha\
%emission-line observations from the Wisconsin \ha\ Mapper.
We find that \msper\ (\ndet/\nsl) of the sightlines show UV absorption 
at Magellanic velocities, indicating that the total cross section of the 
Magellanic System is $\approx$11\,000 square degrees, 
or around a quarter of the entire sky. %*
%This is around five times larger than compared to the 21\,cm %*
%cross section of 2\,700 square degrees (Nidever et al. 2010).
Using observations of the \sit/\siw\ ratio together
with \emph{Cloudy} photoionization modeling, we 
calculate the total gas mass (atomic plus ionized) of the 
Magellanic System to be 
$\approx$2.0$\times$10$^9$\msun\,($d$/55\,kpc)$^2$, %*
with the ionized gas contributing around three times as much mass %*
as the atomic gas. This is larger than the current-day 
interstellar \hi\ mass of both Magellanic Clouds combined, indicating that 
they have lost most of their initial gas mass.
If the gas in the Magellanic System survives to reach the Galactic disk
over its inflow time of $\sim$0.5--1.0\,Gyr, it will represent an average 
inflow rate of $\sim$3.7--6.7\smy, potentially raising %*
the Galactic star formation rate. 
However, multiple signs of an evaporative interaction with the hot 
Galactic corona indicate that the Magellanic gas may not survive its journey to 
the disk fully intact, and will instead add material to (and cool) the corona.
%This evaporated material must subsequently cool and condense to power 
%future star formation.
\end{abstract}

\section{Introduction} 
Like other massive spiral galaxies, the Milky Way has a gas-consumption time 
of a few billion years \citep[e.g.][]{La80}, far shorter than its age. 
Ongoing fuel replenishment (gaseous accretion)
is therefore needed for it to sustain its star formation.
Chemical and kinematic studies of stars in the solar neighborhood
have shown that the Milky Way has been forming
stars at a fairly constant rate for several Gyr \citep{Tw80, Bi00, Ch01},
so we know that fuel replenishment has to be occurring. 
However, several open questions surround {\it how} accretion onto 
the Milky Way and other $\approx\!L_\ast$ galaxies occurs.
Is the accreting gas predominantly cold, warm, or hot? 
Does it arrive smoothly from the unenriched IGM or episodically following 
satellite-galaxy interactions?
And do most accreting streams become disrupted by 
fluid instabilities before reaching the disk? %of the target galaxy? 

Hydrodynamical simulations offer one approach to answering these questions.
They make clear predictions for the 
physical and chemical state of gas accreting onto galaxies
\citep{Ke05, Ke09, Ba10, Fu11, St11, Fe12, VS12, Mu12, Jo12b, Vo12, Sh13}.
However, clear observational evidence for gas accretion in external galaxies 
has been hard to find \citep[see][]{Rb11, Le13, Bu13}, largely due
to a lack of kinematic information.
Unambiguous \emph{spectroscopic} evidence for inflow onto external galaxies
exists in only a small number of cases \citep[e.g.][]{Ru12}, and in these
galaxies the inflow may be metal-enriched.

Fortunately, the Milky Way offers a prime opportunity for studying 
the fueling of an $\approx\!L_\ast$ galaxy. We have detailed knowledge of the 
neutral-gas distribution around the Galaxy from \hi\ 21\,cm 
observations, and information on the ionized-gas distribution
from a large body of absorption-line data taken from 
ultraviolet (UV) spectra of background QSOs.
%which contain information on the 
%neutral and ionized gas content of the Galactic halo.
In both the 21\,cm and UV data, 
gaseous inflow onto the Galaxy can be seen directly 
among the high-velocity clouds (HVCs), interstellar clouds that are not 
co-rotating with the Galactic disk.
First detected by \citet{Mu63}, HVCs are defined as 
having LSR velocities $|v_{\rm LSR}|\!>\!100$\kms\ 
\citep[or sometimes $>$90\kms; see reviews by][]{WW97, Ri06, Pu12}. 
While HVCs trace a variety of interstellar processes,
\emph{infalling} HVCs can be identified by their negative
velocities in the Galactic-Standard-of-Rest (GSR) frame and their 
low metallicities.

Most \hi\ HVCs are now known to be within 5--20\,kpc of the Galactic disk,
based on individual cloud distance measurements 
\citep{Wa01, Wa08, Th08, LH10, Sm11}
and on the observation that the HVC sky-covering fraction $f_{\rm HVC}$ 
measured toward halo stars is similar to $f_{\rm HVC}$ measured 
toward AGN \citep{LH11, Le12}.
However, the Magellanic Stream (MS), anchored by the Magellanic
Clouds, stands as a notable exception. Although the distance to the MS 
is poorly constrained, it likely lies in the interval $\sim$55--200\,kpc.
The SMC distance gives the lower limit, and we use the 
tidal models of \citet{Be12} to bound the upper limit, % 80--230 kpc,
although \citet{JL08} argue the MS distance near the south Galactic pole is 
unlikely to exceed 100\,kpc \citep[see also][]{BH13}.

The Stream's origin in the Magellanic Clouds is supported by its spatial,
kinematic, and chemical properties, but 
the mechanism by which it was removed from the Clouds
has puzzled dynamicists since its discovery in 21\,cm observations
over forty years ago \citep{Di71, WW72, Ma74}\footnote{Parts of the Stream
were detected even earlier by \citet{Di65}, although the association
with the Magellanic Clouds was not realized at that time.}.
Recent models favor the tidal removal of much of the Stream
during a close encounter between the two Magellanic Clouds $\approx$2\,Gyr ago
\citep{GN96, Co06, Be10, Be12, DB11, DB12}, 
although ram-pressure stripping \citep{Me85, MD94, Ma05} and 
supernova-driven blowout of LMC material 
\citep[][hereafter N08]{SS03, LH07, Le09, Ni08} may have also contributed 
to its production.
Tidal forces are strongly favored as the mechanism responsible for creating the 
Leading Arm (LA), the gaseous counterpart to the MS lying in front of the 
direction of motion of the Magellanic Clouds \citep{Wa72,Ma74,Pu98,Lu98}.

%The Stream can be viewed as a case-study of the fueling of a 
%massive star-forming galaxy.
To probe the Stream's physical and chemical conditions across its full
length on the sky, we are conducting an absorption-line survey
with both UV (\hst/Cosmic Origins Spectrograph) and optical (VLT/UVES) 
spectrographs. In Paper I \citep{Fo13a}, we presented new
MS chemical abundance measurements along the sightlines to the AGN 
\object{RBS\,144} and \object{NGC\,7714}, and an upper limit on the 
metallicity toward the QSO \object{PHL\,2525}.
Combined with earlier work on the \object{NGC\,7469} sightline 
\citep[][hereafter F10]{Fo10},
there are now three good measurements of $\approx$0.1 solar metallicity 
in the main body of the Stream, supporting the view that most of
the MS was stripped from the SMC (not the LMC) 1.5--2.5\,Gyr 
ago. This is because the SMC had a metallicity of $\approx$0.1 solar at 
that time according to its age-metallicity relation \citep{PT98, HZ04}.
However, in Paper II \citep[][see also Gibson et al. 2000]{Ri13}, we 
found a much higher metallicity of 0.5 solar along the inner-Stream 
sightline to QSO \object{Fairall\,9}, which passes through a MS filament that 
can be traced kinematically back to the LMC (N08).
This shows that the bifurcation of the Stream,
previously seen in its spatial extent \citep{Pu03a} and kinematics (N08),
is also seen in its metal enrichment.
This supports a dual origin for the Stream, with both the SMC and LMC 
contributing to its origin.

We now turn our attention from the origin of the Stream to its fate,
by observing its ionization level, which encodes information on the physical 
processes occurring as it interacts with the ambient plasma and radiation field.
The MS and the LA contain both warm-ionized and highly-ionized material.
The warm-ionized phase is seen in \ha\ emission \citep{WW96, Pu03b, BH13, Ba14} 
and absorption in low-ionization UV lines 
\citep[][F10, Paper I, Paper II]{Lu94, Lu98, Se03, Fo05a}.
Photoionization and/or shock ionization followed by recombination
may be responsible for exciting
the \ha\ emission seen from the Stream, including the possibility of
ionization by a Seyfert flare at the center of the Milky Way 
$\approx$1--3\,Myr ago \citep{BH13}.
The highly ionized phase in the Stream is seen in
\cf\ \citep[][F10, Paper II]{Lu94, Lu98} and \os\ 
\citep[][F10]{Se03, Fo05a} absorption.
These high ions have column-density ratios consistent with an origin in
the conductive or turbulent interfaces that surround the Stream, where the 
warm low-ionization gas adjoins the hot coronal plasma \citep{Fo05a}.

In this paper, we present the first UV survey of the Stream's ionization 
properties, using \nsl\ COS sightlines passing through or within 30\degr\
of the MS, the LA, or the Magellanic Bridge (MB), the gaseous filament that 
connects the LMC and SMC. For brevity, we use the phrase 
``the Magellanic System'' (MSys) to encompass the MS, MB, and LA.
This is similar to the definition used by \citet{Br05} in their Parkes survey
of 21\,cm emission from Magellanic gas, except that here we include
the ionized outer regions of the MS that are not seen at 21\,cm.
Furthermore, \citet{Br05} define the ``Interface Region'' of \hi-emitting gas
lying in-between the Magellanic Clouds and the Stream proper. In this paper
we include that region as part of the Stream, since the two principal
MS filaments can be traced kinematically through it back to the Magellanic 
Clouds (N08), and hence there is no reason to treat it as a distinct object.
Physically, the MSys refers to all gas that was 
stripped from the Magellanic Clouds at some point in the past
(though not necessarily all at the same time).
We define On-System and Off-System directions as those 
with and without a detection of \hi\ 21\,cm emission at Magellanic velocities.
We also define a subset of ``LMC-Halo'' directions, that lie 
in the outer halo of the LMC; these are not 
part of the MS as traditionally defined, but are clearly Magellanic in origin,
and hence are included in our analysis.

We also include Wisconsin H-alpha Mapper (WHAM) 
observations of \ha\ emission from the Stream in six of the MS sightlines
and three of the MB sightlines.
This allows us to compare the optical, radio, and UV profiles of the 
Magellanic gas in the same directions. 
Throughout the paper we use solar (photospheric) elemental abundances from 
\citet{As09} and absorption-line data (rest wavelengths and oscillator 
strengths) from \citet{Mo03}. We present all velocities in the kinematical 
local standard-of-rest (LSR) frame, and all column densities are given
in units of cm$^{-2}$.

This paper is laid out as follows. 
\S2 describes the sample assembly, observations, data handling procedures, 
and absorption-line measurements.
\S3 presents an overview of the UV absorption-line profiles and 
WHAM emission-line profiles.
\S4 discusses the physical conditions of the Magellanic gas, presenting
new constraints on the ionization level in both the low-ion 
and high-ion gas phases. In \S5 we derive and discuss the total (neutral 
plus ionized) gas mass in the MSys. In \S6 we discuss the accretion rate 
of Magellanic gas onto the Galaxy. In \S7 we briefly discuss the MSys in 
terms of intervening quasar absorption line systems. \S8 summarizes the 
main results.

\section{Observations and Data Handling}

\begin{figure*}[t] 
\epsscale{1.0}\plotone{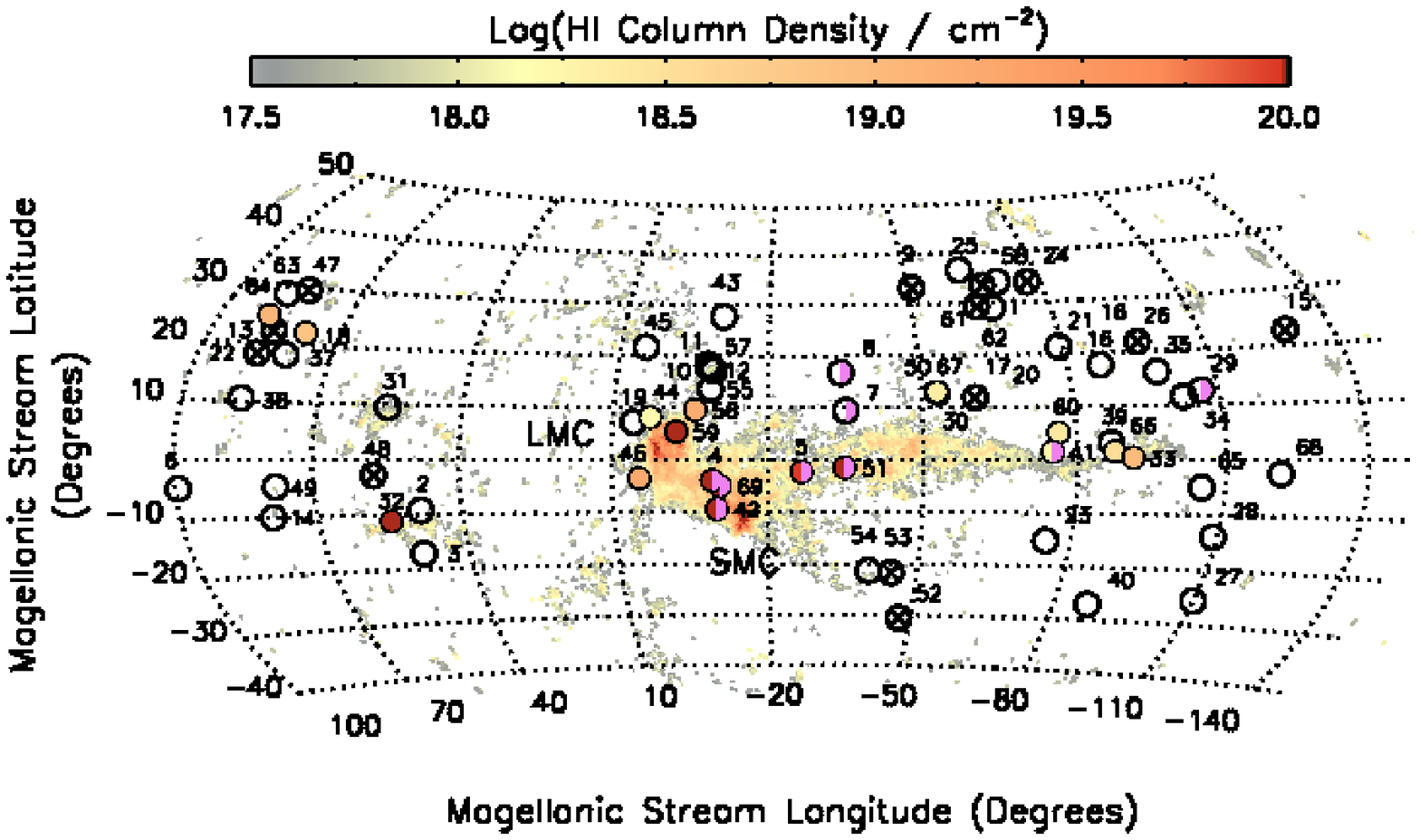} %{f1a.eps}
\epsscale{1.0}\plotone{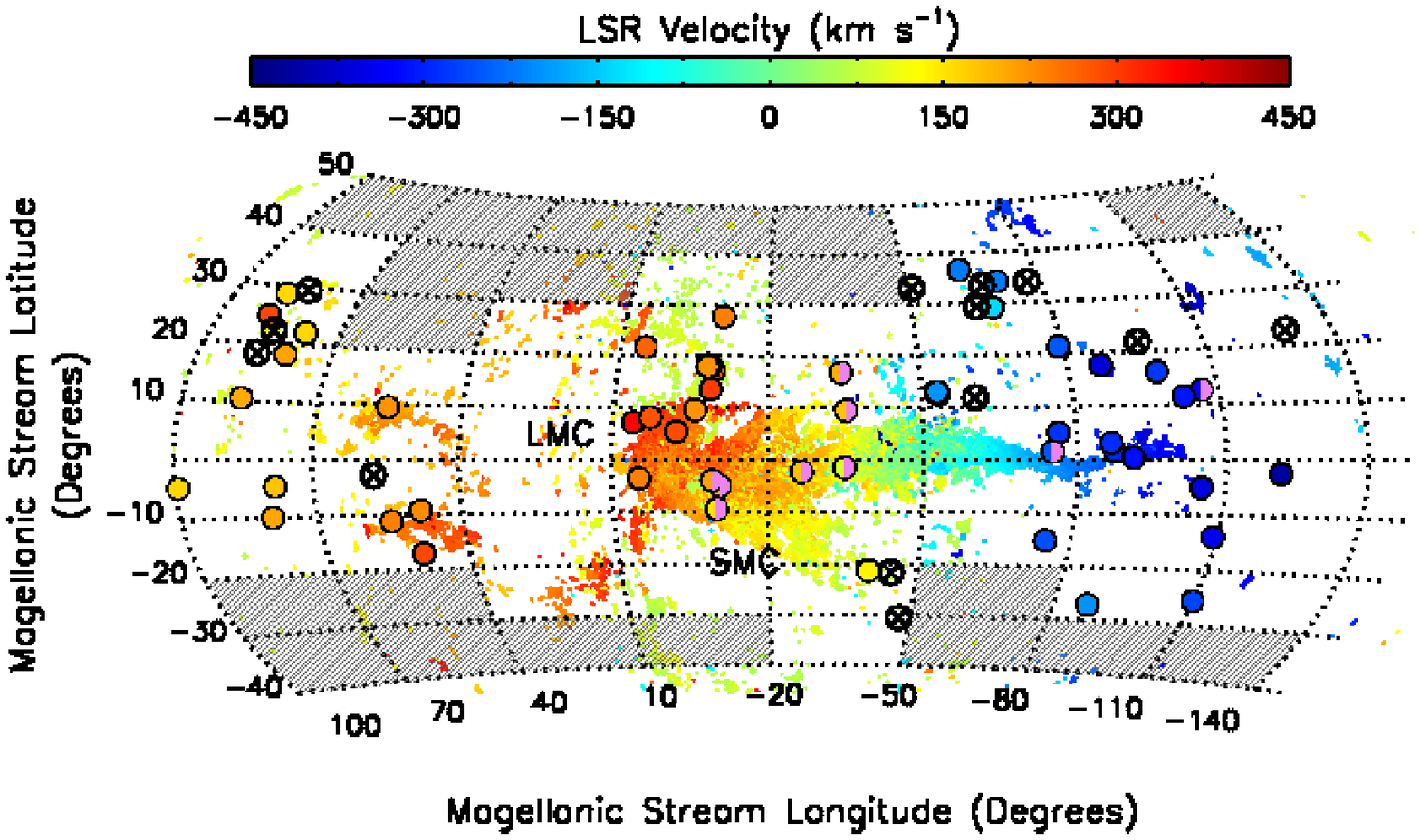} %{f1b.eps}
\caption{Maps of the Magellanic System comparing our \nsl\ COS directions 
(circles) with a Gaussian decomposition (N08) of the \hi\ 21\,cm emission 
from the LAB survey \citep[colored shading;][]{Ka05}. The maps
are plotted in the MS coordinate system of N08.
{\bf Top:} MSys map color-coded by \hi\ column density.
\ndet\ out of \nsl\ COS directions display Magellanic UV absorption;
the non-detections are shown with crossed circles. 
Directions with WHAM data are shown half-shaded in pink.
The numbers next to each sightline identify the
targets in Table 1 and the appendix.
{\bf Bottom:} MSys map color-coded by central LSR velocity.
The coloring of the circles shows the velocity of the UV absorption,
whereas the coloring of the background shading shows the velocity of 
the 21\,cm emission. 
The 21\,cm velocity field is closely reflected in the UV absorption lines,
even when moving from On-System to Off-System directions.
The non-hatched regions show our adopted cross-section of Magellanic gas,
used in our gas-mass and inflow-rate calculations.}
\end{figure*}

\subsection{Sample Selection}

We assembled our sample of \nsl\ targets by identifying
all AGN observed with the \hst/COS spectrograph \citep{Gr12} 
that met the following criteria:\\

\noindent (1) lie within 30\degr\ of the 21\,cm emission from the 
Magellanic System, as defined by the \hi\ contours of \citet{Mo00};\\
(2) have COS data taken with the G130M grating;\\
(3) have data with a signal-to-noise ratio
S/N$\ga$5 per resolution element at 1250\,\AA;\\
(4) lie in directions where Magellanic absorption is at 
$|v_{\rm LSR}|\!>\!100$\kms\ and is therefore unblended with 
Galactic ISM absorption.\\
(5) have data publicly available in the MAST archive as of October 2013;\\

Criterion (1) was adopted to provide an extended search area around
the MSys. This region covers most of the Southern Galactic hemisphere.
Criterion (2) was used since G130M data (covering $\approx$1150--1450\,\AA) 
are needed to cover the \sit\ $\lambda$1206 and \siw\ $\lambda$1193 lines, 
which are used in our ionization analysis. When G160M data are also
available, covering $\approx$1405--1775\,\AA\ and hence the \cf\ 
$\lambda$$\lambda$1548, 1550 doublet, we include them, but we do not include 
sightlines with G160M data only, since they do not contain sufficient 
information for our ionization modeling procedure.
Criterion (3) was adopted to ensure that each spectrum in the sample had
sufficient quality to determine the presence or absence of 
Magellanic absorption in the UV metal lines, and that the apparent
optical depth (AOD) method could be reliably applied 
\citep[this method becomes inaccurate at S/N$\la$5;][]{Fo05b}.
Criterion (4) was necessary to ensure that Magellanic absorption could be 
reliably separated from low-velocity absorption, and implied that 
targets in a small region of sky near the south Galactic pole were excluded 
from the search.
Criterion (5) was used since the holdings of the MAST archive are constantly 
changing, and October 2013 represented the time when we chose to define our 
final sample. 

Targets with available COS spectra that did not meet the above criteria were:
\object{HS\,2154+2228}, \object{SDSS\,J004042.09--110957.7},
\object{LBQS\,0107--0232}, \object{Mrk\,1501}, and \object{NGC\,7469}, 
which have G160M data only (criterion 2);
\object{2dFGRS\_TGS394Z150}, \object{HE\,0946--0500}, \object{LBQS\,0052--0038},
\object{NGC\,3256}, and \object{NGC\,7552}, which have S/N$\la$5 (criterion 3);
\object{HE\,0056--3622} and \object{HE\,2347--4342}, which fall within 
the search area but were removed from the sample since in these directions 
MS absorption overlaps in velocity with Galactic ISM absorption (criterion 4).
%Finally, \object{ESO\,292--G24} was removed due to unusually broad line 
%profiles indicative of target acquisition problems. (not point source)

Six of the targets were observed in 2012 under \hst\ program 
12604 (PI A. Fox), which specifically targeted MS directions. Three were 
observed under \hst\ program 12204 (PI C. Thom), which targeted 
three closely-spaced directions near the compact HVC 224.0--83.4--197
(hereafter, the CHVC), which lies near the edge of the Stream 
\citep{Se02, Ku14}.
Three more were observed under \hst\ program 12263 (PI T. Misawa), which 
targeted QSOs lying behind the MB.
Six were observed under \hst\ program 11692 (PI J. Howk), which 
targeted QSOs lying behind the halo of the LMC.
The remainder were drawn from various programs, including the COS 
Science Team Guaranteed Time Observations (GTO) programs.
The \nsl\ spectra in the sample have a mean S/N per resolution element
at 1200\AA\ of 22 and a median of 18.

Basic information for all the targets in the sample is given in Table 1,
and their distribution on the sky is shown in Figures 1a and 1b, 
where we include \hi\ data from a Gaussian decomposition (N08) to the 
Leiden-Argentine-Bonn (LAB) survey \citep{Ka05}.
Figure 1a is color-coded by \hi\ column density, and Figure 1b is color-coded
by central LSR velocity of Magellanic absorption/emission.
The maps are shown in the MS coordinate system defined by N08,
in which the Stream bisects the equator, the LMC is at 
$L_{\rm MS}, B_{\rm MS}$=0\degr, 0\degr, the LA extends to positive $L_{\rm MS}$, 
and the Stream itself extends to negative $L_{\rm MS}$.
This coordinate system is chosen to display the MSys without the projection 
effects that can distort its appearance when shown in Galactic coordinates.

\begin{deluxetable*}{lllrrrr ccrr}
\tabletypesize{\scriptsize}
\tabcolsep=2.0pt
\tablewidth{0pt}
\tablecaption{Overview of Sightline Properties}
\tablehead{ID\tm{a} & Target & Region\tm{b} & $l$\phs\phs & $b$\phs\phs & $L_{\rm MS}$\tm{c} & $B_{\rm MSys}$\tm{c} & Prog.\tm{d} & 
$v_{\rm min}, v_{\rm max}$\tm{e} & 
$\langle v_{\rm MS} \rangle$\tm{f} & log\,$N$(\hi)\tm{g}\\
 & & & (\degr)\phs & (\degr)\phs & (\degr) & (\degr) &  & (km\,s$^{-1}$)
 & (km\,s$^{-1}$) &  (cm$^{-2}$)}
\startdata
 1 &                      3C57 &   MS-Off & 173.08 & $-$67.26 & $-$64.05 &\phs29.33 &12038 &       $-$260,$-$160 &   \nodata &  $<$17.97\\
 2 &                ESO265-G23 &   LA-Off & 285.91 &\phs16.59 &\phs48.41 & $-$9.20 &12275 &     \phs160,\phs300 &   \phs230 &  $<$18.02\\
 3 &                ESO267-G13 &   LA-Off & 294.10 &\phs18.34 &\phs49.67 & $-$17.11 &12275 &     \phs215,\phs360 &   \phs290 &  $<$18.03\\
 4 &                  ESO31-G8 &   Bridge & 290.33 & $-$40.79 & $-$9.31 & $-$3.93 &12263 &     \phs150,\phs340 &   \phs207 &     20.51\\
 5 &                  FAIRALL9 &    MS-On & 295.07 & $-$57.83 & $-$26.56 & $-$2.30 &12604 &     \phs120,\phs280 &   \phs184 &     19.97\\
 6 &                 H1101-232 &   LA-Off & 273.19 &\phs66.55 &\phs99.29 & $-$4.77 &12025 &     \phs130,\phs230 &   \phs160 &  $<$18.12\\
 7 &               HE0153-4520 &   MS-Off & 271.77 & $-$67.98 & $-$35.10 &\phs9.47 &11541 &     \phs150,\phs240 &   \phs180 &  $<$18.40\\
 8 &               HE0226-4110 &   MS-Off & 253.94 & $-$65.78 & $-$34.34 &\phs16.74 &11541 &     \phs 90,\phs245 &   \phs195 &  $<$18.33\\
 9 &               HE0238-1904 &   MS-Off & 200.48 & $-$63.63 & $-$50.95 &\phs33.21 &11541 &       $-$260,$-$160 &   \nodata &  $<$17.98\\
10 &               HE0429-5343 &  LMC-Off & 262.08 & $-$42.17 & $-$9.19 &\phs17.18 &12275 &     \phs220,\phs325 &   \phs275 &  $<$17.98\\
11 &               HE0435-5304 &  LMC-Off & 261.02 & $-$41.38 & $-$8.45 &\phs18.04 &11520 &     \phs200,\phs380 &   \phs280 &  $<$18.00\\
12 &               HE0439-5254 &  LMC-Off & 260.69 & $-$40.90 & $-$7.98 &\phs18.34 &11520 &     \phs190,\phs340 &   \phs280 &  $<$18.03\\
13 &               HE1003+0149 &   LA-Off & 238.53 &\phs42.79 &\phs85.30 &\phs22.25 &12593 &     \phs200,\phs300 &   \nodata &  $<$18.36\\
14 &               HE1159-1338 &   LA-Off & 285.11 &\phs47.24 &\phs79.51 & $-$9.98 &12275 &     \phs110,\phs240 &   \phs205 &  $<$17.99\\
15 &               HS0033+4300 &   MS-Off & 120.03 & $-$19.51 & $-$129.21 &\phs22.74 &12264 &       $-$400,$-$300 &   \nodata &  $<$18.42\\
16 &                    IO-AND &   MS-Off & 120.80 & $-$59.52 & $-$86.78 &\phs17.66 &11632 &       $-$410,$-$320 &    $-$370 &  $<$18.12\\
17 &            IRAS01003-2238 &   MS-Off & 152.05 & $-$84.58 & $-$60.56 &\phs11.85 &12533 &       $-$280,$-$160 &   \nodata &  $<$18.14\\
18 &           IRASF09539-0439 &    LA-On & 243.33 &\phs37.00 &\phs77.89 &\phs22.18 &12275 &     \phs140,\phs240 &   \phs160 &     18.89\\
19 &           IRASZ06229-6434 &  LMC-Off & 274.31 & $-$27.32 &\phs5.98 &\phs7.14 &11692 &     \phs200,\phs480 &   \phs350 &  $<$18.13\\
20 &             LBQS0107-0233 &   MS-Off & 134.03 & $-$64.78 & $-$79.13 &\phs21.32 &11585 &       $-$330,$-$220 &    $-$260 &  $<$18.03\\
21 &             LBQS0107-0235 &   MS-Off & 134.01 & $-$64.80 & $-$79.12 &\phs21.30 &11585 &       $-$340,$-$150 &    $-$252 &  $<$18.02\\
22 &             LBQS1019+0147 &   LA-Off & 242.16 &\phs46.07 &\phs86.67 &\phs18.28 &11598 &     \phs180,\phs280 &   \nodata &  $<$18.00\\
23 &               MRC2251-178 &   MS-Off &  46.20 & $-$61.33 & $-$74.97 & $-$15.18 &12029 &       $-$350,$-$200 &    $-$260 &  $<$18.12\\
24 &                   MRK1014 &   MS-Off & 156.57 & $-$57.94 & $-$77.04 &\phs33.87 &12569 &       $-$260,$-$180 &   \nodata &  $<$18.21\\
25 &                   MRK1044 &   MS-Off & 179.69 & $-$60.48 & $-$62.67 &\phs36.61 &12212 &       $-$280,$-$130 &    $-$220 &  $<$18.06\\
26 &                   MRK1502 &   MS-Off & 123.75 & $-$50.17 & $-$95.87 &\phs21.78 &12569 &       $-$340,$-$150 &   \nodata &  $<$18.19\\
27 &                   MRK1513 &   MS-Off &  63.69 & $-$29.07 & $-$109.97 & $-$25.55 &11524 &       $-$370,$-$170 &    $-$270 &  $<$18.28\\
28 &                    MRK304 &   MS-Off &  75.99 & $-$34.22 & $-$109.18 & $-$13.91 &12604 &       $-$400,$-$230 &    $-$350 &  $<$18.25\\
29 &                    MRK335 &   MS-Off & 108.76 & $-$41.42 & $-$106.41 &\phs12.64 &11524 &       $-$400,$-$200 &    $-$340 &  $<$18.22\\
30 &               MS0117-2837 & CHVC-Off & 225.73 & $-$83.65 & $-$53.35 &\phs12.53 &12204 &       $-$270,$-$ 95 &    $-$120 &  $<$17.29\\
31 &                   NGC3125 &   LA-Off & 265.33 &\phs20.64 &\phs55.22 &\phs9.52 &12172 &     \phs150,\phs300 &   \phs220 &  $<$18.47\\
32 &                   NGC3783 &    LA-On & 287.46 &\phs22.95 &\phs54.74 & $-$11.13 &12212 &     \phs150,\phs300 &   \phs240 &     20.10\\
33 &                   NGC7714 &    MS-On &  88.22 & $-$55.56 & $-$91.08 &\phs0.43 &12604 &       $-$420,$-$190 &    $-$320 &     19.01\\
34 &                PG0003+158 &   MS-Off & 107.32 & $-$45.33 & $-$102.43 &\phs11.52 &12038 &       $-$430,$-$200 &    $-$315 &  $<$18.21\\
35 &                PG0026+129 &   MS-Off & 114.64 & $-$49.25 & $-$98.04 &\phs16.19 &12569 &       $-$330,$-$250 &    $-$280 &  $<$18.14\\
36 &                PG0044+030 &   MS-Off & 120.80 & $-$59.52 & $-$86.78 &\phs17.66 &12275 &       $-$350,$-$250 &    $-$275 &  $<$18.39\\
37 &                PG1011-040 &   LA-Off & 246.50 &\phs40.75 &\phs80.16 &\phs18.23 &11524 &     \phs185,\phs235 &   \phs208 &  $<$18.08\\
38 &                PG1049-005 &   LA-Off & 252.24 &\phs49.88 &\phs86.72 &\phs10.54 &12248 &     \phs100,\phs280 &   \phs200 &  $<$18.00\\
39 &                PG2349-014 &   MS-Off &  91.66 & $-$60.36 & $-$86.73 &\phs3.16 &12569 &       $-$370,$-$200 &    $-$294 &  $<$18.11\\
40 &                   PHL1811 &   MS-Off &  47.47 & $-$44.82 & $-$87.33 & $-$27.05 &12038 &       $-$380,$-$120 &    $-$200 &  $<$18.00\\
41 &                   PHL2525 &    MS-On &  80.68 & $-$71.32 & $-$75.04 &\phs1.52 &12604 &       $-$280,$-$120 &    $-$256 &     18.27\\
42 &                PKS0202-76 &   Bridge & 297.55 & $-$40.05 & $-$10.09 & $-$9.42 &12263 &     \phs100,\phs360 &   \phs150 &     20.05\\
43 &               PKS0439-433 &  LMC-Off & 247.98 & $-$41.38 & $-$10.41 &\phs27.65 &12536 &     \phs200,\phs340 &   \phs240 &  $<$18.03\\
44 &               PKS0552-640 &   LMC-On & 273.47 & $-$30.61 &\phs2.72 &\phs8.12 &11692 &     \phs200,\phs450 &   \phs279 &     18.23\\
45 &               PKS0558-504 &  LMC-Off & 257.96 & $-$28.57 &\phs4.91 &\phs21.59 &11692 &     \phs220,\phs350 &   \phs265 &  $<$18.07\\
46 &                PKS0637-75 &   LMC-On & 286.37 & $-$27.16 &\phs4.73 & $-$3.51 &11692 &     \phs180,\phs400 &   \phs240 &     19.28\\
47 &                PKS0932+02 &   LA-Off & 232.39 &\phs36.84 &\phs83.08 &\phs29.58 &11598 &     \phs200,\phs300 &   \nodata &  $<$18.10\\
48 &               PKS1101-325 &   LA-Off & 278.12 &\phs24.77 &\phs57.35 & $-$2.79 &12275 &     \phs220,\phs320 &   \nodata &  $<$18.03\\
49 &                PKS1136-13 &   LA-Off & 277.53 &\phs45.43 &\phs77.98 & $-$4.66 &12275 &     \phs150,\phs220 &   \phs180 &  $<$18.09\\
50 &              QSO1220-2859 &  CHVC-On & 227.78 & $-$82.93 & $-$52.69 &\phs12.94 &12204 &       $-$270,$-$120 &    $-$180 &     18.42\\
51 &                    RBS144 &    MS-On & 299.48 & $-$65.84 & $-$34.80 & $-$1.50 &12604 &     \phs 65,\phs210 &   \phs 92 &     20.17\\
52 &                   RBS1795 &   MS-Off & 355.18 & $-$50.87 & $-$47.56 & $-$30.50 &11541 &     \phs100,\phs200 &   \nodata &  $<$18.08\\
53 &                   RBS1892 &   MS-Off & 345.90 & $-$58.37 & $-$44.75 & $-$21.65 &12604 &     \phs160,\phs240 &   \nodata &  $<$18.02\\
54 &                   RBS1897 &   MS-Off & 338.51 & $-$56.63 & $-$40.11 & $-$21.30 &11686 &     \phs100,\phs200 &   \phs135 &  $<$18.29\\
55 &                    RBS542 &  LMC-Off & 266.99 & $-$42.00 & $-$8.76 &\phs13.56 &11686 &     \phs200,\phs400 &   \phs300 &  $<$18.14\\
56 &                    RBS563 &   LMC-On & 272.25 & $-$39.23 & $-$5.94 &\phs9.56 &11692 &     \phs180,\phs400 &   \phs220 &     19.22\\
57 &                    RBS567 &  LMC-Off & 261.22 & $-$40.93 & $-$7.96 &\phs17.94 &11520 &     \phs180,\phs360 &   \phs220 &  $<$17.97\\
58 &            RXJ0209.5-0438 &   MS-Off & 165.99 & $-$60.81 & $-$70.31 &\phs34.10 &12264 &       $-$260,$-$160 &    $-$220 &  $<$18.23\\
59 &            RXJ0503.1-6634 &   LMC-On & 277.18 & $-$35.42 & $-$2.36 &\phs5.41 &11692 &     \phs160,\phs375 &   \phs285 &     20.98\\
60 &   SDSSJ001224.01-102226.5 &    MS-On &  92.32 & $-$70.89 & $-$76.35 &\phs5.07 &12248 &       $-$340,$-$140 &    $-$285 &     18.43\\
61 &   SDSSJ015530.02-085704.0 &   MS-Off & 165.72 & $-$66.34 & $-$67.52 &\phs29.10 &12248 &       $-$180,$-$100 &    $-$140 &  $<$18.07\\
62 &   SDSSJ021218.32-073719.8 &   MS-Off & 171.15 & $-$62.66 & $-$66.72 &\phs33.41 &12248 &       $-$260,$-$160 &   \nodata &  $<$18.12\\
63 &   SDSSJ094331.60+053131.0 &   LA-Off & 230.04 &\phs40.41 &\phs87.57 &\phs28.71 &11598 &     \phs130,\phs200 &   \phs160 &  $<$18.26\\
64 &   SDSSJ095915.60+050355.0 &    LA-On & 233.37 &\phs43.48 &\phs88.38 &\phs24.84 &12248 &     \phs225,\phs325 &   \phs289 &     19.09\\
65 &   SDSSJ225738.20+134045.0 &   MS-Off &  85.28 & $-$40.73 & $-$105.05 & $-$4.97 &11598 &       $-$440,$-$260 &    $-$360 &  $<$18.17\\
66 &   SDSSJ234500.43-005936.0 &    MS-On &  88.79 & $-$59.39 & $-$87.41 &\phs1.56 &11598 &       $-$330,$-$190 &    $-$300 &     18.54\\
67 &                   TONS210 &  CHVC-On & 224.97 & $-$83.16 & $-$53.11 &\phs12.98 &12204 &       $-$280,$-$140 &    $-$200 &     18.36\\
68 &                  UGC12163 &   MS-Off &  92.14 & $-$25.34 & $-$121.31 & $-$2.49 &12212 &       $-$460,$-$350 &    $-$420 &  $<$18.29\\
69 &               UKS0242-724 &   Bridge & 292.17 & $-$42.00 & $-$10.82 & $-$4.98 &12263 &     \phs100,\phs340 &   \phs193 &     20.69\\
\vspace{-4mm}
\enddata
\tn{a}{Numerical ID of target, used to identify source on Figure 1.}
\tn{b}{Region probed by target, either Stream (MS), Bridge, Leading Arm (LA), LMC Halo, or CHVC (near HVC 224.0--83.4--197). On and Off refer to directions with and without Magellanic \hi\ 21\,cm emission.}
\tn{c}{Magellanic Longitude and Latitude, defined in \citet{Ni08}. The LMC is at $L_{\rm MS}$, $B_{\rm MS}$=0\degr, 0\degr.}
\tn{d}{\hst\ Program ID of COS dataset.}
\tn{e}{Minimum and maximum LSR velocities of Magellanic absorption, used for AOD integrations. Errors are $\pm$10\kms\ on each.}
\tn{f}{Central Magellanic LSR velocity, defined by 21\,cm emission when present, and by UV absorption otherwise. "..." denotes no UV detection.}
\tn{g}{\hi\ column density of Magellanic absorption measured from GASS, LAB, or Parkes data. Upper limits are 3$\sigma$, assuming a 40\kms\ wide line.}
\end{deluxetable*}

\begin{deluxetable*}{lcccc ccccc}
\tabletypesize{\scriptsize}
\tablewidth{0pt}\tabcolsep=2.0pt
\tablecaption{Apparent Column Densities and Ion Ratios in Magellanic Gas}
\tablehead{Target & Region & log\,$N$(\siw) & log\,$N$(\sit) & log\,$N$(\sif)
 & log\,$N$(\cw) & log\,$N$(\cf) & log\,$\frac{{\rm Si~III}}{{\rm Si~II}}$ & log\,$\frac{{\rm Si~IV}}{{\rm Si~II}}$ & log\,$\frac{{\rm C~IV}}{{\rm C~II}}$\\
 & & (cm$^{-2}$) & (cm$^{-2}$) & (cm$^{-2}$) & (cm$^{-2}$) & (cm$^{-2})$ & & &}
\startdata
                     3C57 &    MS-Off &                  $<$12.97 &        $<$12.42 &        $<$12.71 &  13.98$\pm$0.04 &        $<$13.25 &              \nodata &              \nodata &           $<$$-$0.73 \\
               ESO265-G23 &    LA-Off &            13.72$\pm$0.10 &  13.35$\pm$0.10 &        $<$13.37 &         \nodata &         \nodata &     $-$0.36$\pm$0.15 &           $<$$-$0.34 &              \nodata \\
               ESO267-G13 &    LA-Off &            13.22$\pm$0.20 &  13.20$\pm$0.20 &        $<$13.58 &        $<$14.40 &         \nodata &     $-$0.03$\pm$0.29 &          $<$\phs0.36 &              \nodata \\
                 ESO31-G8 &    Bridge &            14.63$\pm$0.10 &        $>$13.87 &  13.47$\pm$0.13 &         \nodata &  13.89$\pm$0.14 &           $>$$-$0.76 &     $-$1.17$\pm$0.17 &              \nodata \\
                 FAIRALL9 &     MS-On &            14.53$\pm$0.02 &  13.54$\pm$0.01 &  13.02$\pm$0.09 &        $>$14.88 &  13.49$\pm$0.06 &     $-$0.98$\pm$0.05 &     $-$1.51$\pm$0.10 &           $<$$-$1.39 \\
                H1101-232 &    LA-Off &            13.84$\pm$0.04 &        $>$13.44 &        $<$13.46 &        $>$14.51 &        $<$13.64 &           $>$$-$0.41 &           $<$$-$0.38 &           $<$$-$0.87 \\
              HE0153-4520 &    MS-Off &            12.98$\pm$0.09 &  12.83$\pm$0.04 &        $<$12.72 &  13.82$\pm$0.04 &        $<$13.25 &     $-$0.14$\pm$0.11 &           $<$$-$0.26 &           $<$$-$0.58 \\
              HE0226-4110 &    MS-Off &            13.12$\pm$0.05 &  13.26$\pm$0.01 &  13.03$\pm$0.07 &  14.28$\pm$0.01 &  13.88$\pm$0.02 &    \phs0.15$\pm$0.07 &     $-$0.08$\pm$0.10 &    $-$ 0.39$\pm$0.06 \\
              HE0238-1904 &    MS-Off &                  $<$12.70 &        $<$12.28 &        $<$12.91 &        $<$13.24 &        $<$13.09 &              \nodata &              \nodata &              \nodata \\
              HE0429-5343 &   LMC-Off &            13.42$\pm$0.13 &  13.24$\pm$0.09 &        $<$13.34 &        $<$14.25 &         \nodata &     $-$0.17$\pm$0.17 &           $<$$-$0.07 &              \nodata \\
              HE0435-5304 &   LMC-Off &            13.79$\pm$0.06 &        $>$13.67 &  13.22$\pm$0.12 &         \nodata &  13.62$\pm$0.16 &           $>$$-$0.12 &     $-$0.58$\pm$0.14 &              \nodata \\
              HE0439-5254 &   LMC-Off &            13.62$\pm$0.03 &  13.36$\pm$0.04 &        $<$13.77 &        $<$14.27 &  13.57$\pm$0.13 &     $-$0.27$\pm$0.07 &          $<$\phs0.15 &           $>$$-$0.71 \\
              HE1003+0149 &    LA-Off &                  $<$13.17 &        $<$12.87 &        $<$12.92 &        $<$13.88 &        $<$13.31 &              \nodata &              \nodata &              \nodata \\
              HE1159-1338 &    LA-Off &            13.38$\pm$0.12 &  13.31$\pm$0.07 &        $<$13.16 &         \nodata &         \nodata &     $-$0.07$\pm$0.14 &           $<$$-$0.22 &              \nodata \\
              HS0033+4300 &    MS-Off &                  $<$13.68 &  12.48$\pm$0.20 &         \nodata &  13.49$\pm$0.42 &         \nodata &           $>$$-$1.19 &              \nodata &              \nodata \\
                   IO-AND &    MS-Off &            12.92$\pm$0.08 &  12.85$\pm$0.04 &        $<$12.56 &  13.96$\pm$0.03 &        $<$13.12 &     $-$0.06$\pm$0.10 &           $<$$-$0.36 &           $<$$-$0.85 \\
           IRAS01003-2238 &    MS-Off &                   \nodata &        $<$13.01 &        $<$13.53 &        $<$14.18 &         \nodata &              \nodata &              \nodata &              \nodata \\
          IRASF09539-0439 &     LA-On &                  $>$14.05 &        $>$13.55 &  12.86$\pm$0.10 &        $>$14.86 &         \nodata &              \nodata &           $<$$-$1.19 &              \nodata \\
          IRASZ06229-6434 &   LMC-Off &            14.28$\pm$0.01 &        $>$13.92 &  13.29$\pm$0.05 &         \nodata &        $<$13.36 &           $>$$-$0.36 &     $-$1.00$\pm$0.07 &              \nodata \\
            LBQS0107-0233 &    MS-Off &            12.33$\pm$0.20 &  12.70$\pm$0.12 &        $<$13.11 &        $<$13.59 &  13.29$\pm$0.17 &    \phs0.37$\pm$0.24 &          $<$\phs0.79 &           $>$$-$0.30 \\
            LBQS0107-0235 &    MS-Off &            13.05$\pm$0.19 &  12.87$\pm$0.10 &        $<$13.14 &  13.67$\pm$0.15 &  13.71$\pm$0.07 &     $-$0.18$\pm$0.22 &          $<$\phs0.10 &   \phs 0.05$\pm$0.17 \\
            LBQS1019+0147 &    LA-Off &                  $<$13.38 &        $<$12.85 &        $<$13.15 &        $<$14.17 &        $<$13.72 &              \nodata &              \nodata &              \nodata \\
              MRC2251-178 &    MS-Off &                  $<$12.96 &  12.99$\pm$0.03 &  13.04$\pm$0.06 &  13.29$\pm$0.12 &  13.71$\pm$0.04 &          $>$\phs0.03 &          $>$\phs0.08 &   \phs 0.42$\pm$0.13 \\
                  MRK1014 &    MS-Off &                  $<$12.87 &        $<$12.39 &        $<$12.99 &        $<$13.53 &         \nodata &              \nodata &              \nodata &              \nodata \\
                  MRK1044 &    MS-Off &                  $<$12.96 &  12.73$\pm$0.07 &        $<$12.97 &  13.58$\pm$0.11 &        $<$13.43 &           $>$$-$0.24 &              \nodata &           $<$$-$0.15 \\
                  MRK1502 &    MS-Off &                  $<$13.30 &        $<$13.02 &        $<$13.36 &        $<$13.85 &         \nodata &              \nodata &              \nodata &              \nodata \\
                  MRK1513 &    MS-Off &                  $<$12.91 &  13.00$\pm$0.04 &  12.95$\pm$0.09 &  13.48$\pm$0.10 &  14.04$\pm$0.04 &          $>$\phs0.09 &          $>$\phs0.04 &   \phs 0.56$\pm$0.12 \\
                   MRK304 &    MS-Off &                  $<$13.03 &  13.26$\pm$0.03 &  12.91$\pm$0.10 &  13.70$\pm$0.06 &  13.85$\pm$0.05 &          $>$\phs0.23 &           $>$$-$0.12 &   \phs 0.15$\pm$0.10 \\
                   MRK335 &    MS-Off &                  $<$13.09 &  12.83$\pm$0.05 &        $<$12.90 &        $<$13.42 &  13.62$\pm$0.06 &           $>$$-$0.26 &              \nodata &          $>$\phs0.20 \\
              MS0117-2837 &  CHVC-Off &            13.21$\pm$0.08 &  13.39$\pm$0.03 &  13.05$\pm$0.08 &  14.11$\pm$0.04 &  13.71$\pm$0.06 &    \phs0.18$\pm$0.10 &     $-$0.16$\pm$0.12 &    $-$ 0.40$\pm$0.08 \\
                  NGC3125 &    LA-Off &            13.32$\pm$0.08 &  13.23$\pm$0.19 &        $<$13.34 &        $<$14.43 &         \nodata &     $-$0.09$\pm$0.21 &          $<$\phs0.02 &              \nodata \\
                  NGC3783 &     LA-On &            13.86$\pm$0.02 &  13.02$\pm$0.04 &        $<$12.86 &         \nodata &        $<$13.24 &     $-$0.84$\pm$0.07 &           $<$$-$1.01 &              \nodata \\
                  NGC7714 &     MS-On &            13.65$\pm$0.11 &  13.69$\pm$0.10 &  13.51$\pm$0.11 &  14.46$\pm$0.10 &  13.93$\pm$0.12 &    \phs0.04$\pm$0.16 &     $-$0.14$\pm$0.16 &    $-$ 0.53$\pm$0.16 \\
               PG0003+158 &    MS-Off &            13.36$\pm$0.07 &  13.34$\pm$0.03 &         \nodata &         \nodata &  14.01$\pm$0.03 &     $-$0.02$\pm$0.09 &              \nodata &              \nodata \\
               PG0026+129 &    MS-Off &                  $<$12.95 &  12.71$\pm$0.09 &        $<$12.68 &        $<$13.64 &         \nodata &           $>$$-$0.24 &              \nodata &              \nodata \\
               PG0044+030 &    MS-Off &            13.31$\pm$0.19 &  12.93$\pm$0.19 &        $<$13.25 &  14.02$\pm$0.12 &         \nodata &     $-$0.37$\pm$0.28 &           $<$$-$0.05 &              \nodata \\
               PG1011-040 &    LA-Off &            12.44$\pm$0.08 &  12.57$\pm$0.06 &        $<$12.74 &  13.60$\pm$0.05 &        $<$13.14 &    \phs0.13$\pm$0.11 &          $<$\phs0.30 &           $<$$-$0.46 \\
               PG1049-005 &    LA-Off &            13.88$\pm$0.07 &  13.51$\pm$0.07 &        $<$13.22 &  14.67$\pm$0.06 &  13.59$\pm$0.16 &     $-$0.37$\pm$0.11 &           $<$$-$0.66 &    $-$ 1.08$\pm$0.18 \\
               PG2349-014 &    MS-Off &            13.35$\pm$0.07 &        $>$13.73 &  13.44$\pm$0.04 &  14.44$\pm$0.03 &         \nodata &          $>$\phs0.37 &    \phs0.09$\pm$0.10 &              \nodata \\
                  PHL1811 &    MS-Off &            13.37$\pm$0.04 &  13.60$\pm$0.01 &  13.46$\pm$0.03 &  14.29$\pm$0.02 &  14.30$\pm$0.02 &    \phs0.23$\pm$0.07 &    \phs0.09$\pm$0.07 &   \phs 0.02$\pm$0.06 \\
                  PHL2525 &     MS-On &            13.38$\pm$0.07 &        $>$13.70 &  13.19$\pm$0.09 &  14.38$\pm$0.03 &  14.01$\pm$0.05 &          $>$\phs0.32 &     $-$0.19$\pm$0.12 &    $-$ 0.37$\pm$0.08 \\
               PKS0202-76 &    Bridge &            14.60$\pm$0.07 &         \nodata &  13.18$\pm$0.24 &        $>$15.23 &  13.70$\pm$0.20 &              \nodata &     $-$1.42$\pm$0.26 &           $<$$-$1.53 \\
              PKS0439-433 &   LMC-Off &            14.06$\pm$0.03 &  13.29$\pm$0.03 &  13.16$\pm$0.07 &        $<$13.98 &         \nodata &     $-$0.77$\pm$0.07 &     $-$0.89$\pm$0.09 &              \nodata \\
              PKS0552-640 &    LMC-On &            13.98$\pm$0.03 &        $>$13.86 &        $<$13.10 &         \nodata &  13.83$\pm$0.04 &           $>$$-$0.12 &           $<$$-$0.88 &              \nodata \\
              PKS0558-504 &   LMC-Off &            12.82$\pm$0.20 &  12.68$\pm$0.10 &        $<$13.23 &        $<$14.06 &  13.24$\pm$0.24 &     $-$0.15$\pm$0.23 &          $<$\phs0.41 &           $>$$-$0.82 \\
               PKS0637-75 &    LMC-On &            14.24$\pm$0.03 &  13.82$\pm$0.03 &        $<$13.31 &         \nodata &  13.84$\pm$0.05 &     $-$0.42$\pm$0.06 &           $<$$-$0.93 &              \nodata \\
               PKS0932+02 &    LA-Off &                  $<$13.60 &        $<$13.08 &        $<$13.43 &        $<$14.21 &        $<$13.91 &              \nodata &              \nodata &              \nodata \\
              PKS1101-325 &    LA-Off &                  $<$13.18 &        $<$12.76 &        $<$12.81 &        $<$14.20 &         \nodata &              \nodata &              \nodata &              \nodata \\
               PKS1136-13 &    LA-Off &            12.57$\pm$0.20 &  12.36$\pm$0.17 &        $<$12.77 &  13.62$\pm$0.09 &         \nodata &     $-$0.21$\pm$0.27 &          $<$\phs0.19 &              \nodata \\
             QSO1220-2859 &   CHVC-On &            13.46$\pm$0.06 &  13.32$\pm$0.04 &  13.08$\pm$0.10 &        $<$14.39 &  13.66$\pm$0.11 &     $-$0.15$\pm$0.09 &     $-$0.38$\pm$0.13 &           $>$$-$0.74 \\
                   RBS144 &     MS-On &            13.86$\pm$0.03 &        $>$13.56 &        $<$13.10 &        $>$14.60 &        $<$13.51 &           $>$$-$0.30 &           $<$$-$0.76 &           $<$$-$1.08 \\
                  RBS1795 &    MS-Off &                  $<$12.50 &        $<$12.51 &        $<$12.60 &        $<$13.29 &        $<$13.49 &              \nodata &              \nodata &              \nodata \\
                  RBS1892 &    MS-Off &                  $<$12.53 &        $<$12.37 &        $<$12.80 &        $<$13.57 &        $<$13.26 &              \nodata &              \nodata &              \nodata \\
                  RBS1897 &    MS-Off &            12.60$\pm$0.12 &  12.57$\pm$0.04 &        $<$12.48 &        $<$13.00 &        $<$13.20 &     $-$0.03$\pm$0.14 &           $<$$-$0.12 &              \nodata \\
                   RBS542 &   LMC-Off &            13.29$\pm$0.02 &  13.46$\pm$0.00 &  12.97$\pm$0.08 &         \nodata &  13.53$\pm$0.02 &    \phs0.17$\pm$0.05 &     $-$0.32$\pm$0.10 &              \nodata \\
                   RBS563 &    LMC-On &            14.53$\pm$0.07 &        $>$14.08 &  13.85$\pm$0.04 &         \nodata &  14.36$\pm$0.07 &           $>$$-$0.45 &     $-$0.68$\pm$0.10 &              \nodata \\
                   RBS567 &   LMC-Off &            13.17$\pm$0.12 &  13.38$\pm$0.03 &        $<$13.13 &        $<$14.26 &        $<$13.56 &    \phs0.22$\pm$0.13 &           $<$$-$0.03 &              \nodata \\
           RXJ0209.5-0438 &    MS-Off &                  $<$13.43 &         \nodata &  12.86$\pm$0.14 &  14.03$\pm$0.05 &  13.30$\pm$0.12 &              \nodata &           $>$$-$0.57 &    $-$ 0.73$\pm$0.14 \\
           RXJ0503.1-6634 &    LMC-On &                  $>$14.82 &        $>$14.14 &  13.38$\pm$0.11 &         \nodata &  13.67$\pm$0.14 &              \nodata &           $<$$-$1.45 &              \nodata \\
           SDSSJ0012-1022 &     MS-On &            13.58$\pm$0.16 &        $>$13.79 &  13.50$\pm$0.15 &  14.53$\pm$0.17 &  14.02$\pm$0.13 &          $>$\phs0.21 &     $-$0.08$\pm$0.23 &    $-$ 0.51$\pm$0.22 \\
           SDSSJ0155-0857 &    MS-Off &            13.44$\pm$0.08 &  13.10$\pm$0.05 &        $<$13.08 &  14.13$\pm$0.09 &        $<$13.61 &     $-$0.35$\pm$0.11 &           $<$$-$0.36 &           $<$$-$0.52 \\
           SDSSJ0212-0737 &    MS-Off &                  $<$13.19 &        $<$12.61 &        $<$13.11 &        $<$13.79 &        $<$13.60 &              \nodata &              \nodata &              \nodata \\
           SDSSJ0943+0531 &    LA-Off &            12.81$\pm$0.20 &  12.71$\pm$0.20 &        $<$13.34 &  13.55$\pm$0.30 &        $<$13.77 &     $-$0.11$\pm$0.29 &          $<$\phs0.53 &          $<$\phs0.22 \\
           SDSSJ0959+0504 &     LA-On &                  $<$13.05 &  12.46$\pm$0.15 &        $<$12.99 &        $<$13.99 &  13.52$\pm$0.13 &           $>$$-$0.58 &              \nodata &           $>$$-$0.47 \\
           SDSSJ2257+1340 &    MS-Off &            13.61$\pm$0.16 &        $>$13.63 &  13.76$\pm$0.09 &  14.36$\pm$0.10 &  14.25$\pm$0.09 &          $>$\phs0.02 &    \phs0.15$\pm$0.19 &    $-$ 0.11$\pm$0.14 \\
           SDSSJ2345-0059 &     MS-On &            13.75$\pm$0.11 &  13.52$\pm$0.20 &  13.41$\pm$0.12 &  14.38$\pm$0.11 &        $>$13.87 &     $-$0.23$\pm$0.23 &     $-$0.34$\pm$0.17 &           $>$$-$0.51 \\
                  TONS210 &   CHVC-On &            13.47$\pm$0.03 &  13.40$\pm$0.01 &  13.02$\pm$0.05 &  14.28$\pm$0.01 &  13.79$\pm$0.03 &     $-$0.08$\pm$0.06 &     $-$0.45$\pm$0.07 &    $-$ 0.49$\pm$0.06 \\
                 UGC12163 &    MS-Off &                  $<$13.47 &  13.24$\pm$0.10 &  13.24$\pm$0.15 &  14.02$\pm$0.10 &  13.58$\pm$0.17 &           $>$$-$0.24 &           $>$$-$0.24 &    $-$ 0.44$\pm$0.21 \\
              UKS0242-724 &    Bridge &            14.65$\pm$0.06 &        $>$14.02 &  13.46$\pm$0.10 &        $>$15.08 &  13.86$\pm$0.09 &           $>$$-$0.63 &     $-$1.19$\pm$0.12 &           $<$$-$1.22 \\
\vspace{-3mm}
\enddata
\tablecomments{All measurements of apparent column densities are made in the LSR velocity intervals given in Table 1. Upper limits are 3$\sigma$ and are based on non-detections. Lower limits are 1$\sigma$ and indicate saturated lines. \cw\ column densities are given as upper limits in sightlines where Magellanic absorption in \cw\ $\lambda$1334 blends with Galactic \cw$^*$ $\lambda$1335 absorption. No entry (...) indicates no data coverage. Full names of SDSS quasars are given in Table 1.}
\end{deluxetable*}

\subsection{\hst/COS Observations}
For each direction in our survey, the raw COS spectra were downloaded from the 
MAST archive in the form of {\tt x1d} fits files.
Overlapping exposures were aligned in wavelength space
by matching the centroids of low-ionization interstellar absorption lines,
combined by calculating the total count rate in each pixel,
and converted back to flux after co-addition. Errors were calculated
in the coadded spectra by weighting by inverse variance.
Each coadded spectrum was shifted to the LSR velocity scale
using the centroid of the Galactic 21\,cm emission in that direction
from the LAB survey \citep{Ka05} as the reference point.

In addition, for six directions with Magellanic \hi\ column densities
high enough for \oi\ $\lambda$1302 absorption to be detectable
(\object{RBS\,144}, \object{PHL\,2525}, \object{NGC\,7714}, 
\object{QSO1220--2859}, \object{TON\,S210}, and \object{MS0117--2837}),
we performed an orbital night-only reduction of the COS data to 
remove the geocoronal airglow emission that contaminates this line. 
To do this, the spectra were extracted only using the 
time intervals when the Sun altitude as seen by the telescope was less 
than 20\degr. This criterion selects low-background intervals corresponding
to the night-side portion of the \emph{HST} orbit.
The filtering of the night-only data was done using the timefilter
module in the {\tt costools} package, and the spectral re-extraction 
(for these data only) was done with the standard {\tt CalCOS} pipeline.

The spectra were normalized around each absorption line of interest by fitting
low-order polynomials to the local continuum, using intervals several
100\kms\ wide on either side of the line. 
The spectra have a velocity resolution $R\approx$15\,000--20\,000 
(instrumental FWHM$\approx$15--20\kms) depending on wavelength,
and a native pixel size of 3\kms\ \citep{Ho14}. 
For analysis and display on the figures, the COS spectra were rebinned by 
either three or five native pixels, depending on the S/N ratio.
Spectra with medium or good S/N ($>$10 per resolution element) were rebinned 
by three pixels, giving final spectra that are Nyquist sampled.
Spectra with low S/N ($<$10 per resolution element) were rebinned by five 
pixels, giving final spectra with close to one pixel per resolution element. 

\subsubsection{Identification of Magellanic UV Absorption}
For On-System (21\,cm-bright) sightlines, the Magellanic velocity 
is known from the \hi\ velocity field of the MSys \citep[][N08]{Pu03a}.
In these directions, the Magellanic velocity intervals 
$v_{\rm min}$ to $v_{\rm max}$ 
were identified from the \hi\ profiles, by finding the pixels where the 
21\,cm emission rises above the baseline. 
For Off-System (21\,cm-faint) sightlines, we searched the COS spectra for 
high-velocity ($|v_{\rm LSR}|\!>\!100$\kms) absorption components, and when
such components were found, we compared their velocities 
with the Magellanic velocity \emph{expected} for that sightline's 
Magellanic Longitude based on the 21\,cm velocity field.
When a match was found, the absorption component was labeled as Magellanic,
and $v_{\rm min}$ and $v_{\rm max}$ were defined by the UV absorption.
This approach assumes that the Magellanic velocity intervals do not 
vary over scales of $\la$30\degr\ perpendicular to the Stream, 
which is reasonable 
since the velocity gradient is largest along the Stream, not tangential to it 
\citep[][N08]{Pu03a}. This can be seen in Figure 1b: the \hi\ fragments
to the North and South of the body of the MS reflect the velocity field
of the MS itself.

The MSys map color-coded by central velocity (Figure 1b) illustrates 
how we can be confident that the high-velocity 
absorption features analyzed in this paper are Magellanic in origin.
Notice the close correspondence in color between the velocity field of the 
21\,cm emission (background shading) and the UV absorption (colored circles) 
in the On-System directions.
Also note that as one moves to the North and South (in $B_{\rm MS}$) from the 
Stream, the velocity field of the absorbers matches the velocity along
the axis of the Stream.
Finally, notice the correspondence in color between the 21\,cm emission
and the UV absorption in the LA sightlines.
\emph{In summary, the UV absorption-line detections kinematically trace the 
MSys \hi\ velocity field, even at impact parameters of up to 30\degr\
from directions where 21\,cm emission is detected.}

\subsubsection{Measurement of Magellanic UV Absorption}
Measurements of the apparent column density of each absorbing ion in the COS 
data were made using the AOD technique of \citet{SS91} and are given in 
Table 2. The AOD in each pixel is given by $\tau_a(v)$=ln\,$[F_c(v)/F(v)]$,
where $F_c(v)$ is the continuum level and $F(v)$ is the flux
in that pixel. The apparent column density per pixel is defined by:

\begin{equation}
N_a(v)= \frac{m_e c}{\pi e^2}\frac{\tau_a(v)}{f\lambda} 
= 3.768\times10^{14}\frac{\tau_a(v)}{f\lambda}{\rm cm}^{-2}({\rm km\,s}^{-1})^{-1},
\end{equation}

where $f$ is the transition oscillator strength and $\lambda$ is
the rest wavelength in Angstroms. The apparent column density in the line
is obtained by integrating over the line profile,
$N_a=\int_{v_{\rm min}}^{v_{\rm max}}N_a(v){\rm d}v$. In On-System
directions, the same $v_{\rm min}$ and $v_{\rm max}$ values were used for the 
UV and 21\,cm measurements; in Off-System directions, 
$v_{\rm min}$ and $v_{\rm max}$ were defined by the UV data only.

The AOD technique returns accurate estimates of
the true column density of a given species provided the absorption-line 
profiles are resolved and unsaturated. 
While resolved saturation is readily apparent, unresolved saturation can be 
difficult to detect. When resolved saturation is seen in a given line, which
we define as occurring whenever the normalized flux falls below 0.1 anywhere
within the line profile, we treat the apparent column density of that ion
as a lower limit in the analysis. Table 2 also presents the 
apparent-column-density ratios between several key ions; these ratios are 
used in the ionization modeling (\S4.1).

\subsection{21 cm \hi\ spectra}
We use publicly available 21\,cm \hi\ spectra from the 
LAB survey \citep{Ka05} and the Galactic All-Sky Survey 
\citep[GASS;][]{MG09, Ka10} to derive \hi\ column densities (or upper limits) 
in each Magellanic sightline. When both LAB and GASS data
are available (which is true for sightlines with southern declinations),
we use the GASS data because of their smaller beam size.
In the case of the three sightlines labeled CHVC in Table 1, we instead
use additional data obtained from the Parkes radio telescope. 
The LAB data have 1.3\kms\ velocity resolution and 30--36\arcmin\
angular resolution, the GASS data have 0.8\kms\ 
velocity resolution and 14.4\arcmin\ angular resolution, and
the additional Parkes data have 0.4\kms\ velocity 
resolution and 14.4\arcmin\ angular resolution. 
For both the LAB and the GASS data we adopt the closest pointing to each
target direction, % Bart: is that right, or are they interpolated?
whereas the additional Parkes data were taken precisely 
along the three CHVC sightlines.
The Magellanic \hi\ columns were determined by integrating the 21\,cm 
profiles using the equation:

\begin{equation}
N({\rm H\,I})=1.823\times10^{18}{\rm cm}^{-2}\int^{v_{\rm max}}_{v_{\rm min}} T_B\,{\rm d}v, 
\end{equation}

where $T_B$ is the brightness temperature in K and
the line is assumed to be optically thin.
%These velocities are identified by eye upon inspection of the 21\,cm profiles.
In Off-System directions, the 21\,cm non-detections are used to place 
3$\sigma$ upper limits on the Magellanic \hi\ column, assuming a line width 
of 40\kms.

\subsection{WHAM \ha\ emission-line spectra}
WHAM measurements of the \ha\ emission
in a one-degree beam surrounding \nwham\ of our sample directions 
(six MS directions and three MB directions) were taken in 2011 and 
2012. The MS and MB observations are described in detail in \citet{Ba14}
and \citet{Ba13}, respectively. 
WHAM \citep{Ha03} is a dual-\'etalon Fabry-Perot spectrometer with 
a velocity resolution of 12\kms\ (FWHM). The bandpass 
was tuned to $\pm$100\kms\ around the central MS velocity in each
sightline. The \ha\ profiles are presented in the form of % in Figure 3
intensity in milli-Rayleighs (mR) against LSR velocity, where 
1 Rayleigh = 10$^6$/4$\pi$\,photons\,s$^{-1}$\,cm$^{-2}$\,sr$^{-1}$,
which is $\approx$10$^{-6}$\,erg\,s$^{-1}$\,cm$^{-2}$\,sr$^{-1}$ at \ha.

\section{Overview of Magellanic absorption}
To illustrate the quality and nature of the data analyzed in this paper, 
in Figure 2 we present the COS absorption-line profiles of either \sw\ 
$\lambda$1253 or \sit\ $\lambda$1206 and the \hi\ 21\,cm emission profiles 
for six sightlines selected to represent the various components of the MSys.
The \object{Fairall\,9} spectrum shows multiple-component UV absorption 
from the MS at positive velocities with clear accompanying 21\,cm emission 
(see Paper II).
Toward \object{Mrk\,1513}, strong negative-velocity UV absorption is observed
from the MS \emph{without} accompanying 21\,cm emission, defining this as an 
Off-Stream direction.
The \object{3C57} spectrum shows no Magellanic UV absorption, even in the highly
sensitive \sit\ $\lambda$1206 line.
Toward \object{B0242--0724}, multiple-component UV absorption from the MB
is observed at positive velocities with a profile closely following the 
21\,cm emission.
The \object{NGC\,3783} spectrum displays multiple-component positive-velocity 
UV absorption from the LA, with strong 21\,cm component at the same velocity.
The \object{TON\,S210} spectrum shows strong UV absorption and 21\,cm 
emission at --205\kms\ from the CHVC \citep{Se02, Ri09, Ku14}.

\begin{figure*}
\epsscale{1.2}\plotone{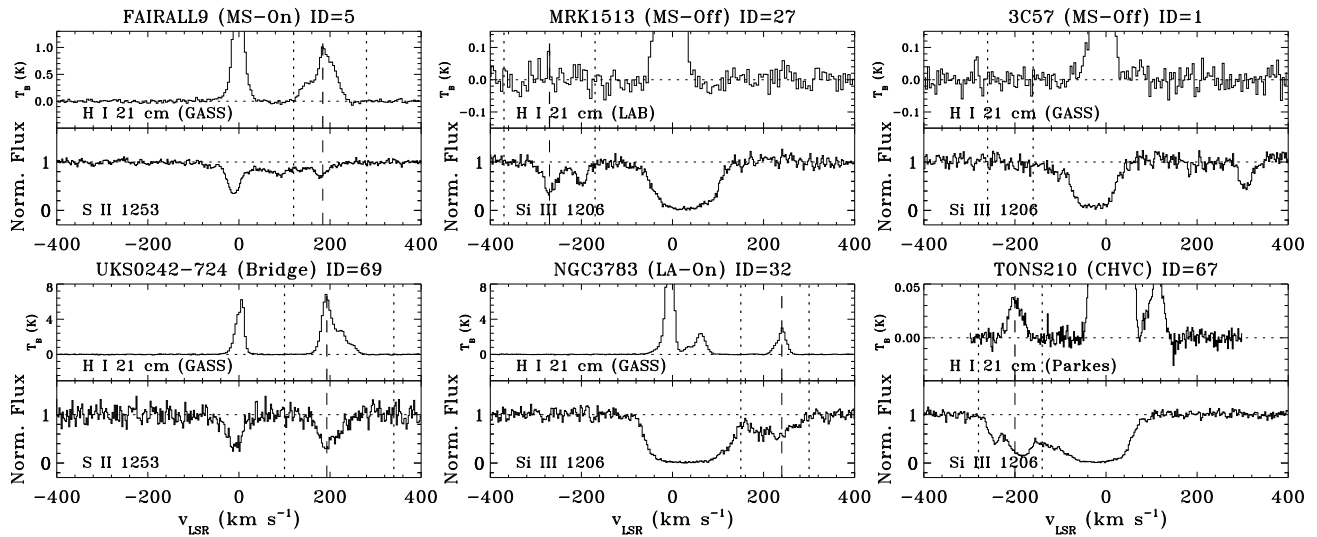}
\caption{Example of COS absorption-line and 21\,cm profiles
in six MSys directions. 
The chosen sightlines sample different Magellanic regions: 
On-Stream (upper left), Off-Stream with a UV detection (upper middle), 
Off-Stream without a UV detection (upper right),
Magellanic Bridge (lower left), Leading Arm (lower middle), and CHVC 
(lower right). The numerical ID in the title of each panel can be used 
to locate the sightline in Figure 1.
The COS data are shown as normalized flux against LSR velocity, in each
case showing a line that is unsaturated at Magellanic velocities.
The 21\,cm profiles are shown as brightness temperature against LSR velocity
and are taken from the LAB survey, the GASS survey, or the Parkes telescope. 
Dashed vertical lines show the central Magellanic velocity in each direction, 
and dotted lines show the velocity integration range.
Note how the MS velocities can be highly positive (as toward Fairall\,9)
or highly negative (as toward Mrk\,1513), due to the large LSR velocity 
gradient along the Stream.}
\end{figure*}

A larger set of absorption-line profiles for each sightline in our sample 
is shown in the appendix, together with the \hi\ 21\,cm emission profile.
The default UV lines included are \cw\ $\lambda$1334, \siw\ $\lambda$1193, 
\sit\ $\lambda$1206, \sif\ $\lambda$1393, and \cf\ $\lambda$1548. 
These lines were chosen because they are among the strongest available 
indicators of low-ion (\cw\ and \siw), intermediate-ion 
(\sit), and high-ion (\cf, \sif) absorption from the MSys.
\siw\ $\lambda$1193 is used instead of the stronger line 
\siw\ $\lambda$1260, because in directions where MSys absorption is at 
negative LSR velocities, MSys absorption in \siw\ $\lambda$1260 is blended with 
Galactic ISM absorption in \sw\ $\lambda$1259.
However, in directions where MSys absorption is at positive LSR velocities,
we use \siw\ $\lambda$1260 when it gives a better indication of the \siw\ 
profile. In cases where the default lines are blended or saturated, we 
use \cf\ $\lambda$1550 instead of \cf\ $\lambda$1548, 
\sif\ $\lambda$1402 instead of \sif\ $\lambda$1393, and 
\siw\ $\lambda$1526 instead of \siw\ $\lambda$1193.
When MSys absorption in \cw\ $\lambda$1334 is blended with Galactic ISM 
absorption in \cw$^*$ $\lambda$1335 (which occurs when the MSys absorption is
near $v_{\rm LSR}$=+260\kms), only an upper limit on the MSys \cw\ column 
density can be derived, unless the absorption is saturated, in which case no 
limit on $N$(\cw) can be made.

In a substantial fraction of sightlines, the Magellanic absorption consists
of multiple components (see Figure 2, top-left panel).
This indicates that the gas is clustered in velocity and hence is likely
to be clustered spatially as well. In the extreme case of \object{Fairall\,9},
seven components are seen in high-resolution VLT/UVES data of \caw\ (Paper II). 
In our analysis, we present the \emph{integrated} column densities, rather than 
component-by-component column densities, and our ionization modeling treats
the Magellanic gas in each direction as arising in a single cloud.
This is mainly because the COS spectral resolution is not high enough to 
fully resolve the cool-gas component structure, and our UVES observations
only cover a subset of 13 sightlines (Paper I, Paper II). 

\subsection{Detection Rate of Magellanic Absorption}
Of the \nsl\ sightlines in our \hst/COS sample,
\ndet\ (\msper) show a Magellanic detection, which we
define as significant absorption (3$\sigma$) at Magellanic 
velocities in a UV metal line. 
We refer the reader back to Figure 1b for an illustration of how
the Magellanic velocities are identified, and how the UV velocity
field closely follows the 21\,cm velocity field.
The lines detected vary between absorbers (see the appendix), 
but almost always include \sit\ $\lambda$1206, the most sensitive
UV tracer of ionized gas available, and \cw\ $\lambda$1334. 
\siw\ $\lambda$1260 or \siw\ $\lambda$1193 are often detected.

Given that we selected sightlines lying within 30\degr\ of the \hi\ emission, 
and that the \hi\ cross-section of the MSys $A$(\hi)$_{\rm MS}$ is well known, 
we can use our UV detection rate to estimate the total cross-section of the 
MSys. $A$(\hi)$_{\rm MS}$ is 2\,701 square degrees down
to a limiting column density $N$(\hi)=1$\times$10$^{18}$\sqcm\ \citep{Ni10}.
By finding the area on the sky in which Magellanic UV absorption 
is detected (determined by summing the area of the non-shaded grid cells in 
Figure 1b), and multipling by the covering fraction of \msper, the 
total cross-section of the MSys is $\approx$11\,000 square degrees, %*
of which $\approx$three-quarters is only seen in UV absorption. %*
\emph{The MSys therefore covers approximately one quarter of the 
entire sky of 41\,253 square degrees}.

There is no clear break in the detection rate at 
impact parameters of 30\degr\ from the Magellanic \hi\ contours, so it is 
possible that the true cross section is even higher, but we notice
that non-detections are more likely to be found at at high impact parameter.
The UV detection rate can be broken down by region:
among the MB, CHVC, and LMC halo sightlines, the detection rate is 100\% 
(3/3 for the MB, 3/3 for the CHVC, and 12/12 for the LMC halo). 
Among the MS sightlines, the UV detection rate is 74\% (26/35). %*
Among the LA sightlines, the UV detection rate is 75\% (12/16). %*

The UV non-detections are shown in Figure 1 as circles with black crosses,
and they all lie in Off-System directions, as expected, since the \hi\ 
column densities in the On-System directions are high enough that many 
UV metal lines are detectable in those sightlines. The MS non-detections 
tend to be found either to the North or South (in MS coordinates, as used 
in Figure 1) of the main body of the Stream. In particular, two clusters of  
non-detections are seen, one pair of sightlines 
(\object{RBS\,1795} and \object{RBS\,1892})
near $L_{\rm MS}$=--40\degr\ $B_{\rm MS}$=--25\degr, 
and a group of four sightlines 
(\object{3C57}, \object{HE0238--1904}, \object{SDSSJ021218.32--073719.8},
and \object{Mrk\,1014})
near $L_{\rm MS}$=--65\degr, $B_{\rm MS}$=+30\degr. 
Notably, three MS \emph{detections} are also seen in this latter region, 
indicating the gas there is clumpy with a non-unity covering fraction. 

We report four LA sightlines with UV non-detections 
(\object{HE1003+0149}, \object{LBQS1019+0147}, \object{PKS0932+02}, 
and \object{PKS1101--325}).
High-velocity absorption between +100 and +200\kms\ \emph{is} detected in 
each of these directions, but these velocities are consistent with the 
foreground HVC complexes WA and WB \citep{Wa01}, rather than the LA at 
$\ga$200\kms. While we cannot rule out an association between these 
complexes and the LA, we conservatively classify these sightlines
as LA non-detections for the purposes of identifying Magellanic material.
We also note that the sightlines toward \object{IO--AND} and 
\object{HS0033+4300}, which lie in the MS tip region,
pass through the extended halo of \object{M31}, 
which has a systemic velocity of $-$300\kms\ and a known velocity field
\citep{Ch09}. For \object{IO--AND}, multiple HVC components are seen.
We identify the $-$370\kms\ component as the MS component, since
this is the MS velocity in this part of the sky, and the absorption between
$-$300 and $-$130\kms\ as \object{M31} components. Therefore we count 
this sightline as an MS detection.
For \object{HS0044+4300}, high-velocity absorption is seen between $-$270 and
$-$130\kms. Since genuine MS absorption in this part of the sky would
be closer to $-$400\kms, these components could be entirely due to the M31 halo
and its surrounding HVCs \citep{Th04}, and so we classify
this sightline as an MS non-detection.

\subsection{Kinematic Comparison of Neutral and Warm-Ionized Phases}

\begin{figure*}
\epsscale{1.2}\plotone{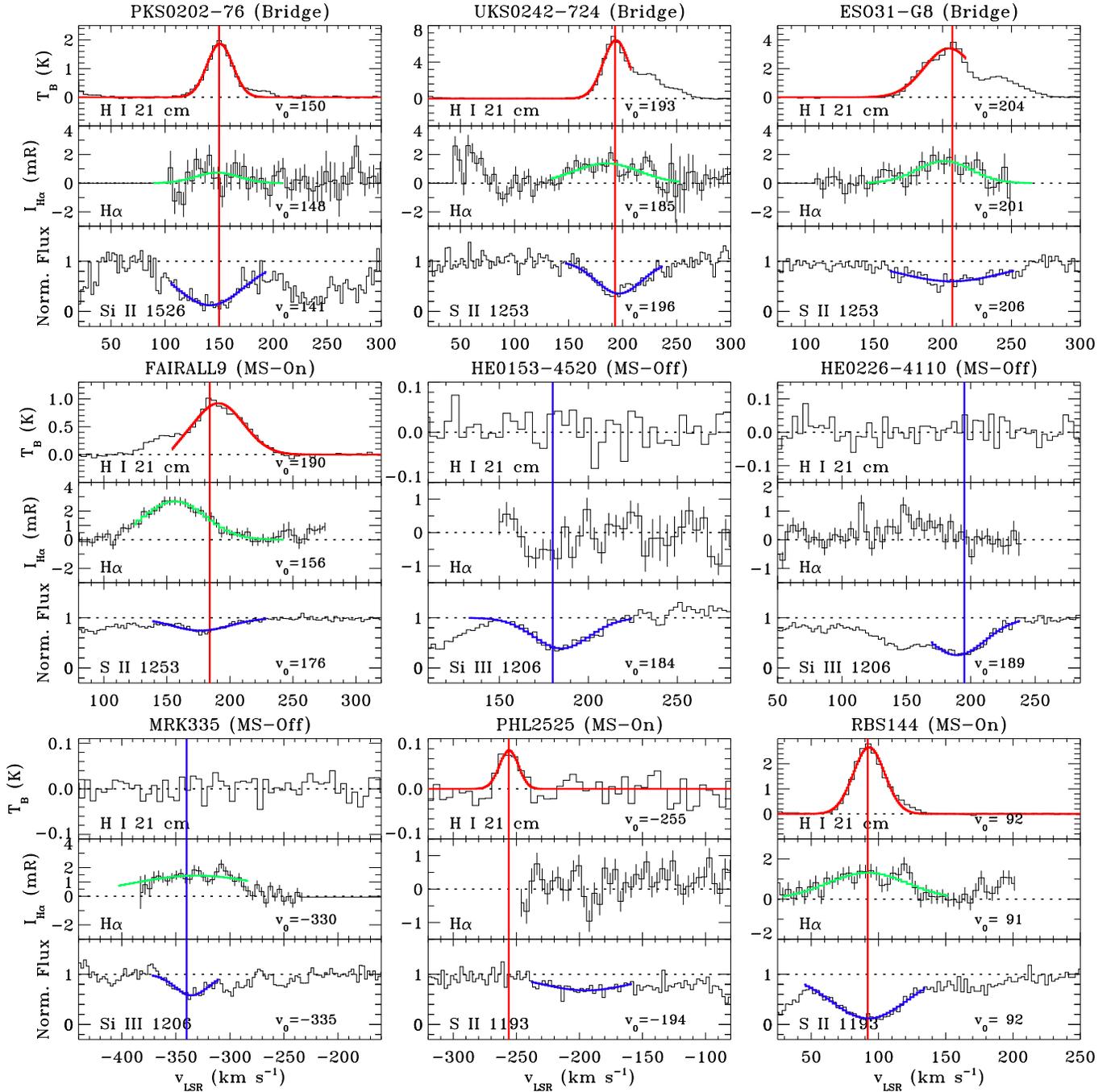}
\caption{Comparison of 21\,cm emission (GASS or LAB), \ha\ emission (WHAM), 
and UV metal-line absorption (COS) profiles in the nine sightlines with
\ha\ data covering Magellanic velocities.
In each panel, solid colored lines show Gaussian fits to the strongest 
component of emission or absorption. The velocity centroids of the fits 
are annotated on each panel, in units of km\,s$^{-1}$. 
The UV metal line shown in each panel is chosen to be
unsaturated through the MS velocity interval.
In each direction, a larger selection of UV absorption lines is shown in the 
appendix. The WHAM \ha\ intensities are in units of milli-Rayleighs (mR).
Red vertical lines indicate the centroid of the Magellanic 21\,cm emission; 
blue lines indicate the centroid of the Magellanic UV absorption.}
\end{figure*}

Our multi-wavelength dataset allows us to inter-compare the profiles
of emission or absorption from the various phases of the MSys in the 
same directions. In Figure 3, we present such a comparison for the nine 
sightlines with WHAM data covering Magellanic velocities.
The plot compares the profiles of \hi, \ha, and an unsaturated low-ionization 
UV metal-absorption line. 

In five directions, namely the three MB sightlines \object{PKS0202--76}, 
\object{UKS0242--724}, and \object{ESO\,31--G08}, and the high-column-density
sightlines \object{Fairall\,9} and \object{RBS\,144}, 
\hi\ emission, \ha\ emission, and UV metal-line
absorption are \emph{all} detected at Magellanic velocities.
These five directions are the most useful for assessing 
the connection in velocity between the various phases of the Magellanic gas.
In each direction, Gaussian fits show that the
\ha\ line widths are broader than the 21\,cm \hi\ line widths.
This reflects two effects: (1) the ionized gas is warmer and/or more 
kinematically disturbed; (2) the lower WHAM spatial resolution causes 
multiple components to blend together within the 1\degr\ beam.
For the three MB directions, the \hi, \ha, 
and \sw\ profiles are all centered within 10\kms\ of each other.
For the \object{RBS\,144} sightline, the \hi, \ha, and \sw\ profile centroids 
agree to within 1\kms. 

For the \object{Fairall\,9} direction, the profiles are more complex.
A significant offset (34\kms) is seen between the strongest 21\,cm component 
centered at 190\kms\ and the \ha\ component centered at 156\kms, with the \sw\ 
$\lambda$1253 centroid falling in-between at 178\kms\ 
%(see K. A. Barger et al. 2014, in prep.)
\citep[see][]{Ba14}.
In Paper II, we reported \cw$^*$ $\lambda$1335 absorption at 150\kms, 
\emph{very close to the velocity where the \ha\ emission peaks}.
This confirms the presence of ionized gas that is offset
in velocity from the neutral gas. The strong \ha\ emission and \cw$^*$ 
absorption indicate that the ionized gas has a relatively high electron density 
and recombination rate. This could indicate that the ionized gas exists 
in a boundary layer being compressed by ram pressure as the neutral core moves 
through an external medium. The presence of a dense neutral core 
in the \object{Fairall\,9} direction is known from observations of
H$_2$ absorption \citep{Ri01} and \nao\ absorption (Paper II).

\subsection{Observed Ion Ratios}

In Table 2, we assess the ionization level in the Magellanic gas 
using three column-density ratios: \sit/\siw, \sif/\siw, and \cf/\cw.
The \sit/\siw\ ratio probes the ionization level in
the warm-ionized phase of the gas, because the two ions
\siw\ and \sit\ generally show aligned absorption profiles and therefore 
co-exist in the same phase of gas.
The \sif/\siw\ and \cf/\cw\ ratios probe the relative amounts of highly-ionized
and warm-ionized gas, since \sif\ and \siw\ (and \cf\ and \cw)
are thought to largely arise in different gaseous phases (see \S4).
These ratios are plotted against \hi\ column density in the upper panels
of Figure 4, and against MS Longitude $L_{\rm MS}$ in the lower panels.

In the upper panels, all three ratios (\sit/\siw, \sif/\siw, 
and \cf/\cw) show anti-correlations with $N$(\hi), although the trends
are complicated by the fact that many of the ion ratios and \hi\ columns 
are limits, not measurements.
Kendall-tau correlation tests show that the anti-correlations are significant
at the 97.5\% level for \sit/\siw\ (number of data points $N$=10), %*
96.4\% level for \sif/\siw\ ($N$=11), and %*
96.1\% level for \cf/\cw\ ($N$=6), where upper and lower limits were 
excluded from the analysis. For the \cf/\cw\ ratio, including the limits
strengthens the correlation, with
all directions with log\,$N$(\hi)$\ga$20 having log\,(\cf/\cw)$<$--1,
and all directions with log\,$N$(\hi)$\la$19 having log\,(\cf/\cw)$>$--1.1.

In the lower panels of Figure 4,
a large scatter ($\approx$1\,dex) is seen in each ion ratio
at any $L_{\rm MS}$. \sif/\siw\ (middle panel) and \cf/\cw\ (right panel)
each show a mild tendency for higher ratios at highly negative $L_{\rm MS}$,
corresponding to MS directions far from the SMC and LMC, and lower ratios
at less negative $L_{\rm MS}$.
After excluding the upper and lower limits, \sif/\siw\ and \cf/\cw\ %*
show weak anti-correlations with $L_{\rm MS}$ at the 92.3\% ($N$=37) and 
95.9\% ($N$=43) confidence levels, respectively. %*
The trends shown in Figures 4 reflect the declining \hi\ column
density along the body of the Stream \citep{Ni10}, following an
approximately exponential decline from the Magellanic Clouds towards the tip.

\begin{figure*}[t]
\epsscale{1.2}\plotone{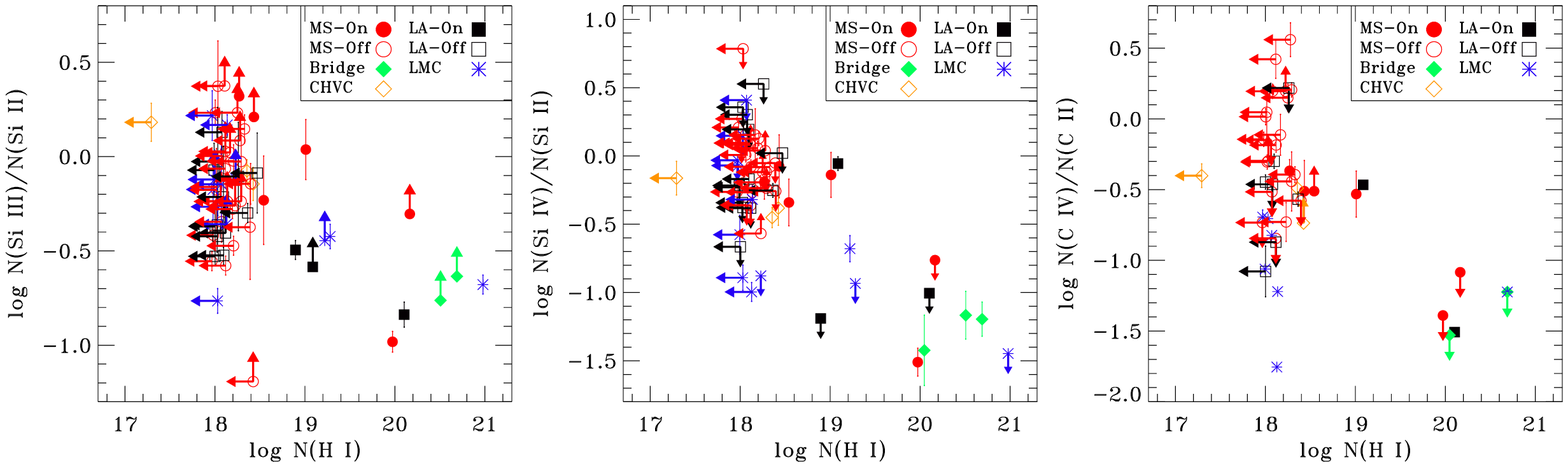}\plotone{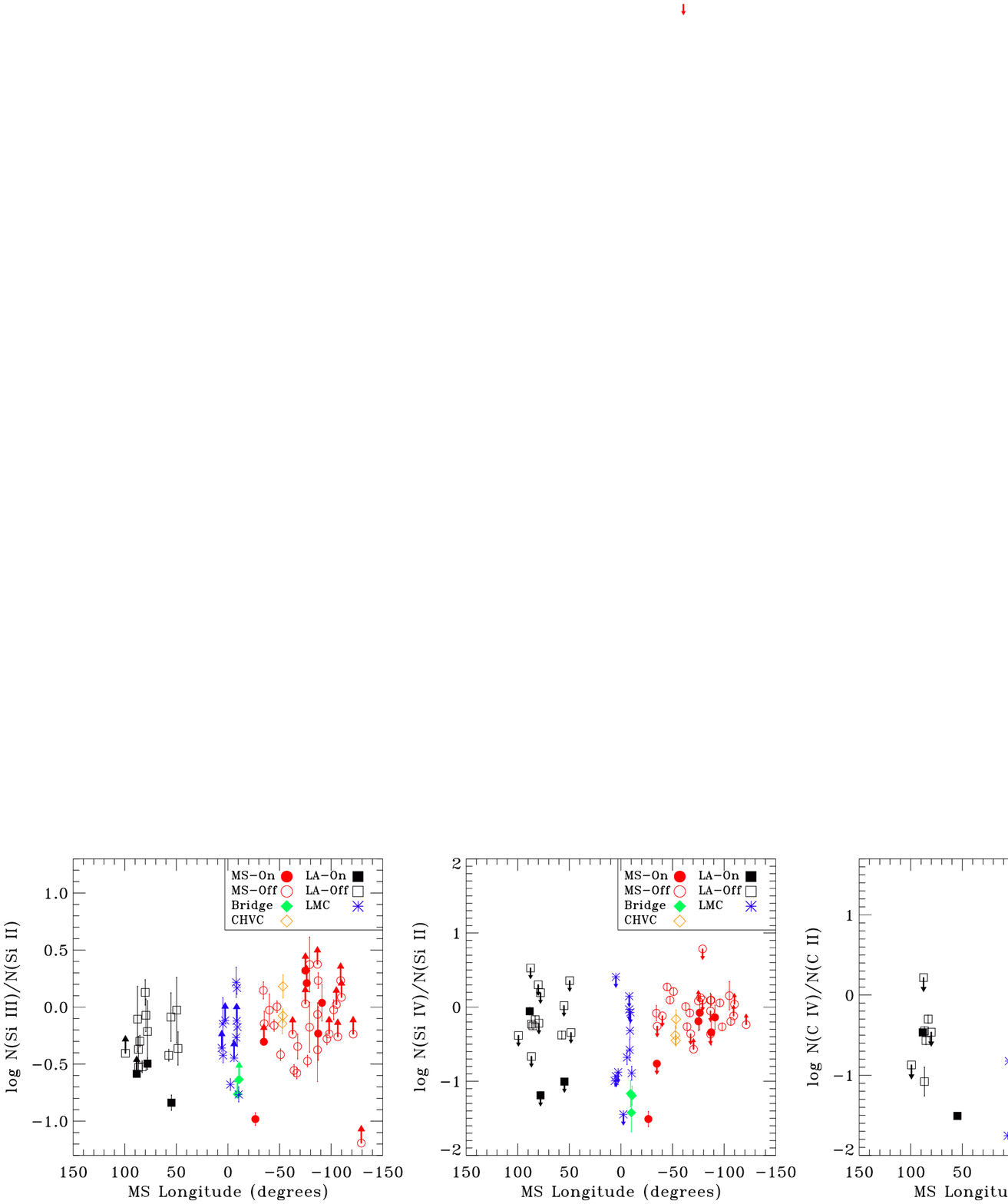}
\caption{{\bf Upper panels:} observed ion ratios against \hi\ column density 
in Magellanic gas. The left, central, and right panels show
the \sit/\siw, \sif/\siw, and \cf/\cw\ ratios, respectively.
Upper and lower limits account for non-detections and saturation.
The ratios were formed from apparent column densities
measured in the same velocity integration range for each ion and for \hi.
{\bf Lower panels:} observed ion ratios against MS Longitude $L_{\rm MS}$,
as defined by N08. The LMC is at $L_{\rm MS}$=0\degr, the
MS is at negative $L_{\rm MS}$, and the LA is at positive  $L_{\rm MS}$.
Note that $L_{\rm MS}$ here increases to the left, 
to match the format of the other figures.}
\end{figure*}

\section{Physical Conditions in the Magellanic Gas}
In this section, we use ionization modeling to interpret the 
column densities of the UV ions in the MSys and to derive 
information on its physical conditions. These conditions are closely related 
to the System's interaction with the Galactic corona and with the incident 
ionizing radiation field.

We model the Magellanic gas as a two-phase plasma,
with the \hi, low ions (singly-ionized), and intermediate ions (doubly-ionized)
existing in a photoionized (PI) phase, and 
the high ions (triply-ionized and above: \cf, \sif, and \os)
existing in a separate, collisionally ionized (CI) phase.
This approach is taken because photoionization models of the Stream, 
the Bridge, and other HVCs under-predict the observed high-ion column 
densities by orders of magnitude, particularly for \cf\ and \os\
\citep[][F10]{Se03, Tr03, Co04, Co05, Ga05, Fo05a, Mi09}.
Further evidence for a two-phase structure is given by the different 
velocity-component structure seen in the low-ion and high-ion absorption 
profiles in many HVCs 
\citep{Se99, Tr03, Fo04, Sh11, We11, TS12, He13}, and by high-resolution 
observations of \hi\ line widths in the MS \citep{Ng12}.
The placement of the \hi\ in the same phase as the
low and intermediate ions, which is central to our modeling procedure,
is supported by their similar ionization potentials and
the similar kinematic structure of the \hi\ and low-ion 
absorption profiles in directions where they have been compared 
\citep{Fo06, Ze08}. 
In our sample, the \hi\ emission and UV absorption profiles are well-aligned
for all On-System directions except \object{PHL\,2525} and \object{J2345--0059}.

\subsection{Low Ions: {\it Cloudy} Photoionization Modeling}

\subsubsection{Methodology}
We ran a set of photoionization models to reproduce the low-ion column
densities observed in the MSys using the spectral synthesis code 
\emph{Cloudy} \citep[v08.00][]{Fe98}. 
These models assume the gas exists in a uniform-density plane-parallel
slab exposed to an incident radiation field (described below).
We followed the same procedures described in F10 and Paper I, in which
the key observational constraint on the models is the \sit/\siw\ ratio.
For a given direction, the measured \sit/\siw\ ratio, 
the observed \hi\ column density $N$(\hi), 
the metallicity (Z/H), and the 
ionizing photon density $n_{\gamma}$ in the incident
radiation field are used as inputs to the models,
with $N$(\hi) acting as the stopping condition.
A grid of models at varying gas density $n_{\rm H}$ was constructed
for each direction on a 0.1\,dex grid, and matching the observations 
solved for many physical parameters in the Magellanic gas, including
the ionization parameter $U\!\equiv\!n_{\gamma}/n_{\rm H}$, 
the warm-ionized hydrogen column density $N$(\hw),
the hydrogen ionization fraction $x_{\rm H~II}$=$N$(\hw)/[$N$(\hi)+$N$(\hw)]
the line-of-sight cloud size $l$=$N_{\rm H}$/$n_{\rm H}$, and
the thermal pressure $P/k$=(1+$x_{\rm H~II}$)$n_{\rm H}T$ (see Table 3). %2.17

The models are only applied to the \ncloudy\ sightlines with measured
\hi\ columns and \sit/\siw\ ratios, because these two quantities are needed
as inputs to \emph{Cloudy}. If \sit\ and \siw\ are both undetected or both 
saturated, the ratio is undefined.
The advantage of solving for the ionization parameter from the 
\sit/\siw\ ratio alone is that it involves adjacent ionization states 
from the same element. This eliminates the possibility of non-solar 
abundance variations or differential dust depletions impacting the results.

The ionizing radiation field used in the models has two components:\\
(1) the Galactic radiation field taken from \citet{Fo05a}, which was 
based on \citet{BM99, BM02}, and which has an escape fraction of 1--2\% 
averaged over solid angle and $\approx$6\% normal to the disk;\\
(2) the ionizing radiation field emerging from the LMC and SMC, which is 
constrained by the \ha\ intensity of the Magellanic Bridge \citep{Ba13};\\
%(3) the $z$=0 extragalactic UV background (UVB) incorporated within
%\emph{Cloudy} \citep{HM12}.\\%of components (1) and (2) 
The combined field is shown in Figure 5,
a contour plot showing the lines of constant hard (hydrogen-ionizing) and
soft (non-hydrogen-ionizing) flux.
Because this field is anisotropic, we extracted the hard and soft fluxes 
at the Magellanic Longitude $L_{\rm MS}$ of each sightline modeled, assuming 
an MS distance of 50\,kpc. We then normalized the Galactic
radiation field presented in \cite{Fo05a} to these fluxes. 
This procedure gave the radiation
field appropriate for a sightline at a given $L_{\rm MS}$, which was 
used as an input to the corresponding \emph{Cloudy} model.
For the six MS directions, we ran a second set of models appropriate
for a MS distance of $d$=100\,kpc, to see the effect of varying the distance.
This was unnecessary for the LA, MB, and LMC Halo directions, since the
Magellanic gas in those sightlines is clearly not at 100\,kpc.
We did not include the possibility of radiation from the vicinity of
the super-massive black hole at the center of the Milky Way in a period of 
past Seyfert activity \citep{BH13}, since that radiation
was incident on the MSys $\approx$1--3\,Myr ago, not the current epoch.

One potential source of systematic error in our modeling procedure
is the beam-size mismatch between the radio data (finite beam) and UV data
(infinitesimal beam). However, the magnitude of this effect 
on metallicity measurements along pencil-beam MS sightlines 
when using radio data with beam sizes of $\approx$14\arcmin\ or less
(such as GASS data)
has been shown to be small ($\la$0.15\,dex; F10, Paper II).

\begin{figure}[t]
\epsscale{1.2}\plotone{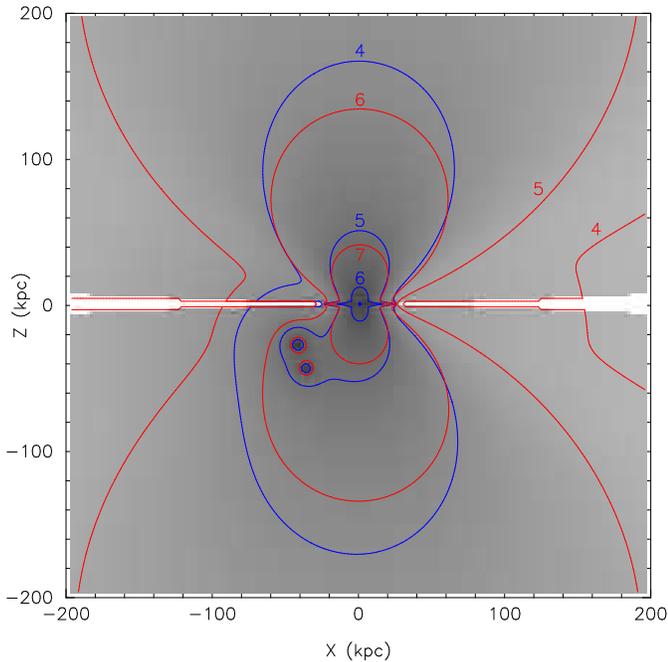}
\caption{Galactic ionizing radiation field used in our \emph{Cloudy} 
photoionization modeling, including contributions from the LMC and SMC.
The plot shows a 400$\times$400\,kpc slice through the Galactic halo
viewed from outside.
The blue and red contours show the contours of constant hydrogen-ionizing flux 
($E\!>\!13.6$\,eV) and non-hydrogen-ionizing flux ($E\!<\!13.6$\,eV), 
respectively. The numbers next to the contours give the logarithm of the flux 
in units of photons\,cm$^{-2}$\,s$^{-1}$. 
This model is an updated version of the radiation field
presented in \citet{BM99, BM02} and \citet{Fo05a}.
The LMC and SMC contributions are normalized by the \ha\ intensity
from the MB \citep{Ba13}; they are important for the 
MB and LA ionization models but less important for the MS models.
For comparison, the UV background at $z$=0 (not shown)
has an ionizing flux of $\approx$10$^4$ photons\,cm$^{-2}$\,s$^{-1}$
\citep{HM12}.}
\end{figure}

\subsubsection{Results}

\begin{deluxetable*}{lll ccc ccccr}
\tabletypesize{\scriptsize}
\tabcolsep=2.0pt
\tablewidth{0pt}
\tablecaption{Results of \emph{Cloudy} Photoionization Simulations: Physical Conditions in Magellanic Gas}
\tablehead{Target & \multicolumn{4}{c}{\underline{~~~~~~~~~~~~~~~~~~~~~~~~Input~~~~~~~~~~~~~~~~~~~~~~~~}} & \multicolumn{6}{c}{\underline{~~~~~~~~~~~~~~~~~~~~~~~~~~~~~~~~~~Output~~~~~~~~~~~~~~~~~~~~~~~~~~~~~~~~~~}}\\
 & log\,$\frac{\rm Si~III}{\rm Si~II}$\tm{a} & log\,$N$(\hi) & [X/H]\tm{b} & log\,$n_\gamma$\tm{c} & log\,$U$\tm{d} & log\,$n_{\rm H}$\tm{e} & log\,$N$(\hw)\tm{f} & $x_{\rm H II}$\tm{g} & $r_{\rm los}$\tm{h} & $P/k$\tm{i}\\
  & & (cm$^{-2}$) & & (cm$^{-3}$) & & (cm$^{-3}$) & (cm$^{-2}$) & & (kpc) & (cm$^{-3}$\,K)}
\startdata
{\bf Leading Arm}\\
                 NGC3783 &    $-$0.84 & 20.17 & $-$0.6 & $-$5.27 & $-$3.5 & $-$1.8 & 19.68 & 0.24 & 3.7 & 210 \\
{\bf LMC Halo}\\
             PKS0552-640 & $>$$-$0.12 & 18.23 & $-$1.0 & $-$4.65 & $-$3.8 & $-$0.9 & 19.40 & 0.94 & 0.1 & 2740 \\
              PKS0637-75 &    $-$0.44 & 19.29 & $-$1.0 & $-$4.90 & $-$3.8 & $-$1.1 & 19.44 & 0.59 & 0.2 & 1260 \\
{\bf Bridge}\\
             UKS0242-724 & $>$$-$0.63 & 20.69 & $-$1.0 & $-$5.07 & $-$3.1 & $-$2.0 & 20.15 & 0.22 & 19 & 130 \\
                ESO31-G8 & $>$$-$0.76 & 20.51 & $-$1.0 & $-$4.97 & $-$3.3 & $-$1.7 & 19.94 & 0.21 & 6.2 & 260 \\
{\bf CHVC}\\
            QSO1220-2859 &    $-$0.11 & 18.42 & $-$1.0 & $-$5.41 & $-$3.7 & $-$1.7 & 19.52 & 0.93 & 0.6 & 380 \\
                 TONS210 &    $-$0.08 & 18.36 & $-$1.0 & $-$5.41 & $-$3.7 & $-$1.7 & 19.52 & 0.94 & 0.6 & 380 \\
{\bf Stream (50 kpc)}\\
                FAIRALL9 &    $-$0.98 & 19.97 & $-$0.3 & $-$5.33 & $-$3.5 & $-$1.8 & 19.60 & 0.30 & 2.9 & 190 \\
          SDSSJ0012-1022 & $>$+0.21 & 18.43 & $-$1.0 & $-$5.54 & $-$3.4 & $-$2.1 & 19.83 & 0.96 & 3.1 & 140 \\
          SDSSJ2345-0059 &    $-$0.23 & 18.54 & $-$1.0 & $-$5.67 & $-$3.8 & $-$1.9 & 19.42 & 0.88 & 0.7 & 250 \\
                 NGC7714 &    +0.01 & 19.01 & $-$1.0 & $-$5.74 & $-$3.5 & $-$2.2 & 19.74 & 0.84 & 3.7 & 110 \\
                 PHL2525 & $>$+0.32 & 18.12 & $-$1.0 & $-$5.52 & $-$3.3 & $-$2.2 & 19.91 & 0.98 & 4.4 & 120 \\
                  RBS144 & $>$$-$0.32 & 20.17 & $-$1.0 & $-$5.40 & $-$3.2 & $-$2.2 & 20.04 & 0.43 & 13 & 89 \\
{\bf Stream (100 kpc)}\\
                FAIRALL9 &    $-$0.98 & 19.97 & $-$0.3 & $-$5.36 & $-$3.5 & $-$1.9 & 19.60 & 0.30 & 3.1 & 180 \\
          SDSSJ0012-1022 & $>$+0.21 & 18.43 & $-$1.0 & $-$5.99 & $-$3.4 & $-$2.6 & 19.83 & 0.96 & 8.9 & 50 \\
          SDSSJ2345-0059 &    $-$0.23 & 18.54 & $-$1.0 & $-$6.17 & $-$3.8 & $-$2.4 & 19.42 & 0.88 & 2.3 & 80 \\
                 NGC7714 &    +0.01 & 19.01 & $-$1.0 & $-$6.26 & $-$3.5 & $-$2.8 & 19.74 & 0.84 & 12 & 32 \\
                 PHL2525 & $>$+0.32 & 18.12 & $-$1.0 & $-$5.96 & $-$3.3 & $-$2.7 & 19.91 & 0.98 & 12 & 43 \\
                  RBS144 & $>$$-$0.32 & 20.17 & $-$1.0 & $-$5.60 & $-$3.2 & $-$2.4 & 20.04 & 0.43 & 20 & 56 \\
\vspace{-4mm}
\enddata
\tablecomments{Two sets of models are given for the MS directions, one assuming $d$=50\,kpc and one assuming $d$=100\,kpc. Full names for SDSS QSOs are given in Table 1.}
\tn{a}{Logarithm of apparent-column-density ratio measured in LSR velocity interval given in Table 1.}
\tn{b}{Metallicity of Magellanic gas, taken to be 0.1 solar (Lehner et al. 2008, Fox et al. 2010; Paper I) for all directions except Fairall~9 (0.5 solar; Gibson et al. 2000, Paper II) and NGC~3783 (0.25 solar; Sembach et al. 2001).}
\tn{c}{Logarithm of ionizing photon density in radiation field. Value is variable since field is non-isotropic (see Figure 5).}
\tn{d}{Logarithm of ionization parameter $U\!\equiv\!n_\gamma/n_{\rm H}$.}
\tn{e}{Logarithm of gas density.}
\tn{f}{Logarithm of \hw\ column density in the low-ion phase; this does not include the high-ion \hw\ living with the \cf\ and \sif.}
\tn{g}{Ionization fraction $x_{\rm H II}$=$N$(\hw)/[$N$(\hi)+$N$(\hw)].}
\tn{h}{Line-of-sight cloud size, $r_{\rm los}$=$N_{\rm H}/n_{\rm H}$ where $N_{\rm H}$=$N$(\hi)+$N$(\hw).}
\tn{i}{Thermal pressure, $P/k$=(1+$x_{\rm H~II})n_{\rm H}T$ and $T$=10$^4$\,K.}
\end{deluxetable*}

The results from our photoionization models are given in Table 3.
The target direction is given in the first column,
the next four columns define the model inputs specific to that direction,
and the final six columns give the model outputs specifying %seven
the physical conditions in the Magellanic gas.
Among all MSys sightlines, and considering the $d$=50\,kpc models,
the best-fit ionization parameters log\,$U$ range from
--3.8 to --3.1 with a median value of --3.5. %*
The same range in log\,$U$ is observed among MS sightlines only 
(i.e. not considering CHVC, MB, LMC Halo, or LA sightlines),
in close agreement
to values derived in earlier work on the Stream (F10, Paper I, Paper II).
The median log\,$U$=$-$3.5 corresponds to an average Magellanic gas density %*
log\,($n_{\rm H}$/cm$^{-3}$)$\approx$--1.8 given the %*
calculated density of the ionizing radiation field.
In turn, this implies line-of-sight cloud sizes for the Magellanic gas 
ranging from a few tenths of a kpc to several kpc (Table 3). %*

The hydrogen ionization level in the low-ion phase of the MSys
depends closely on the \hi\ column density, as expected;
it ranges from $\approx$20\% for the MB directions with log\,$N$(\hi)$\ga$20 to 
$>$80--90\% for directions with log\,$N$(\hi)$\la$19.5. %*
%As expected, the ionization level is low (40--50\%) %*
%in the two MB directions modeled, which have high neutral-gas columns,
%log\,$N$(\hi)$\ga$20.5.
Since the low-$N$(\hi) directions occupy a much larger cross-section 
than the high-$N$(\hi) directions (\S3.1), 
\emph{the gas is predominantly ionized in most 
directions through the MSys}, and even the ``low-ion'' gas is predominantly 
ionized when log\,$N$(\hi)$<$20. 
In terms of low-ion \hw\ columns (warm H$^+$ columns), all six MS models 
have log\,$N$(\hw) in a fairly narrow range between 19.4 and 20.0,
even though the \hi\ columns vary by over two orders of magnitude.
This finding allows us to use an average log\,$N$(\hw) for On-System directions
when calculating the MSys mass (\S5) and inflow rate onto the Galaxy (\S6).
The one LA direction modeled (\object{NGC\,3783}) shows a similar 
\hw\ column as the MS models. %*

The final six rows in Table 3 show the results from the 
MS models assuming a distance $d$=100\,kpc.
Here the ionizing photon density is lower than in the $d$=50\,kpc case 
(Figure 5), although the degree by which it is lower depends on 
Magellanic Longitude, with sightlines close on the sky to the Magellanic 
Clouds (\object{Fairall\,9} and \object{RBS\,144})
showing less reduction in flux than those farther away.
As expected, the same best-fit ionization parameter 
is found for each sightline for the $d$=100\,kpc models as for the 
$d$=50\,kpc models (because the shape of the radiation field does not change).
This in turn implies that the $d$=100\,kpc MS models
have lower gas density, lower thermal pressure, and larger line-of-sight
cloud sizes. The \hw\ columns show no dependence on distance, which
is important for our mass calculation (Section 5).

The thermal pressure $P/k$ for the six MS %* factor 2.17?
directions (considering both the $d$=50\,kpc and $d$=100\,kpc cases)
lies in the range $\approx$30--250\,cm$^{-3}$\,K. %*
This can be compared with the thermal pressure exerted by the 
hot Galactic corona thought to confine the Stream.
In the isothermal hot corona models of \citet{Sg02}, which
have a plasma temperature of 2$\times$10$^6$\,K, 
the coronal pressure is $\approx$250\,cm$^{-3}$\,K at 50\,kpc,
and $\approx$100\,cm$^{-3}$\,K at 100\,kpc,
corresponding to a coronal density log\,($n_{\rm hot}$/cm$^{-3}$)=$-$3.9 
at 50\,kpc and $-$4.3 at 100\,kpc.
Therefore, we find that if the Stream is at 50--100\,kpc (as suspected
by LMC/SMC orbital considerations),
it would be close to pressure equilibrium with the hot corona.

\subsection{High Ions: Collisional Ionization Modeling}
\cf\ $\lambda\lambda$1548,1550 is the most highly-ionized doublet
detected in the COS data, because \nf\ $\lambda\lambda$1238,1242 is 
generally too 
weak to be detected in Magellanic gas and \os\ $\lambda\lambda$1031,1037 
falls below the bandpass of our COS observations.
Therefore, we use \cf\ as the tracer of the highly-ionized phase of the MSys.
\cf\ ions are created from \ion{C}{3} with an energy of 47.9\,eV and can 
be produced by either photoionization or collisional ionization.
The \cf\ absorption profiles generally cover the same velocity interval
as the low-ion absorption profiles, but often show important 
kinematic differences in their detailed component structure.

The highly-ionized hydrogen column traced by the \cf\ absorption in
each direction is given by 

\begin{equation}
N({\rm H~II})_{\rm C~IV}=N({\rm C~IV})/[f({\rm C~IV}){\rm (C/H)_{MS}}],
\end{equation}
 
where the \cf\ ionization fraction $f$(\cf)$\equiv\!N$(\cf)/$N$(C) never 
rises above a maximum value 0.3 in either collisional ionization or 
photoionization \citep{SD93, GS07, OS13}.
In collisional ionization equilibrium, \cf\ peaks in abundance at a temperature 
of 2.0$\times$10$^5$\,K, so the values of $N$(\hw)$_{\rm C~IV}$ we derive
only represent the H$^+$ column existing at or below that temperature; they do 
not include any hotter plasma regions that may be present.
With two exceptions (discussed below), 
the carbon abundance (C/H)$_{\rm MS}$ is taken to be 
0.1(C/H)$_\odot$, which assumes that the low-ion and high-ion phases of 
the Stream share the same metallicity of 0.1 solar (F10, Paper I). 
This assumption is valid if the high-ion phase
evaporated out of the cool gas, which is believed to be the case for the 
MS as it interacts with the hot corona \citep[][F10]{Se03}. 
The two sightlines where we adopt a higher carbon abundance are
\object{Fairall\,9}, where we use 0.5 solar (Paper II), and
\object{NGC\,3783}, where we use 0.25 solar \citep{Se01}, because
these are the directions where higher abundances have been measured.
Our calculation also assumes that carbon atoms
in the high-ion phase are undepleted onto dust grains, and ignores
any deviation of the C/$\alpha$ ratio from solar.
Because our adopted value of $f$(\cf) is an upper limit,
the high-ion \hw\ columns given by Equation (3) are lower limits.

\section{Total Mass of Magellanic Gas}

\begin{figure*}[t]
\epsscale{1.25}\plotone{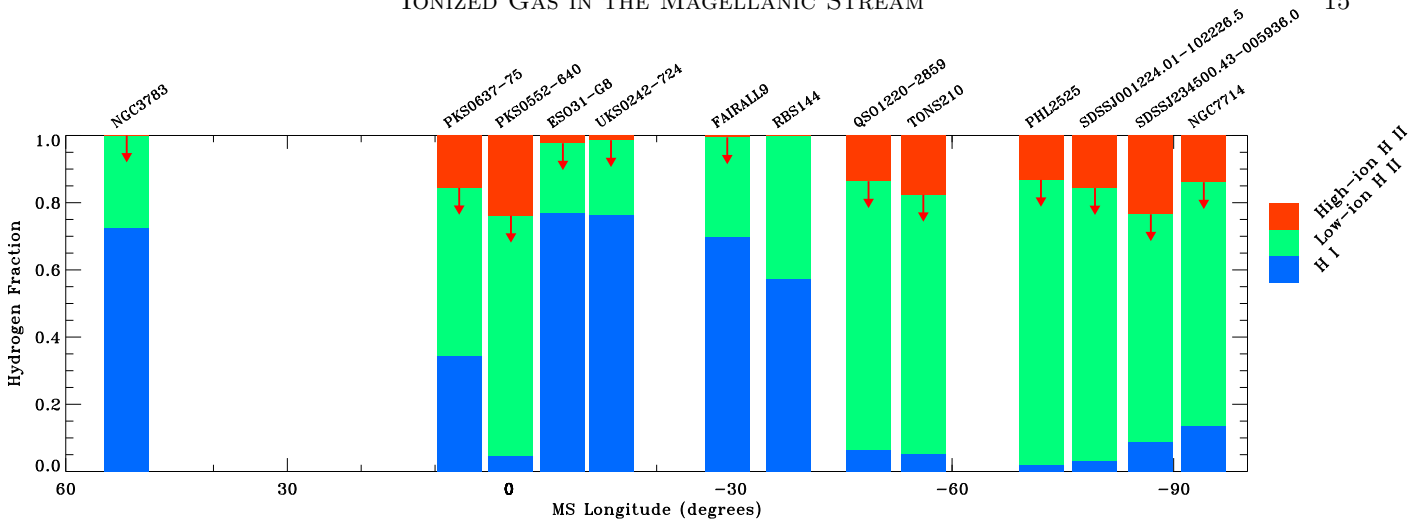}
\caption{Hydrogen ionization breakdown in the Magellanic gas in each 
sightline where 21\,cm emission is detected. 
The height of each bar indicates the fraction of the total
hydrogen column density in the atomic (blue), warm-ionized (green), and 
highly-ionized (red) phase. The bars are presented
in order of decreasing Magellanic Longitude, so LA directions
are on the left, Bridge and LMC Halo directions are in the center, 
and Stream directions are on the right.
The low-ion \hw\ columns are derived from the
photoionization models discussed in \S4.1. The high-ion \hw\ fractions
are lower limits derived from the \cf\ observations (see \S4.2)
when \cf\ is detected.}
\end{figure*}

In Figure 6, we combine the results from the ionization modeling 
(\S4.1 and \S4.2) to illustrate how the mass in the MSys is distributed
between the atomic, warm-ionized, and highly-ionized phases. 
For each direction modeled, the plot shows
the fraction of the total hydrogen column in each phase.
The number of sightlines shown (\ncloudy) is determined by the number
of sightlines with 21\,cm measurements of the Magellanic \hi\ column
(translating to directions with $N$(\hi)$\ga$10$^{18}$\sqcm) and a 
\sit/\siw\ ratio.
The height of the green and red bars relative to the blue bars indicates the
dominant contribution of the ionized gas to the mass budget, particularly when 
considering that only sightlines with 21\,cm \hi\ measurements are included on
this plot. If included, Off-System directions 
would be dominated by the green and red bars (low-ion and high-ion \hw). 
Note that among the sightlines modeled, the MB directions (\object{ESO31--G8}
and \object{UKS0242--724}) show substantial \hi\ fractions, as do the
inner-Stream sightlines toward \object{RBS\,144} and \object{Fairall\,9}.
This is expected given the high \hi\ columns in these directions.
However, the MS sightlines (at negative MS Longitude) are dominated
by low-ion \hw. The LA sightlines (at positive MS Longitude) show a lower
ionization level than the MS sightlines. Particularly interesting is that
the single LA sightline modeled (\object{NGC\,3783}) shows no \cf\ absorption.
Only one out of sixteen LA sightlines shows \cf\ absorption (Table 2),
hence the LA shows a much less developed high-ion phase than the MS.

The ionization fractions shown on Figure 6
can be used to estimate the total gas mass of the MSys.
Among the \ncloudy\ Magellanic directions modeled, 
the median value of $N$(\hw)$_{\rm total}$/$N$(\hi) is 6.4, where %*
$N$(\hw)$_{\rm total}$=$N$(\hw)$_{\rm low}$+$N$(\hw)$_{\rm C~IV}$.
If we ignore the high-ion \hw\ and just consider the low-ion \hw, 
which is better constrained, the median value of 
$N$(\hw)$_{\rm low}$/$N$(\hi) is 5.4, %*
only slightly lower, indicating that the low-ion phase dominates the \cf\ 
phase in the mass budget of the MSys. However, our data has not constrained
the mass of hotter plasma at $T\!\ga$3$\times$10$^5$\,K in the MSys, which
is seen in \os\ absorption in both On-System and Off-System directions
\citep{Se03}. 

To estimate the total gas mass (\hi+\hw) of the MSys
M(MSys), we need to account for the \hw\ mass in the 
directions with and without 21\,cm emission,
which we write as M(\hw, on) and M(\hw, off), respectively.
Splitting up the calculation in this way is necessary to identify
where most of the ionized gas mass resides, because of two competing effects:
the \hw\ columns are higher in the directions with high \hi\ columns,
but the \hw\ cross-section is higher in the directions with low \hi\ columns.
M(MSys) can be written as a sum of three parts:

\begin{equation}
{\rm M(MSys)}={\rm M(H~I)}+{\rm M(H~II, on)}+{\rm M(H~II, off)}.
\end{equation}

M(\hi) in the MSys is known from 21\,cm measurements
to be 4.87$\times$10$^{8}$($d$/55\,kpc)$^2$\msun\ \citep{Br05}, 
summing over the MS, MB, and LA,
and treating the ``Interface region'' as part of the MS.

M(\hw, on) depends on the product of the cross-sectional area of the 
21\,cm-bright MSys and the average \hw\ column density in those directions:

\begin{equation}
{\rm M(H~II, on)}=m_{\rm H} A({\rm H~II, on})\langle N({\rm H~II, on})\rangle.
\end{equation}

where $m_{\rm H}$ is the mass of a hydrogen atom,
$A$(\hw, on)=$A$(\hi, on) by definition, and 
$\langle N$(\hw, on)$\rangle$ is the median H$^+$ column
in the 21-cm bright directions.
The angular cross-section of the MSys of 2\,701 square degrees (N08) 
corresponds to $A$(\hi, on)=2.4$\times$10$^{46}$\,cm$^{2}$($d$/55\,kpc$^2$), %*
where $d$ is conservatively assumed to be constant 
along the Stream. Tidal models \citep{Co06, Be12} predict that $d$ increases 
approximately linearly along the MS as a function of MS Longitude, which would 
serve to increase the cross-section.
From our {\it Cloudy} modeling to the low ions in
\ncloudy\ On-System directions, we adopt a median
$\langle N$(\hw, on)$\rangle$=19.68 (Table 3), %*
and this is not strongly dependent on $N$(\hi).
Equation (5) then gives M(\hw, on)=9.5$\times$10$^8$\msun. %*
Comparing this to M(\hi)=4.87$\times$10$^{8}$\msun, we see that the majority
of the Magellanic gas is ionized \emph{even in On-System sightlines.}

M(\hw, off) can be written in an analogous way, combining the 
cross-sectional area of 21\,cm-faint ionized gas in the MSys with an 
estimate of the mean \hw\ column $\langle N$(\hw, off)$\rangle$ in these 
Off-System directions:

\begin{equation}
{\rm M(H~II, off)}=m_{\rm H} A({\rm H~II, off})\langle N({\rm H~II, off})\rangle.
\end{equation}

Using our derived \hw\ cross-section of the MSys of 
$\approx$11\,000 square degrees (\S3.1)
and a distance of 55\,kpc, we derive 
$A$(\hw)$\approx$7.6$\times$10$^{46}$\,cm$^{2}$. %*
Only two published derivations of $N$(\hw, off) exist, in directions
with far-UV data covering the sensitive high-order \hi\ Lyman series 
absorption lines.
In the \object{HE0226--4110} direction, passing 10.8\degr\
from the 2.0$\times$10$^{18}$\sqcm\ 21\,cm contour of the MS, \citet{Fo05a} 
measured an MS \hi\ column of log\,$N$(\hi)=17.05 and derived 
%an MS ionization level $x_{\rm H}\!\approx\!0.98$, corresponding to 
log\,$N$(\hw, off)=18.75. 
In the \object{Mrk\,335} direction (part of the current sample),
F10 measured log\,$N$(\hi)=16.67$\pm$0.05 and derived 
%an MS ionization level $x_{\rm H}$=98.9\%--99.5\%, corresponding to
log\,$N$(\hw, off)=18.64--18.96, very 
similar to the level toward \object{HE0226--4110}.

With the large number of Off-System directions present in our sample, we can
calculate $N$(\hw, off) by summing the \siw, \sit, and \sif\ columns to 
derive a total ionized silicon column in each direction, then using a 
0.1 solar metallicity to convert this to an ionized hydrogen column:

\begin{equation}
N({\rm H~II,~off})=\frac{N({\rm Si~II+Si~III+Si~IV})}{0.1({\rm Si/H})_\odot}
\end{equation}

This does not include the contribution from unobserved highly-ionized states of 
silicon (\ion{Si}{5} and above), but that would only 
serve to increase $N$(\hw, off).
Using 40 Off-System directions where at least one of \siw, \sit, and \sif\ 
was detected, we find a mean and standard deviation 
log\,$N$(\hw, off)=18.93$\pm$0.50. Using this mean value in Equation 5,
we derive M(\hw, off)$\approx$5.5$\times$10$^8$\msun, 
smaller than M(\hw, on) but slightly larger than M(\hi). %*

Using Equations 4--7 above, we calculate that the total 
(neutral, warm ionized, and highly ionized) gas mass of the MSys is 
M(MSys)$\approx$2.0$\times$10$^9$\,($d$/55\,kpc)$^2$\msun. %*
If much of the Stream is at $\approx$100--150\,kpc, a scenario predicted by
recent tidal models \citep{Be12} and supported by other arguments 
\citep{JL08, BH13}, then this mass estimate would increase by a 
factor of (100/55)$^2$ to (150/55)$^2$ or $\approx$3--7.
As such our mass estimate is conservatively low.

An alternative, simpler method for estimating M(MSys) is to multiply
the \hi\ mass of 4.87$\times$10$^{8}$($d$/55\,kpc)$^2$\msun\ with the 
median ionized-to-neutral ratio derived in the \ncloudy\ directions with 
photoionization models, $\langle N$(total \hw)/$N$(\hi)$\rangle$=6.4. %*
This method gives M(MSys)$\approx$3.1$\times$10$^9$\,($d$/55\,kpc)$^2$\msun,
approximately 50\% higher than the value derived from Equations 4--7. 
However, we adopt the lower value in our analysis below, since it accounts
for the different ionization conditions in On- and Off-System directions.

We can calculate the contribution of the MS to M(MSys)
%the total MSys gas mass
by reevaluating Equation (4) without the contributions of
the MB, LA, CHVC, and LMC Halo. 
Using M(\hi)$_{\rm MS}$=2.74$\times$10$^8$\msun\ 
\citep[including the Interface Region;][]{Br05}, 
$A$(\hi)$_{\rm MS}\!\approx\!0.5A$(\hi)$_{\rm MSys}\!\approx$1\,350 %*
square degrees, and $A$(\hw)$_{\rm MS}\!\approx$5\,500 square degrees %*
(Figure 1b), 
we derive M(MS)$\approx$1.0$\times$10$^9$\,($d$/55\,kpc)$^2$\msun, or 
approximately half of M(MSys). %*
Our mass calculations can be summarized in simple terms: the
MSys gas mass is approximately two billion solar masses and the
MS gas mass is approximately one billion solar masses.

Our calculated value of M(MSys) is approximately twice
the combined interstellar \hi\ mass of the LMC 
\citep[4.4$\times$10$^8$\msun;][]{Br05,Be08} and SMC 
\citep[4.0$\times$10$^8$\msun;][]{St99,Br05}. 
Since ionized gas within the Magellanic Clouds is thought to be a minor 
contributor to their interstellar mass, a comparison
of M(MSys) (the mass of gas stripped out of the Clouds) 
with the ISM mass remaining in the Magellanic Clouds 
shows that that \emph{the MSys represents
over two-thirds of the initial ISM mass of its two parent galaxies.} %*
The MSys is therefore a striking example of the powerful nature of 
galaxy interactions and their ability to disperse the majority of a dwarf 
galaxy's gas (or two dwarf galaxies' gas) over large distances.
The Milky Way will be the beneficiary of this interaction, since
the MSys is now bound to the Galactic halo, and so the net result of the MSys
creation is the transfer of gas out of the Magellanic Clouds and 
into the Galaxy's potential well.

The total MSys gas mass 
M(MSys)$\approx$2.0$\times$10$^9$\,($d$/55\,kpc)$^2$\msun\ %*
can also be compared to
(1) the stellar mass of the LMC disk, which is 2.7$\times$10$^9$\msun\
\citep{vdM06}, only $\approx$50\% more than the MSys gas mass;
(2) the total mass of the ionized HVCs close the Milky Way disk, which is
$\approx$1.1$\times$10$^8$\msun\ \citep{LH11}, indicating that the MSys
is by far the most massive HVC complex around the Milky Way;
(3) the total mass of the hot ($\sim$10$^6$\,K) Galactic corona out to 50\,kpc, 
which for a uniform density can be written as
M$_{\rm hot}\!\approx\!1.7\!\times$10$^9$\msun($r_{\rm hot}$/50\,kpc)$^3$($n_{\rm hot}$/10$^{-4}$\,cm$^{-3}$).
While $r_{\rm hot}$ and $n_{\rm hot}$ are poorly constrained 
\citep{Br07, Gu12, Fa13, MB13},
the order-of-magnitude similarity of M(MSys) %the total Magellanic gas mass 
and the coronal mass has implications for the thermodynamics of the interaction
between the MSys and the corona, because the corona cannot be 
viewed as an infinite heat reservoir. Instead, the passage of the MSys 
through the corona will simultaneously heat the Magellanic gas and cool 
the corona as the ensemble heads to energy equipartition. A similar
``refrigeration'' mechanism for the cooling of coronal gas by cool 
embedded clouds has been proposed for other (nearby) Galactic HVCs 
\citep{Fr13, Ma13}.

\section{Inflow Rate of Magellanic Gas}

Combining the average MS inflow velocity onto the Milky Way of 
$\approx$100\kms\ \citep[galactocentric;][]{Ma77} with a conservative 
MS distance $d$=55\,kpc gives an inflow timescale of $\approx$540\,Myr.
If we instead use $d$=100\,kpc, as may apply to the tip of the MS, 
then this timescale rises to $\approx$1.0\,Gyr.
Using %our total MSys gas mass estimate of 
M(MSys)$\approx$2.0$\times$10$^9$($d$/55\,kpc)$^2$\msun, %*
we then derive a \emph{current-day} MSys inflow rate 
$\dot{\rm M}$(MSys)=M(MSys)/$t$ of 
$\approx$3.7--6.7\smy\ for $d$=55--100\,kpc. %*

It is of interest to calculate the MS contribution to the MSys inflow rate, 
because the gas in the MB and LMC halo may be bound to the LMC and hence
less likely to reach the Milky Way (unless stripped in the future). 
Using the value M(MS)$\approx$1.0$\times$10$^9$\,($d$/55\,kpc)$^2$\msun\ %*
derived in the previous section,
we derive $\dot{\rm M}$(MS)=M(MS)/$t$=1.9--3.4\msun\ for $d$=55--100\,kpc, %*
or about half of the MSys inflow rate.
Since one end of the MS is anchored to the Magellanic Clouds at 55\,kpc,
but the tip is likely to be closer to 100\,kpc, the true MS
inflow rate likely lies somewhere in-between these two values.

The MSys inflow rate is larger than the current 
Galactic star-formation rate (SFR), which recent measurements place between 
0.68--1.45\smy\ \citep{RW10} and 
1.9$\pm$0.4\smy\ \citep{CP11}. \emph{Therefore, the 
MSys is bringing in fuel at a rate sufficient to raise the Galactic SFR,} 
although whether or not it survives the journey to the disk
depends on evaporative processes, as we discuss below.

The MSys inflow rate is also larger than the total (neutral plus ionized)
inflow rate represented by all nearby ($d\!\la$10--15\,kpc) HVCs combined. 
Estimates for the HVC inflow rate vary, ranging from 
0.08--0.4\smy\ \citep{Pu12} based on
a fit to the inflow rate represented by the largest 21\,cm complexes, 
to 0.45--1.4\smy\ \citep{LH11} based on 
measuring the incidence of (predominantly ionized) HVCs seen in UV absorption 
lines toward background AGN and distant halo stars
\citep[see also][]{Wa99, Wa08, Sh09, Co09}. The MSys inflow rate
surpasses both these estimates, and understanding the fate of the MSys is 
therefore crucial to understand the fueling of future star formation in the 
Galaxy.

A key question in determining the fate of the MSys
is whether it will survive to reach the disk. Parts of the LA appear to have 
already reached the disk in the outer Galaxy \citep{MG08}, where the LA gas 
shows morphologies indicative of interaction with the ambient ISM, but the MS
is further away, particularly the tip which may lie at $\ga$100\,kpc 
\citep{Be12}. At 100\,kpc the inflow time is $\sim$1\,Gyr, which is much
longer than the evaporation timescale for a small gas cloud being disrupted
by its interaction with a hot corona: simulations find that
clouds with \hi\ masses $<$10$^{4.5}$\msun\ survive for $\approx$100\,Myr 
\citep[][see also Bland-Hawthorn et al. 2007]{HP09}. 
The survival of the MSys therefore
depends on which part of the structure is being considered, and on whether
the gas fragments into smaller clumps or is stabilized against fragmentation
\citep[e.g. by magnetic fields;][]{Kw09, Hi13}.
Clear fragmentation and small-scale structure is seen in deep 21\,cm 
observations of the MSys, particularly in the LA and MS tip regions 
\citep{Wa02, St02, St08, WK08}. 
Observations of high-ion absorption in the MS 
\citep[][F10, Paper II, this paper]{Se03, Fo05a} are also indicative
of MS/corona interactions, since the high ions are thought
to trace the boundary layers between the \hi\ and the hot corona.

The MS/corona interaction is disrupting the Stream's passage to the disk, 
and the evaporated gas must re-condense and cool to participate in 
the Galactic inflow process instead of simply replenishing the hot corona
with fresh material.
Therefore the \emph{current-day} gas inflow rate derived here will not 
necessarily equal the rate at which that fuel will be deposited in the disk 
in the future. However, an equilibrium scenario is possible in which infalling 
gas in the MSys evaporates into the corona at the same rate at which 
neutral gas condenses out of other (denser or more metal-enriched) regions,
perhaps triggered by thermal instability 
\citep{Fi65, Oo66, MB04, Jo12a}.
In this case, the MSys would be effectively fueling the halo 
rather than directly fueling star formation in the disk. We note that
a population of \hi\ clouds has been identified in the inner Galaxy,
which could represent objects that have condensed out of the hot corona
\citep[][see also Peek et al. 2008]{Lo02, St06}.

Despite the evidence for an evaporative encounter 
between the Stream and the Milky Way corona, the fact that the MS has survived
to reach its current age of $\approx$2\,Gyr is itself an important clue
to its fate. \citet{Mu00} argued that the continued existence
of the MS constrains the density of the surrounding hot corona to be
log\,($n_{\rm H}$/cm$^{-3}$)$<$--5.0 at the distance to the Stream, 
because otherwise the Stream would have been destroyed within $\approx$500\,Myr 
by the ram pressure exerted by the hot corona.
Although this density is lower
than expectations for the coronal density at $\approx$50\,kpc 
\citep[e.g.][]{St02,GP09}, recent proper-motion observations show that 
the Magellanic Clouds may be on their first 
infall into the Galactic halo \citep{Be10, Ka13}. In this case,
the density of the surrounding gas (and hence the strength of the 
ram pressure) was lower in the past, when the Stream
was at larger Galactocentric radius, allowing the Stream to survive for longer
than would otherwise be the case.

Our study of the MSys has indicated that the true role of tidal gas streams 
may be more subtle than the \emph{direct} transport of fuel to the disks of 
massive galaxies. Instead, we
favor a scenario where tidal streams serve to replenish 
the hot coronae of massive galaxies with new material 
at large galactocentric radii, and where cool gas clouds condense out of other, 
denser regions of these coronae via the thermal instability.
In this picture, gaseous tidal streams represent one evanescent component of 
galaxy ecosystems in which hot coronae are equally important players.
This illustrates the crucial and 
complex role of the gas cycle in the evolution of galaxies.

\section{Connection of Magellanic System to Quasar Absorption Line Systems}
Given its large cross-section and gas mass, the MSys provides 
an interesting case-study for understanding the origin of 
intervening quasar absorption line (QAL) systems.
If the MSys was observed as an intervening absorber along a 
random sightline through the Galactic halo, it would present a UV metal-line 
covering fraction of $\approx$1/4 out to 55 kpc (based on the covering %*
fraction of $\approx$1/4 observed at the solar circle).
Since most of this material is below the detection threshold for 21\,cm 
emission, it has \hi\ column densities of 
10$^{16}\!<$log\,$N$(\hi)$<$10$^{19}$\sqcm\ that classify it as a 
Lyman Limit System (LLS).
Low-redshift LLSs have a bimodal metallicity distribution \citep{Le13} and a 
substantial highly-ionized phase \citep{Fo13b}, and their relationship to 
galaxies is only known in a small number of cases 
\citep{Je05, Ra11, Rb11, Le13}. Our results show that 
tidally-stripped material can produce a LLS at impact parameters of 
$\approx$50--100\,kpc from a star-forming ($\approx\!L_\ast$) spiral galaxy, 
and that such material can have a substantial covering fraction.

The $10^4$\,K ionized material in the MSys
is effectively part of the Milky Way's cool circumgalactic medium (CGM).
\citet{We14} recently calculated the mass of the cool CGM
around low-redshift $\sim\!L_\ast$ galaxies to be $>$6$\times$10$^{10}$\msun\ 
out to 150\,kpc. Our results show that the MSys dominates the mass budget of 
the Galactic cool CGM, and hence that tidally stripped material can be the 
dominant form of such gas around $\approx\!L_\ast$ galaxies. That being said, 
the fact that the Milky Way possesses two dwarf irregular satellite galaxies 
within $\approx$50\,kpc makes it highly aptyical \citep{To11}, so although 
the Milky Way's gaseous halo offers a fertile test bed for CGM studies, 
caution is needed about using it to draw general conclusions about $L_\ast$ 
galaxy halos.

Optical absorption-line measurements of \caw, which
serves as tracer for neutral gas with log $N$(\hi)$\ge$17.4, 
demonstrate that the neutral-gas cross-section of galaxy halos 
exceeds that of galaxy disks by a factor of 2--3 \citep{Ri11}.
These measurements further indicate that the characteristic neutral-gas extent 
of galaxy halos at low redshift is $\approx$55\,kpc, in line with the spatial 
distribution of neutral gas around the Milky Way and M31 \citep{Ri12}. 
Direct 21\,cm searches for neutral gas in the extended halos of Milky-Way-type
galaxies indicate that only a relatively small fraction contain \hi\ 
structures detectable in emission beyond the disk-halo interface \citep{He11}.
One exception is the nearby spiral galaxy \object{NGC\,6946}, which
is connected to its nearest satellites by diffuse, filamentary \hi\ 
\citep{Pi14}.
A clumpy \hi\ filament has also been detected between \object{M31} and 
\object{M33} \citep{Wo13}. Many such (possibly tidal) gas streams may await 
their detection in more sensitive 21\,cm observations. 
 
%For a better understanding of the origin and fate of such neutral gas 
%features around other low-redshift galaxies and their similarities with the 
%MSys, multi-wavelength observations combining absorption and 
%emission measurements are required. 

\section{Summary}
We have presented an absorption-line survey of ionization
in the MSys (MS, MB, LMC Halo, and LA)
using a sample of medium-resolution (18\kms\ FWHM)
\hst/COS UV spectra of \nsl\ background AGN lying within 30\degr\
of the 21\,cm emission from the MSys.
These data are supplemented by Wisconsin H$\alpha$ Mapper (WHAM) 
\ha\ emisssion-line observations in 
\nwham\ directions, and 21\,cm emission spectra from the LAB survey, GASS
survey, and additional Parkes telescope observations.
The sightlines sample approximately four orders of magnitude of \hi\ column 
density, ranging from the dense cores of the MB to the 
diffuse outer layers of the MS and LA.
This dataset has allowed us to characterize the physical
properties of the MSys and its relationship to the extended 
Galactic halo. Our main results are the following.

\begin{enumerate}

\item{\bf Detection Rate}. 
UV absorption at Magellanic velocities is detected in \ndet\ out of \nsl\ 
directions (\msper\ detection rate). The detected lines include 
one or more of \sit\ $\lambda$1206, \siw\ $\lambda$1190, \cw\ $\lambda$1334, 
\sif\ $\lambda$1393, and \cf\ $\lambda$1548, and occasionally other lines
including \oi\ $\lambda$1302.
Since the 21\,cm-emitting regions of the MSys cover 2\,701 square degrees 
\citep{Ni10}, the total cross-section of the MSys is enormous, at 
$\approx$11\,000 square degrees, which is around a quarter of the 
entire sky. %* 
The ionized (UV-absorbing) regions occupy $\approx$four times as much %*
area as the neutral (21\,cm-emitting) regions.

\item{\bf Line ratios}. 
We measure the column-density ratios \sit/\siw, \sif/\siw, and \cf/\cw\ 
in the Magellanic gas. Among directions with 21\,cm detections 
(with log\,$N$(\hi)$\ga$18), all three ratios show 
significant anti-correlations with $N$(\hi) (Figure 4).
In addition, \sif/\siw\ and \cf/\cw\ each show weak (but significant)
anti-correlations with MS Longitude, such that the ionization level
increases along the Stream from the Magellanic Clouds toward the tip.
Given our current understanding of the Stream's orbit, in which the tip 
is the most distant region, this result is equivalent to an increase in 
ionization level with Galactocentric distance, consistent with
the observed decline in $N$(\hi) along the MS.

\item {\bf Ionization parameter and gas density.}
Using \emph{Cloudy} photoionization models applied
to \ncloudy\ Magellanic directions with measured \hi\ columns and
\sit/\siw\ ratios, we derive ionization parameters log\,$U$
in the Magellanic gas ranging from --3.8 to --3.1 with a median 
value of --3.5. %*
This corresponds to an average Magellanic gas density 
log\,($n_{\rm H}$/cm$^{-3}$)$\approx$--1.8 given the %*
calculated density of the ionizing radiation field (including
both Milky Way and Magellanic photons).
%Among the six MS directions, log\,($n_{\rm H}$/cm$^{-3}$) 
%lies between --2.4 and --1.9. %*

\item {\bf Ionization level and warm \hw\ column.}
The hydrogen ionization level $x_{\rm H~II}$ varies considerably
between MSys regions, depending on the \hi\ column.
In the MB, where log\,$N$(\hi)$\ga$20, $x_{\rm H~II}\!\approx\!20$\%,
whereas the directions with log\,$N$(\hi)$\la$19.5 have $x_{\rm H~II}$
up to $\approx$98\%. %*
Since the Off-Stream directions have a larger cross-section than the On-Stream 
directions, the gas is predominantly ionized in most sightlines through 
the MSys, and even the phase referred to as ``low-ion'' phase 
is predominantly ionized.
The warm \hw\ columns in the low-ion phase of the MSys take a
fairly narrow range of values, between log\,$N$(\hw)$\approx$19.4--20.1, %*
even though the \hi\ columns in the same directions cover several orders of 
magnitude.

\item {\bf Thermal pressure.} 
The MS thermal pressure $P/k$ calculated from the photoionization models
varies from $\approx$30--250\,cm$^{-3}$\,K in the six Stream directions %*
analyzed. This places the Stream close to pressure equilibrium
in a two-million-degree hydrostatic Galactic corona at 50--100\,kpc, where 
$P_{\rm corona}/k$=100--250\,cm$^{-3}$\,K in the isothermal models
of \citet{Sg02}. 

\item {\bf Total mass}.
We calculate the total (neutral plus ionized) gas mass of the MSys
to be M(MSys)$\sim$2.0$\times$10$^9$\,($d$/55\,kpc)$^2$\msun, %*
by combining the \hi\ mass of 4.9$\times$10$^8$\msun\ \citep{Br05} 
with estimates of the \hw\ mass in 21\,cm-bright regions 
(9.5$\times$10$^8$\msun) and 21\,cm-faint regions (5.5$\times$10$^8$\msun). %*
The MS accounts for about half of the mass in the MSys.
M(MSys) is over twice as large as 
the remaining interstellar \hi\ mass of the LMC 
and SMC combined, indicating \emph{these two dwarf galaxies 
have lost over two-thirds of their initial gas mass.}
M(MSys) is also comparable to the mass of the hot 
Galactic corona out to 50\,kpc,
M$_{\rm hot}\!\approx$1.7$\times$10$^9$\msun($r_{\rm hot}$/50\,kpc)$^3$($n_{\rm hot}$/10$^{-4}$\,cm$^{-3}$), 
although $r_{\rm hot}$ and $n_{\rm hot}$ are poorly constrained.
The similarity between the mass of the MSys and the 
corona indicates that energetically, both components will be affected by 
their mutual interaction, causing the corona to cool as the Magellanic gas 
heats up.

\item {\bf Mass flow rate onto the Galaxy.}
The present-day infall rate of the MSys onto the Milky Way
is $\approx$3.7--6.7\smy, %*
using an average galactocentric infall velocity of 100\kms,
a MS distance of 55--100\,kpc, and the total gas mass derived above. %*
%The MS inflow rate is 1.9--3.4\smy.
This is considerably larger than the inflow rate of
all nearby HVCs combined and larger than
the current SFR of the Galaxy ($\approx$1--2\smy). 
\emph{The MSys therefore has the potential 
to raise the global Galactic SFR}. However, this gaseous fuel faces 
a tortuous path to reach the disk: multiple signs of an evaporative 
encounter with the hot corona suggests that the MSys is disintegrating 
and replenishing the hot coronae with fresh material at large 
galactocentric radii. For the inflow rate across the entire halo to be 
in or close to equilibrium, cooler clouds must condense out of other, 
denser or more metal-enriched regions of the corona.
The MSys can therefore be thought of as fueling the Galactic halo rather than
directly fueling star formation in the disk.

\end{enumerate}

{\it Acknowledgments.}
We thank David Nidever for providing the total area and velocity field of 
the \hi\ emission from the Magellanic System.
Support for programs 11692, 12204, 12263, and 12604 was provided by NASA 
through grants from the Space Telescope Science Institute, which is 
operated by the Association of Universities for Research 
in Astronomy, Inc., under NASA contract NAS~5-26555.
WHAM science and ongoing operations are supported by NSF award AST~1108911.
The Parkes radio telescope is part of the Australia Telescope National 
Facility which is funded by the Commonwealth of Australia for operation as a 
National Facility managed by CSIRO.
K.A.B. is supported through NSF Astronomy and Astrophysics Postdoctoral 
Fellowship award AST~1203059.
The research was partially supported by the Japan Society 
for the Promotion of Science through Grant-in-Aid for Scientific 
Research 23740148.

\appendix
The figures below show the \hst/COS absorption-line profiles for all 
sightlines in the sample, sorted alphabetically. 
The title of each stack includes a numerical ID and the Magellanic coordinates
of the target, which can be used to cross-reference with Figure 1,
and a label identifying the region of the MSys probed by the sightline 
(MS-On, MS-Off, LA-On, LA-Off, Bridge, LMC Halo, or CHVC). 
The top panels of each stack show the \hi\ 21\,cm emission profile from 
either the Parkes telescope, the LAB survey, or the GASS survey,
with brightness temperature plotted versus LSR velocity, and with
the y-range scaled to the strength of the Magellanic emisison, 
regardless of the strength of the (usually stronger) Galactic emission.
All other panels show the COS spectra, with normalized flux 
plotted against LSR velocity. The dotted vertical lines show the central 
velocity (solid) and velocity integration range (dashed) 
of Magellanic absorption and emission.
In sightlines where Magellanic 21\,cm emission is detected, 
these vertical lines are red and the solid line shows the centroid 
of the \hi\ emission. 
In sightlines without Magellanic 21\,cm emission but with UV absorption, 
these lines are blue and the solid line shows the velocity of the 
strongest UV absorption. 
In sightlines with neither 21\,cm emission nor UV absorption at Magellanic 
velocities, the velocity integration range is taken from the predicted 
velocity field of the MSys in that direction (Figure 1b), and no centroid 
is given. The shaded gray regions denote the 
velocity interval of Magellanic absorption.

\clearpage
\begin{figure}\epsscale{1.1}
\plottwo{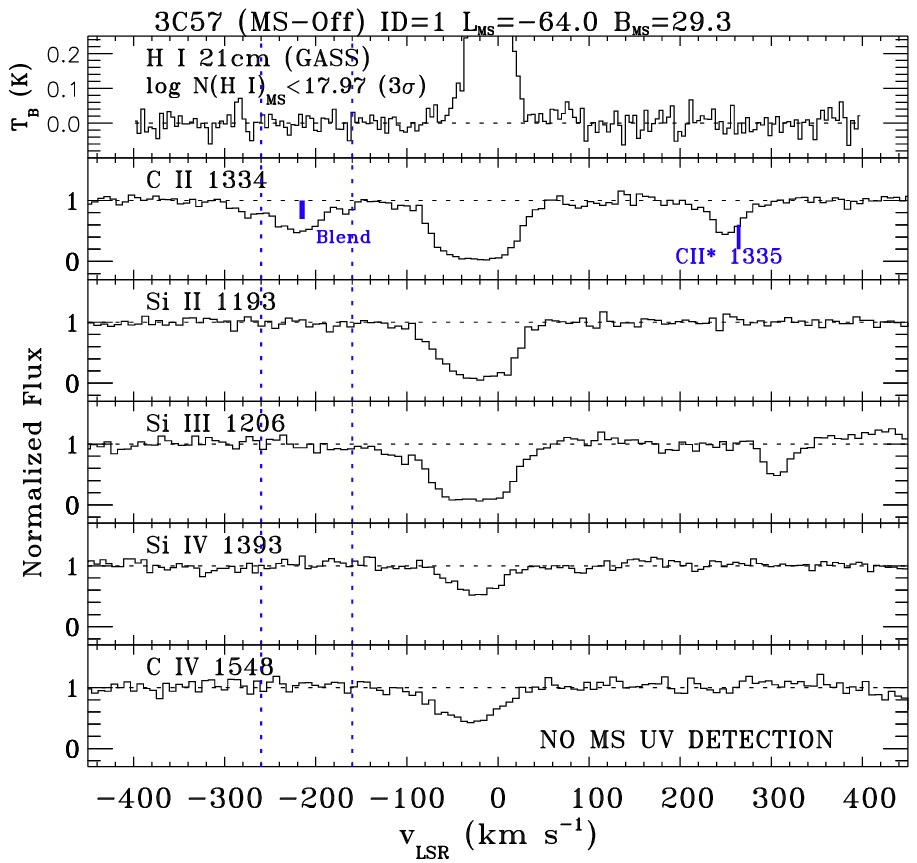}{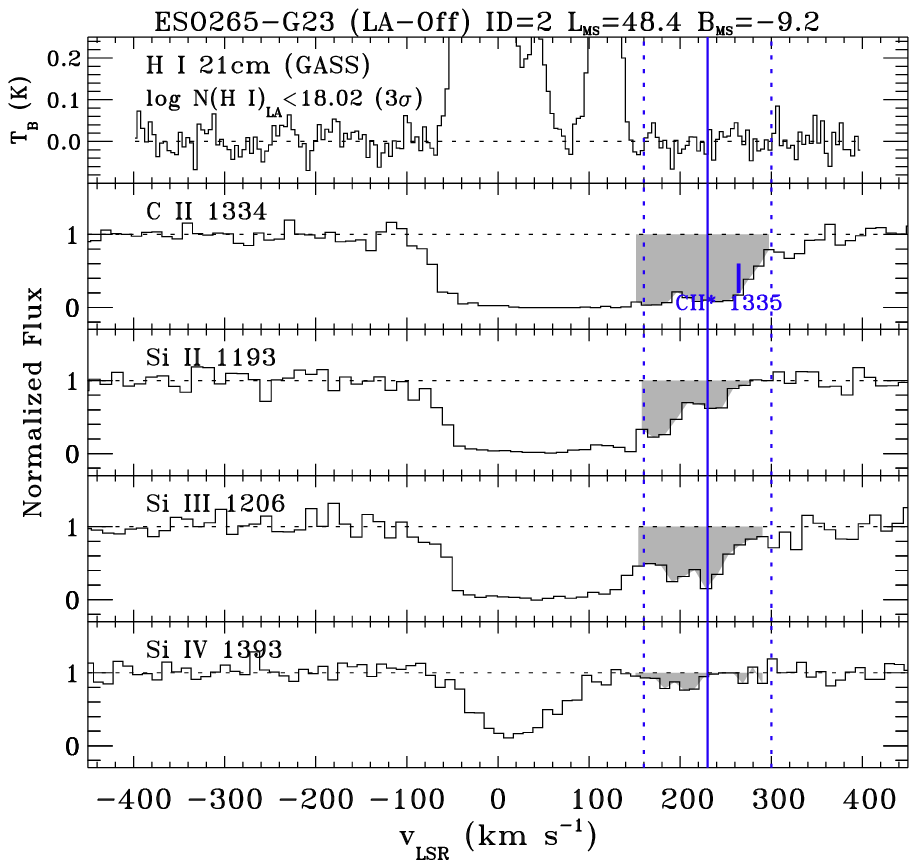}\end{figure}
\begin{figure}\epsscale{1.1}
\plottwo{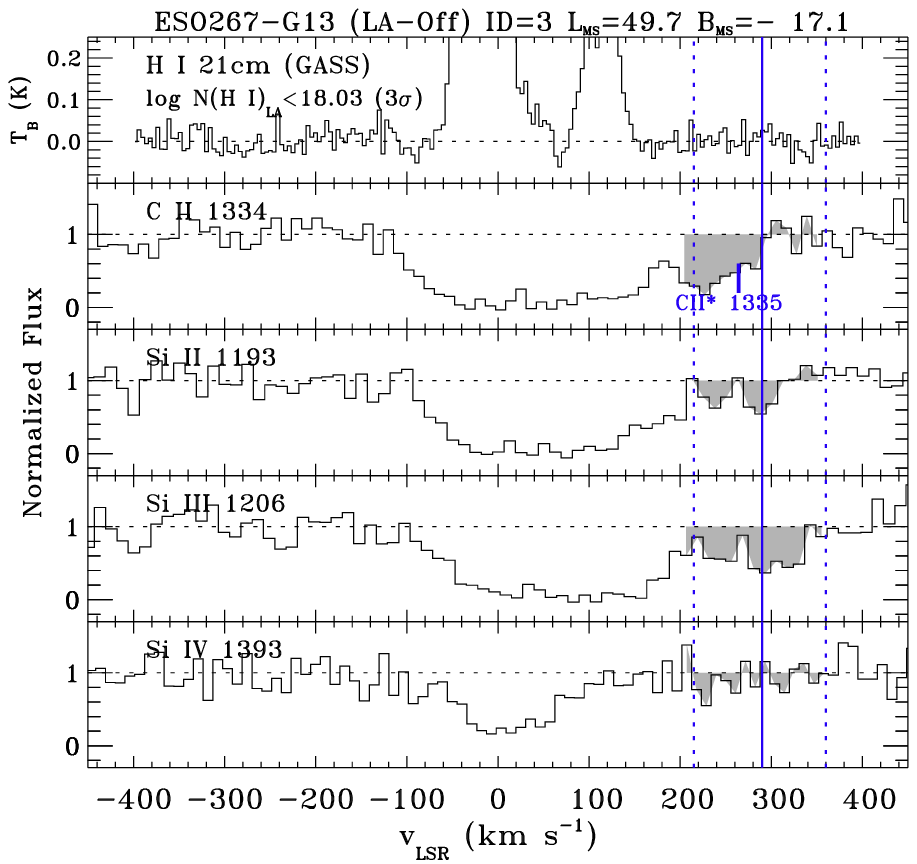}{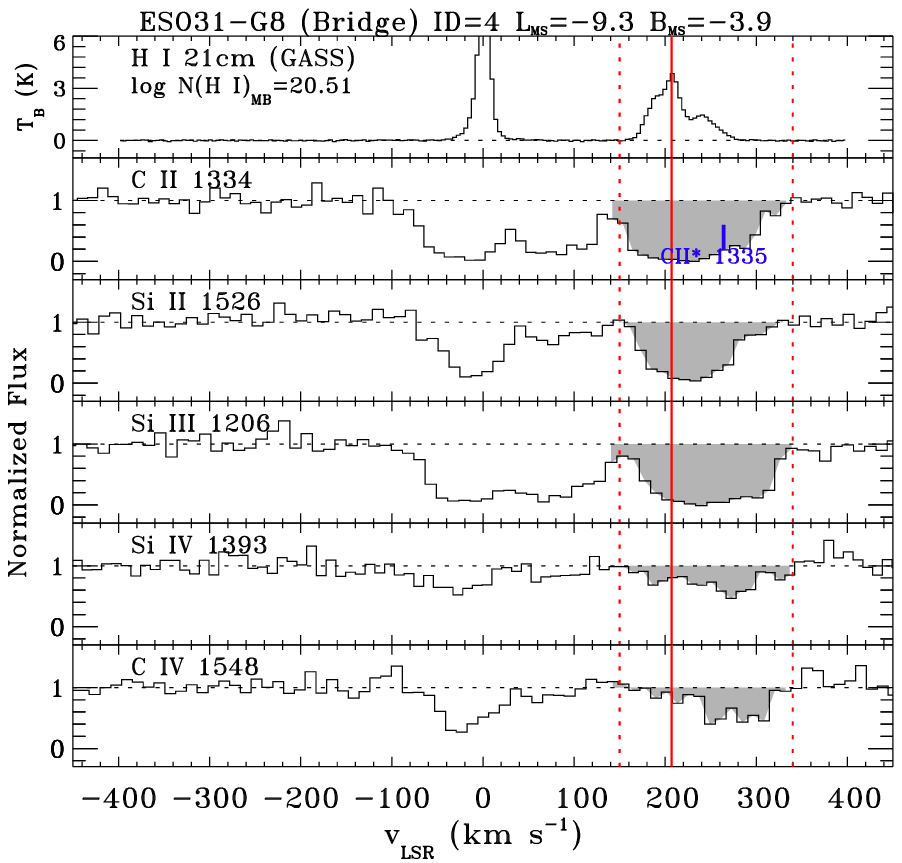}\end{figure}
\begin{figure}\epsscale{1.1}
\plottwo{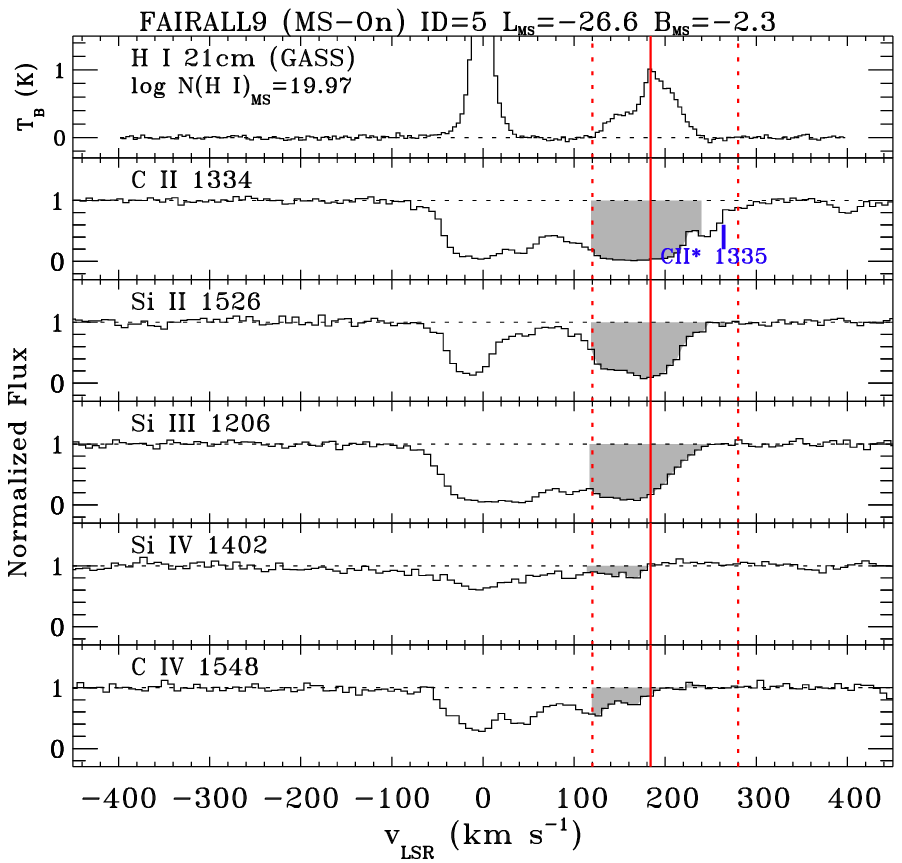}{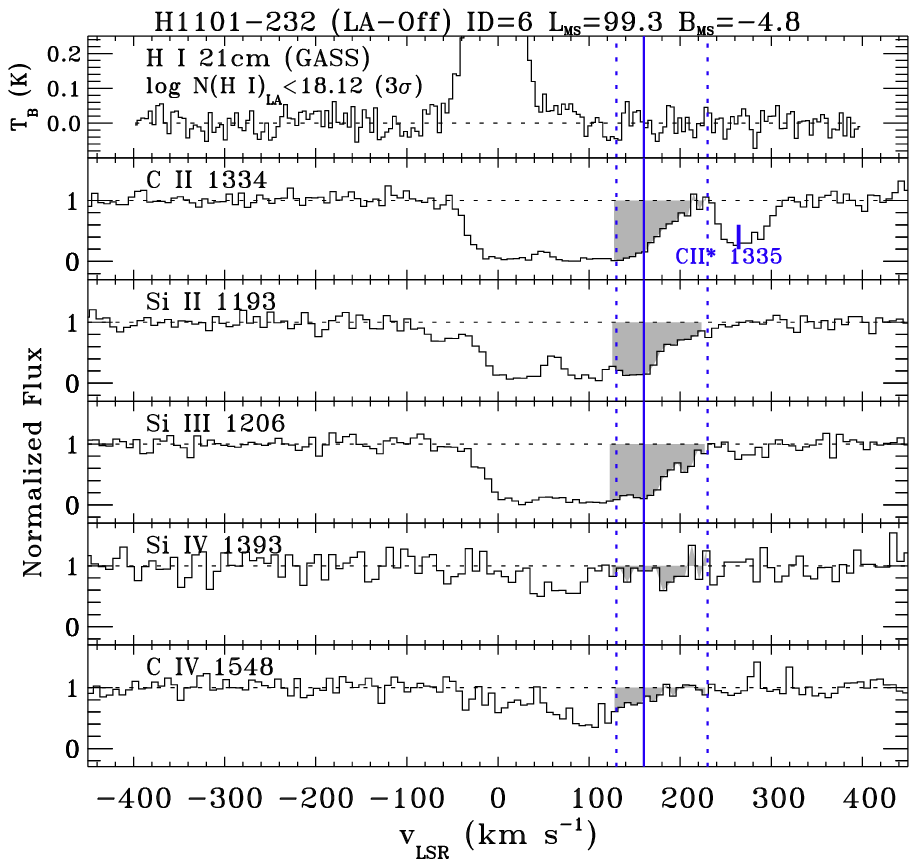}\end{figure}
\begin{figure}\epsscale{1.1}
\plottwo{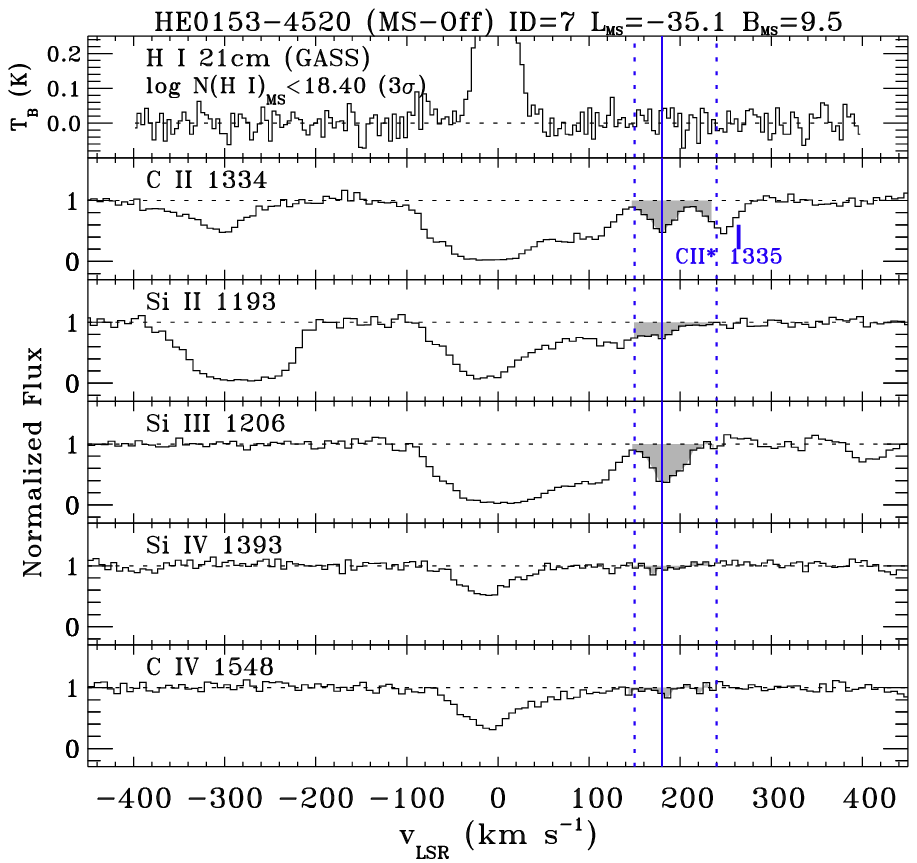}{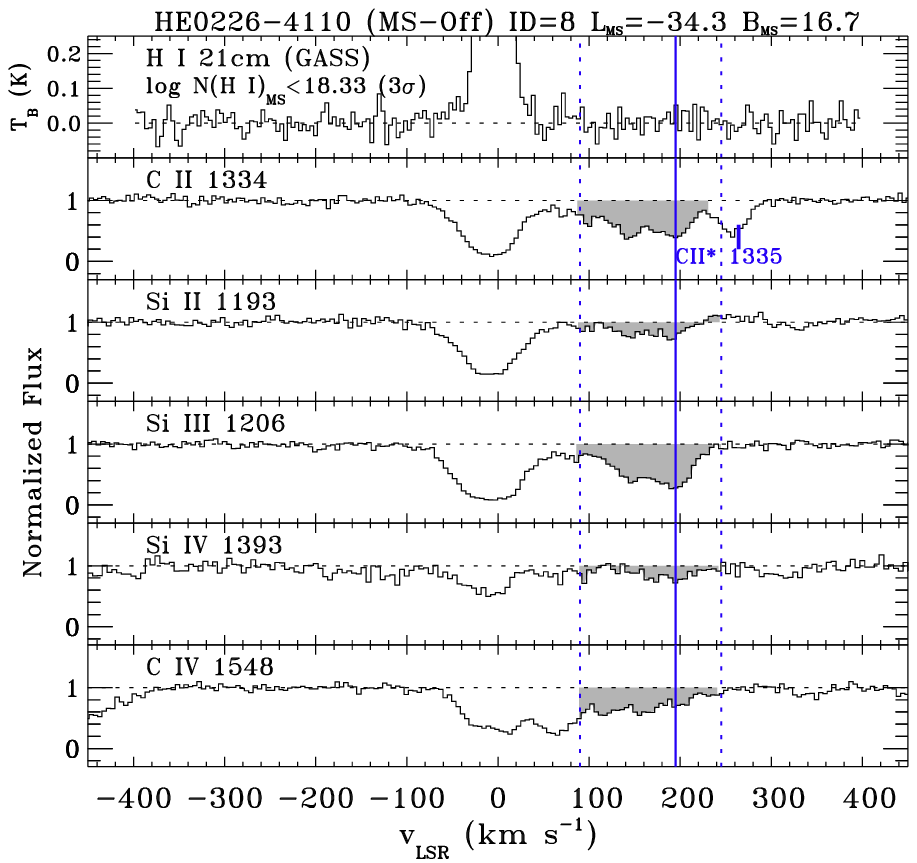}\end{figure}
\begin{figure}\epsscale{1.1}
\plottwo{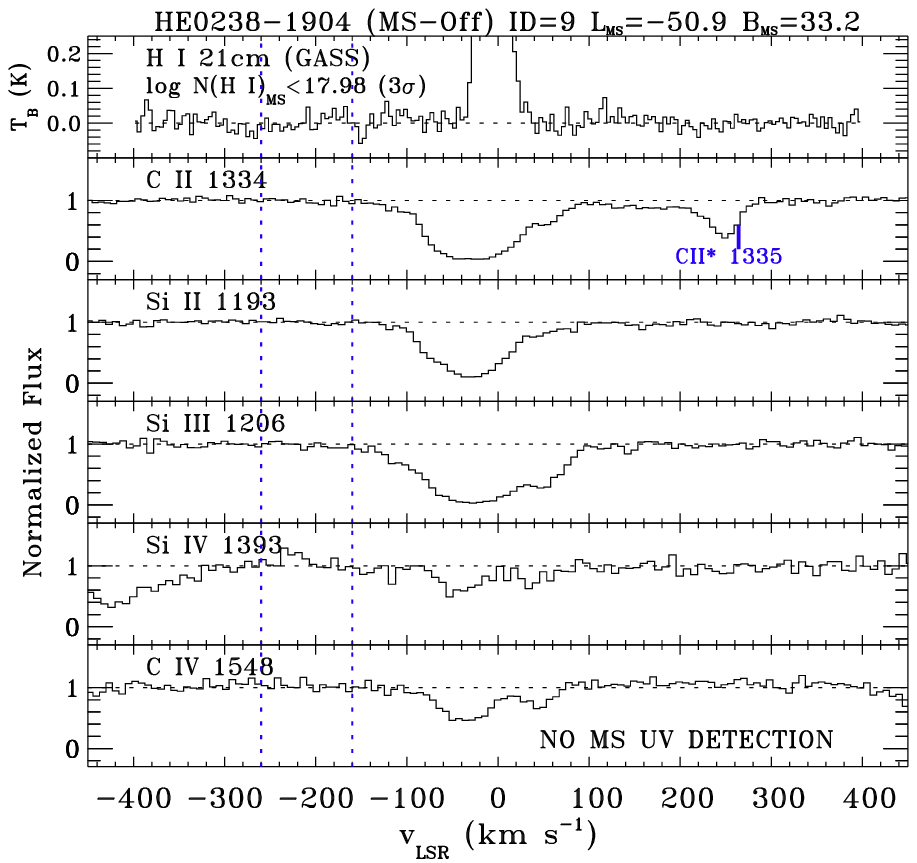}{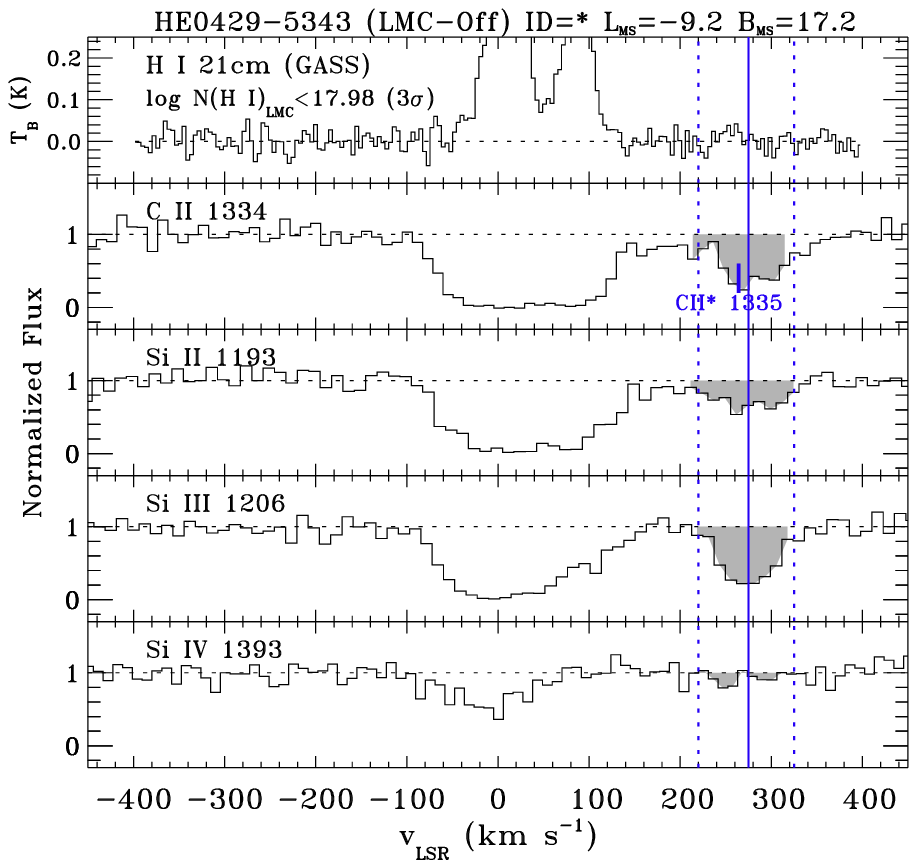}\end{figure}
\begin{figure}\epsscale{1.1}
\plottwo{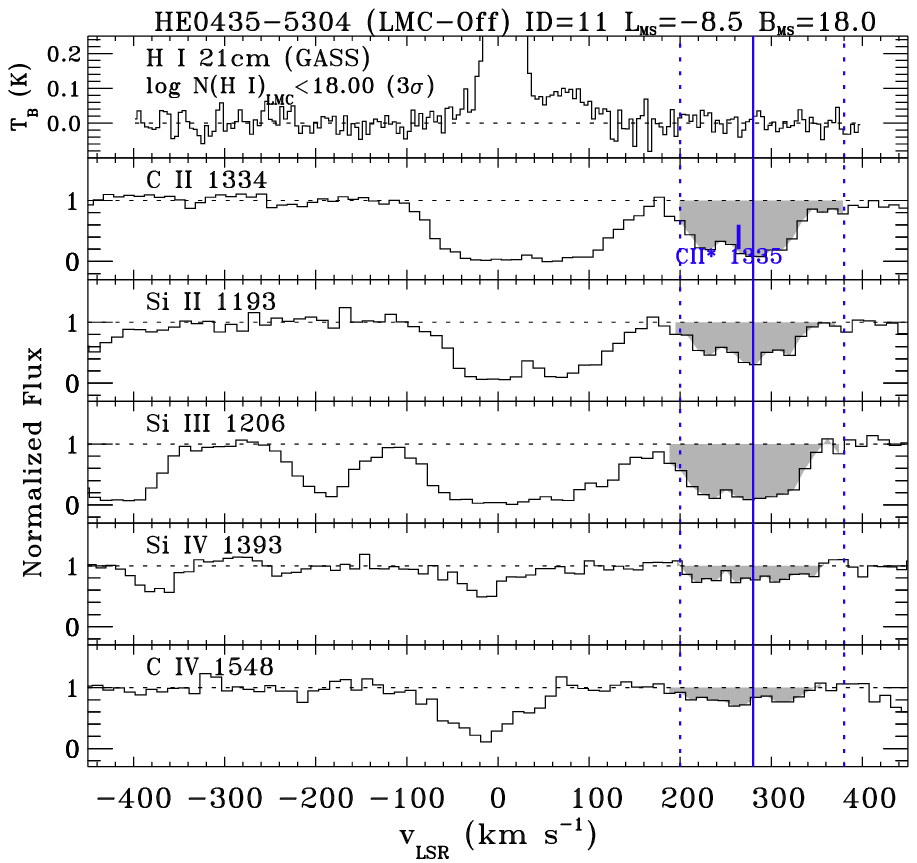}{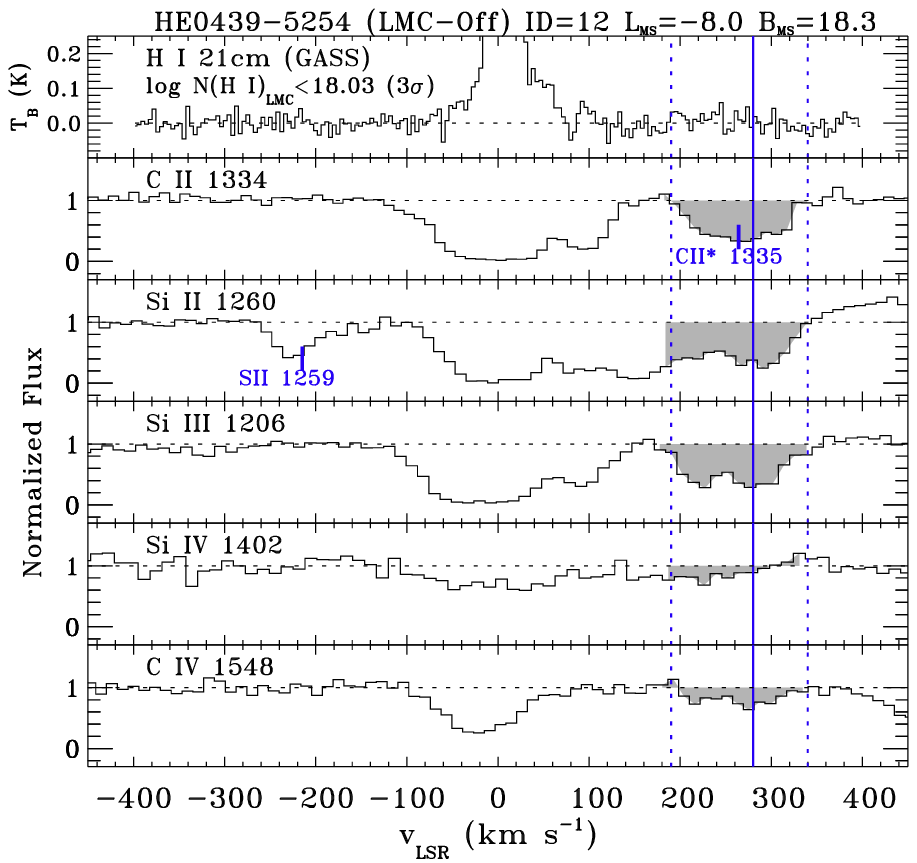}\end{figure}
\clearpage
\begin{figure}\epsscale{1.1}
\plottwo{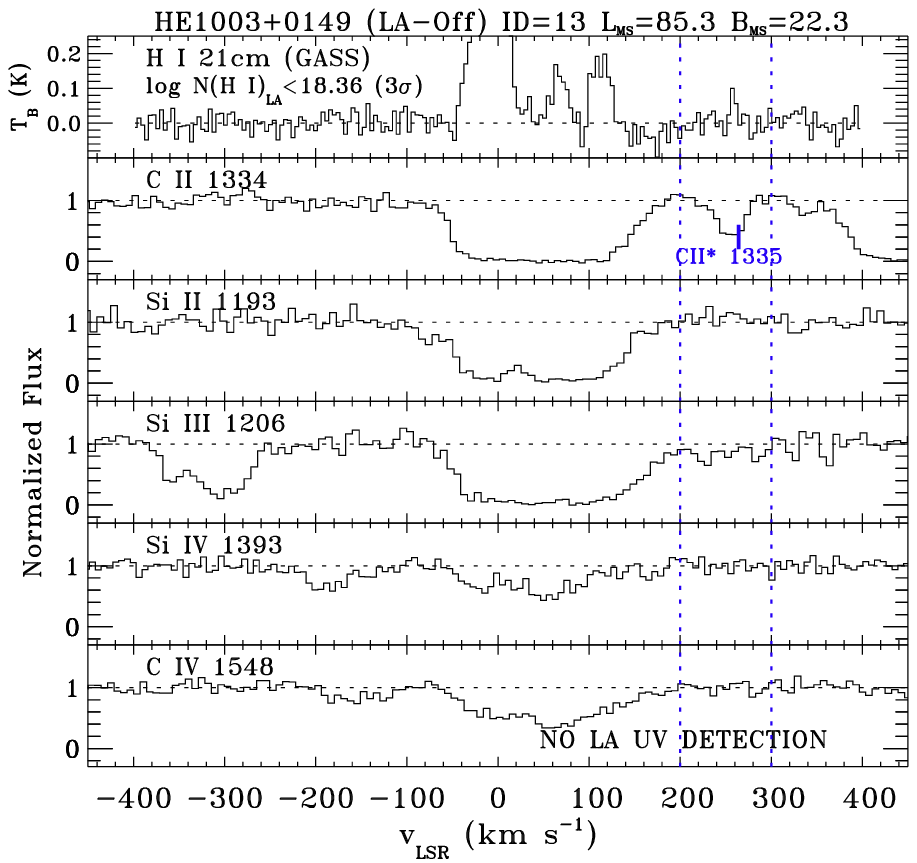}{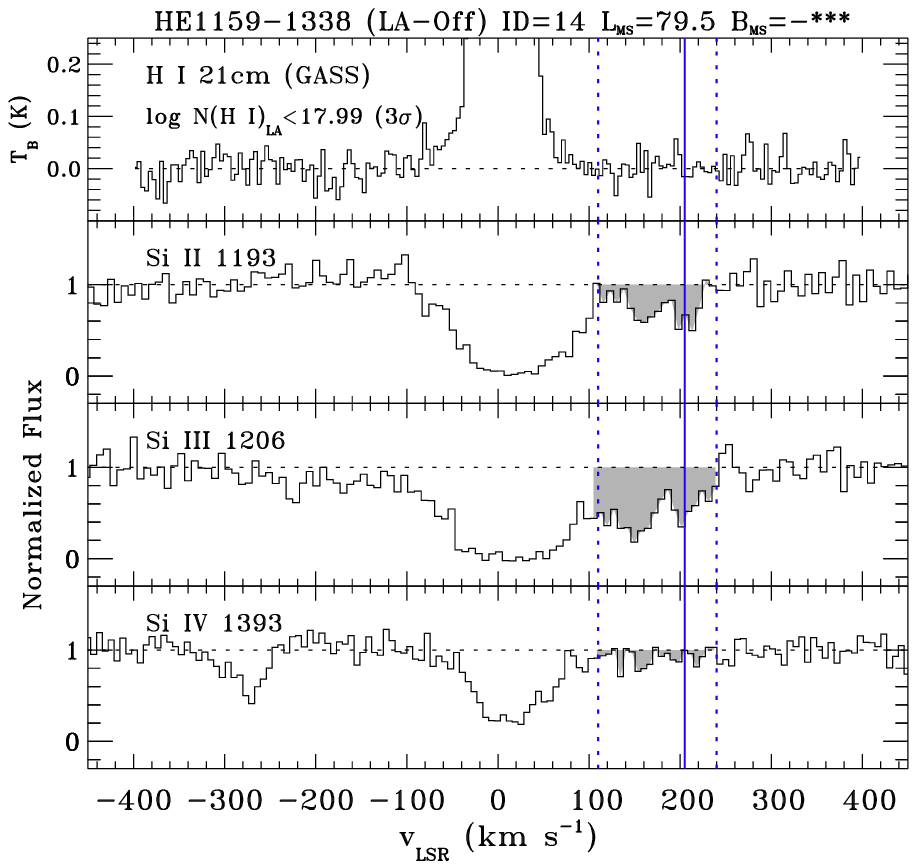}\end{figure}
\begin{figure}\epsscale{1.1}
\plottwo{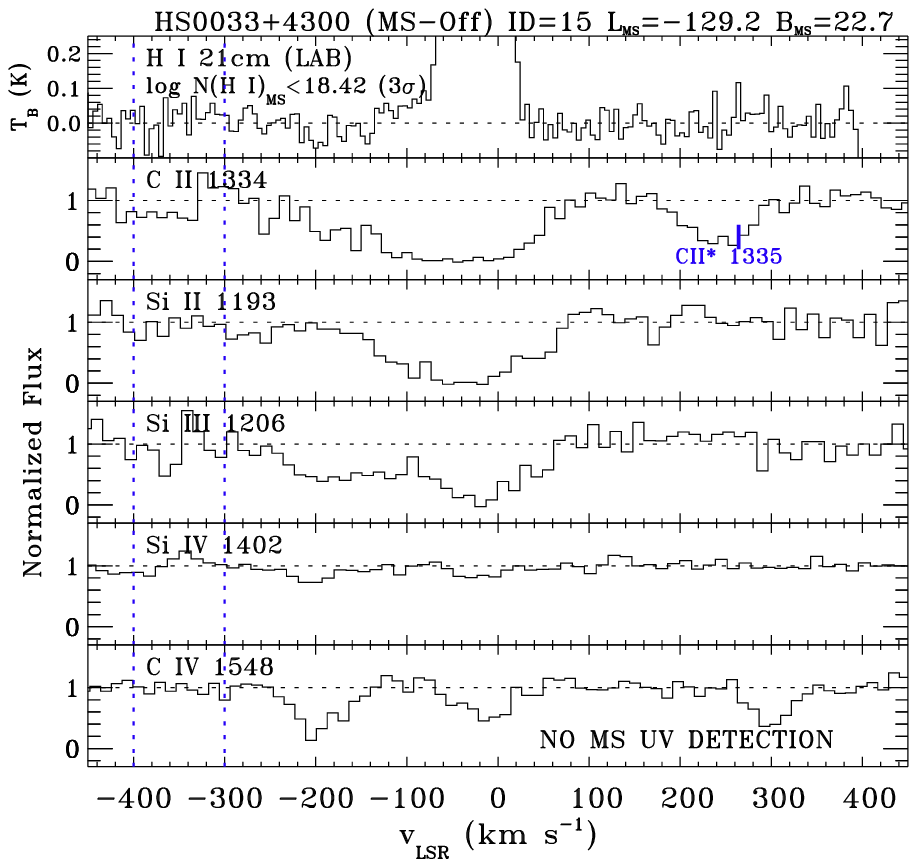}{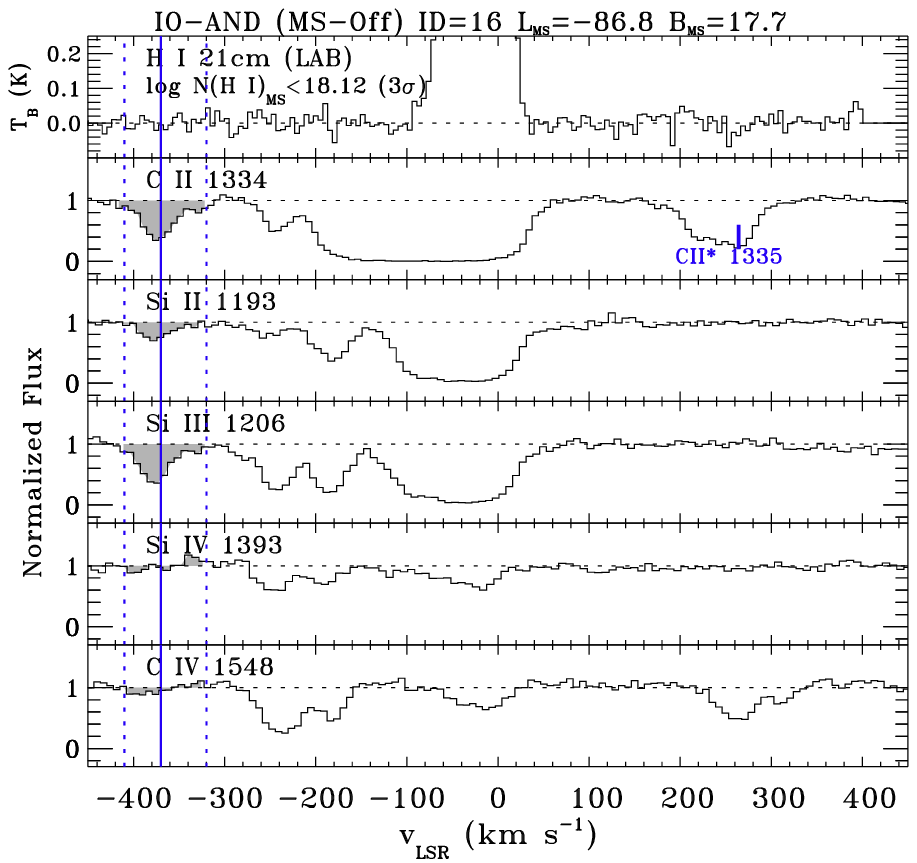}\end{figure}
\begin{figure}\epsscale{1.1}
\plottwo{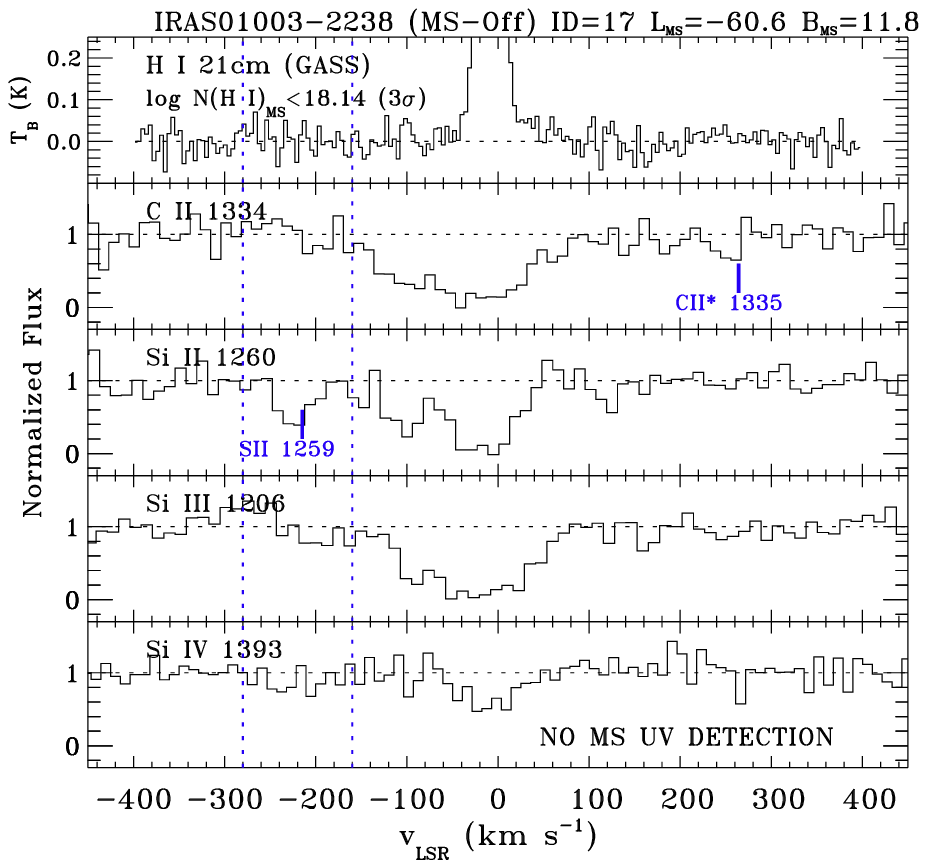}{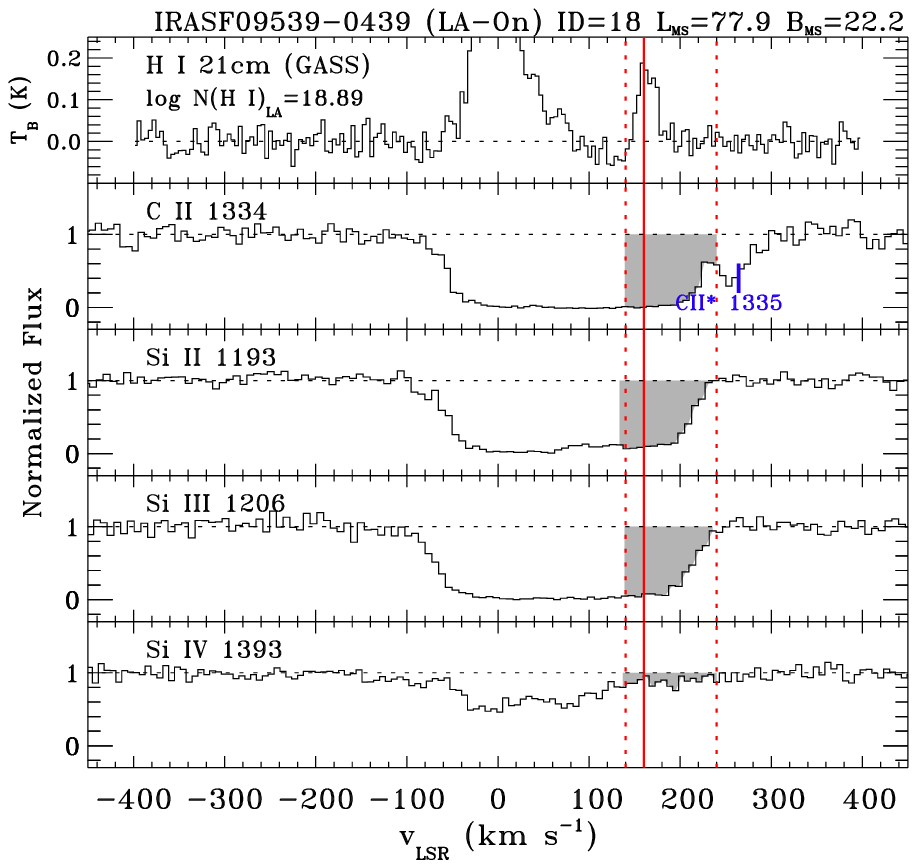}\end{figure}
\begin{figure}\epsscale{1.1}
\plottwo{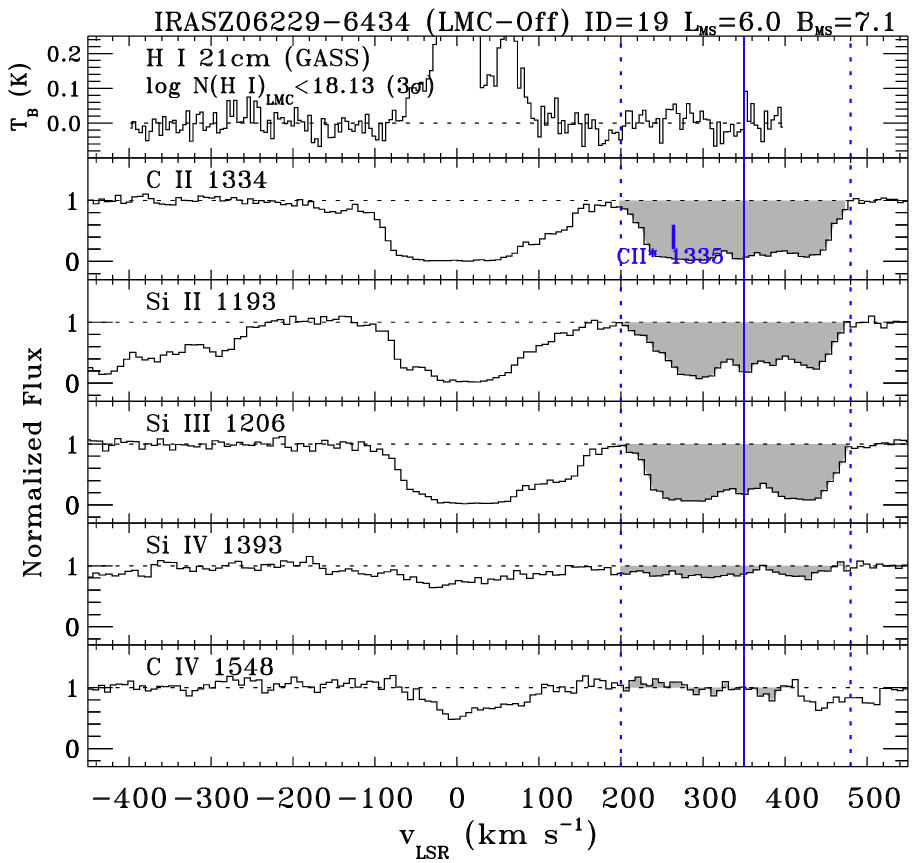}{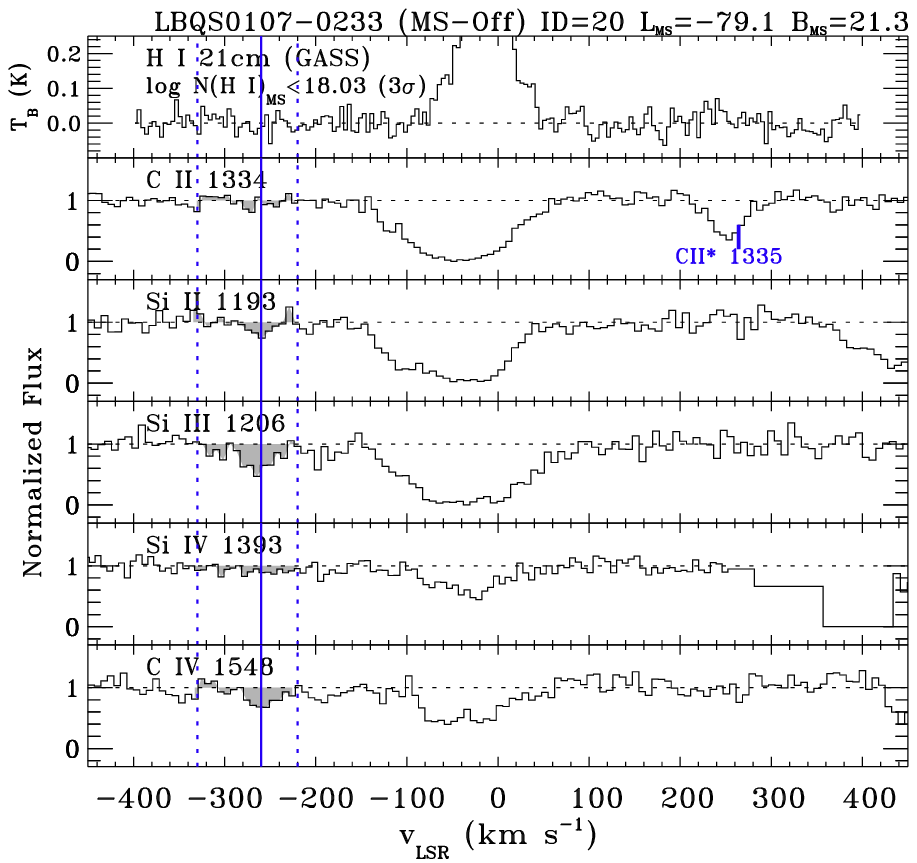}\end{figure}
\begin{figure}\epsscale{1.1}
\plottwo{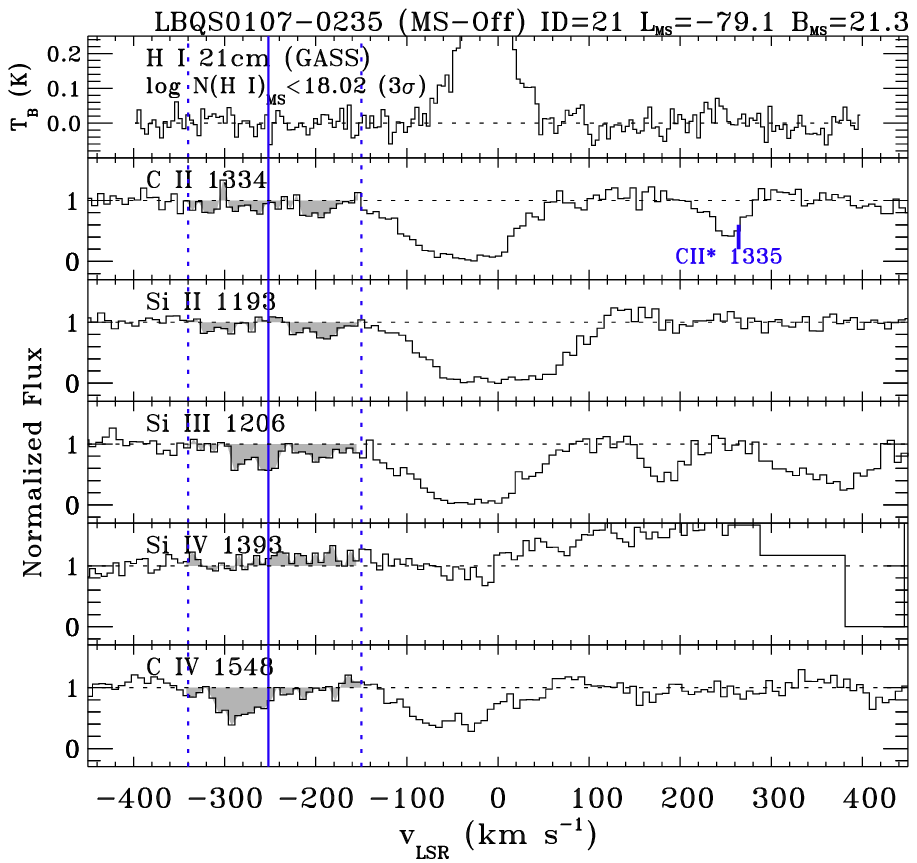}{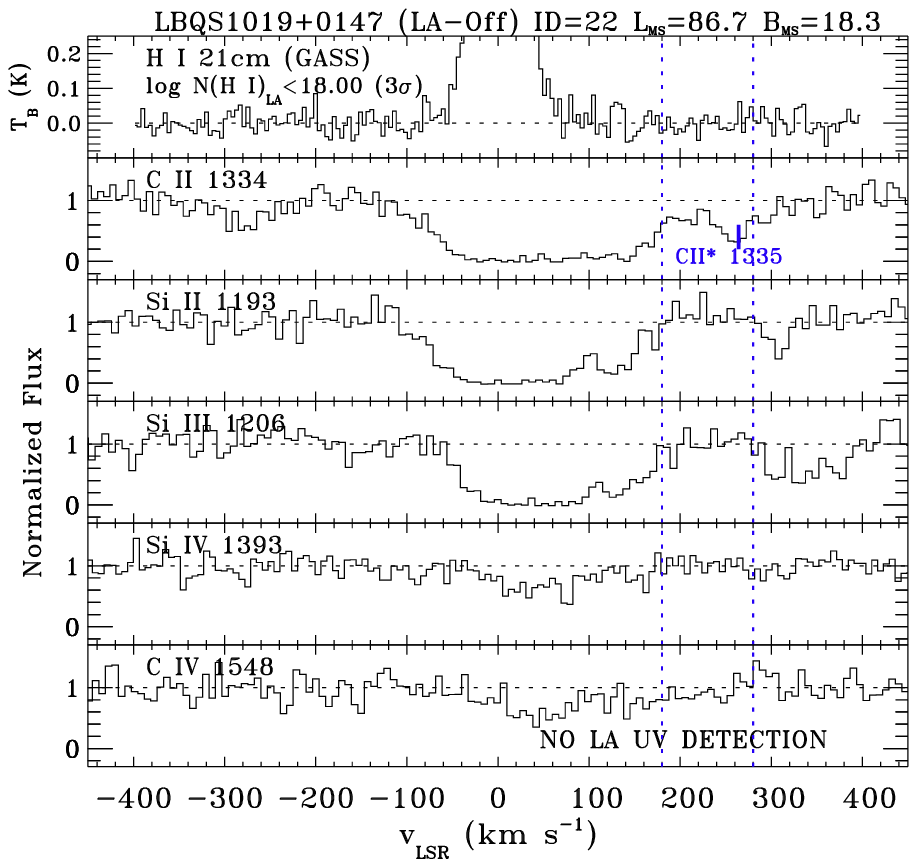}\end{figure}
\begin{figure}\epsscale{1.1}
\plottwo{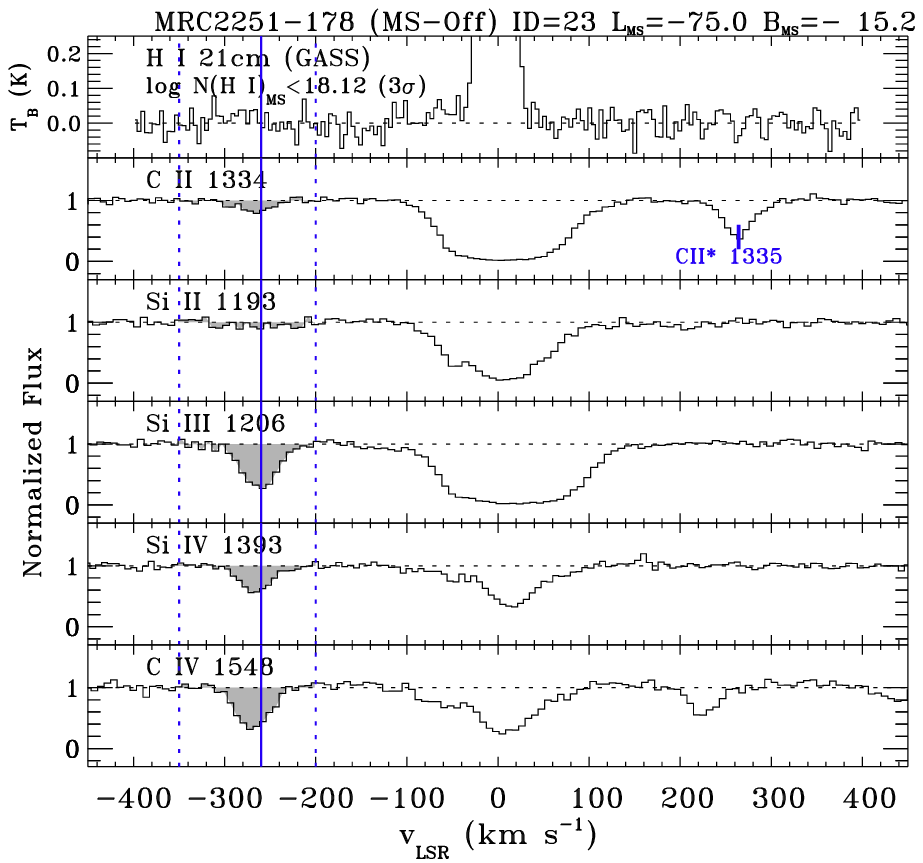}{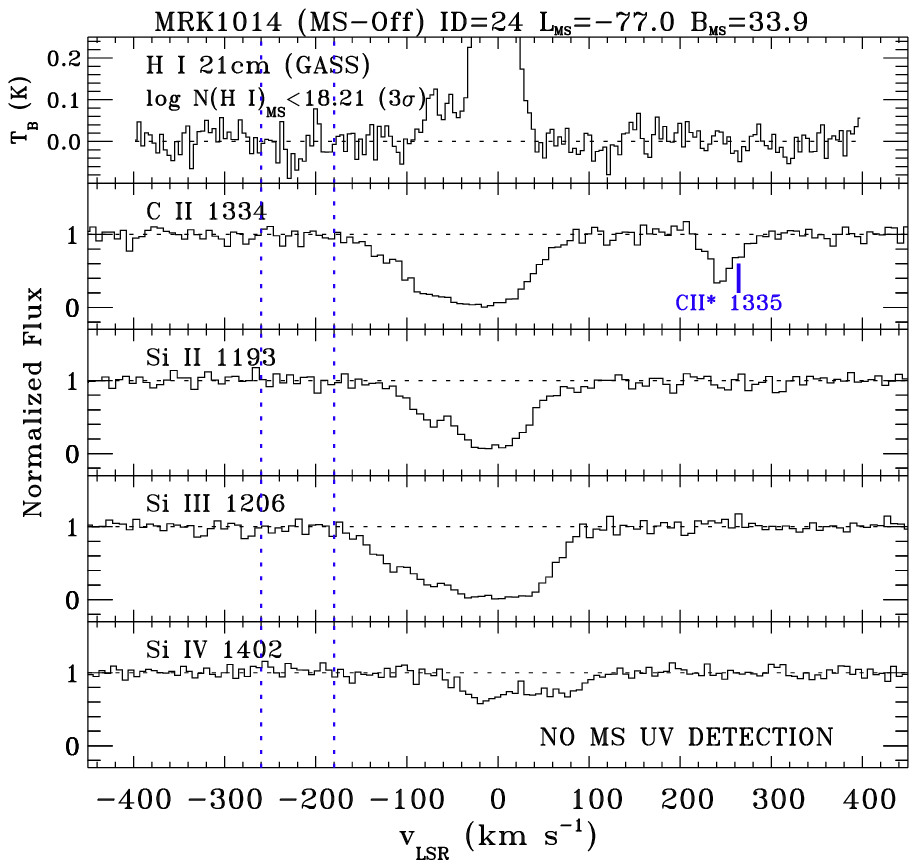}\end{figure}
\clearpage
\begin{figure}\epsscale{1.1}
\plottwo{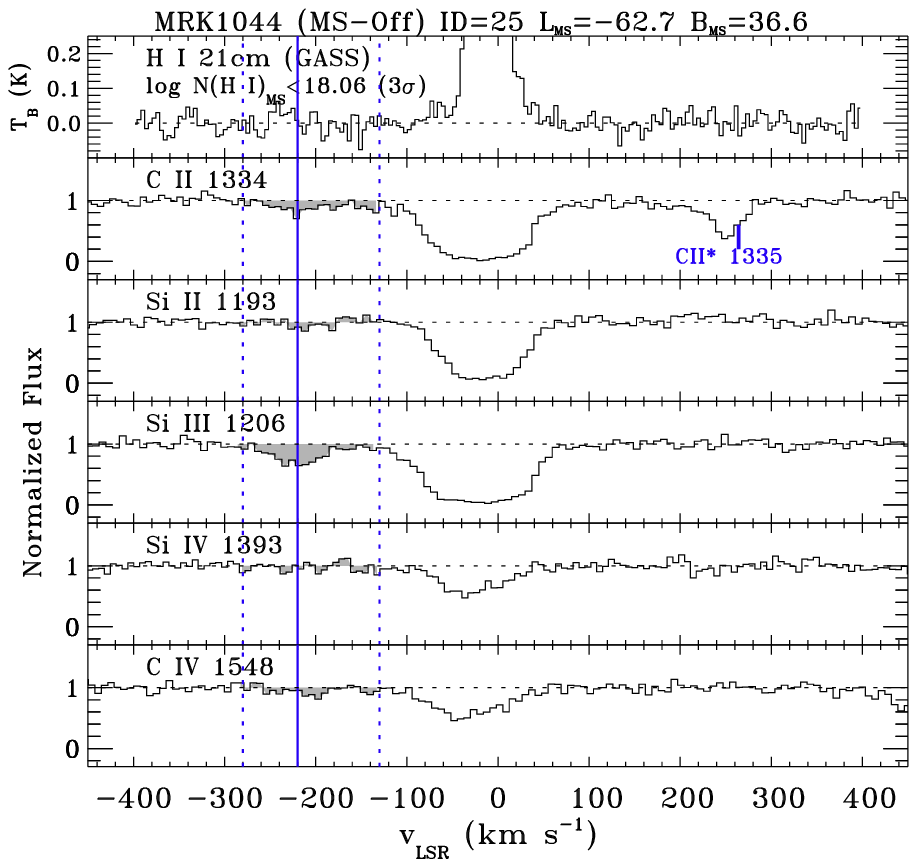}{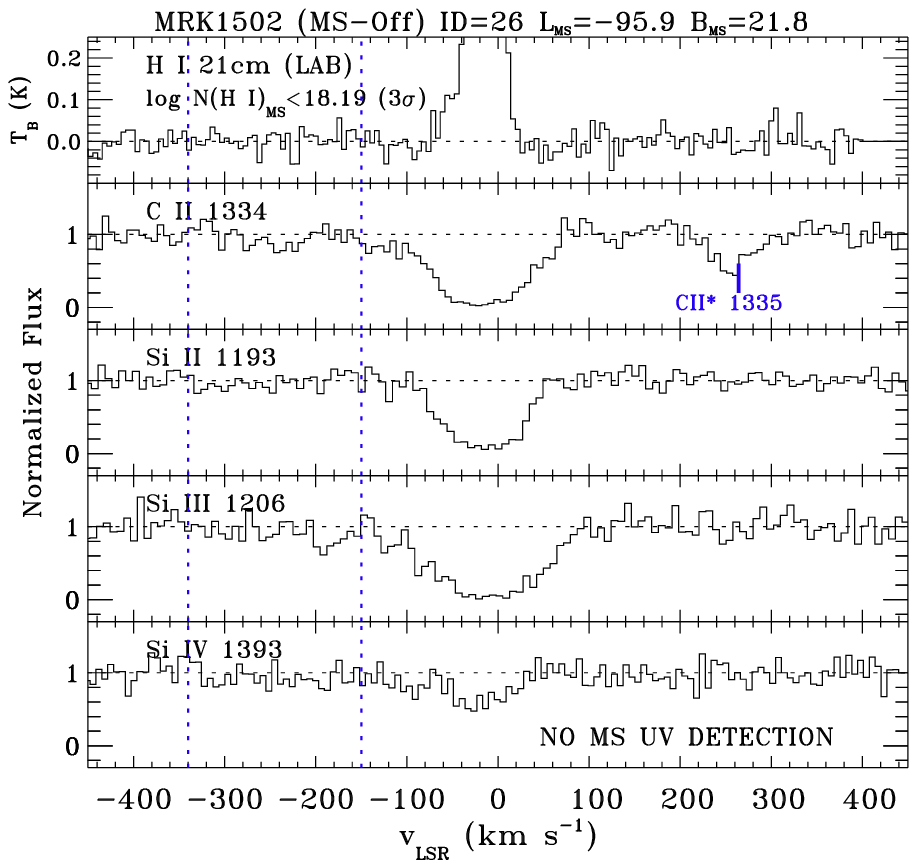}\end{figure}
\begin{figure}\epsscale{1.1}
\plottwo{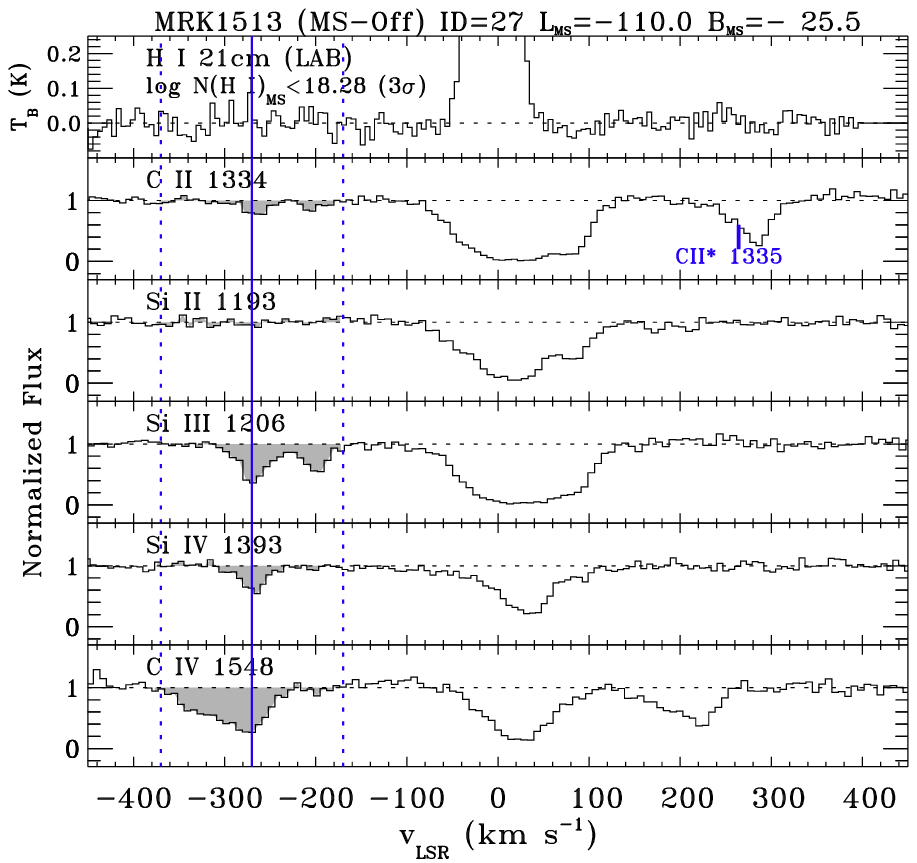}{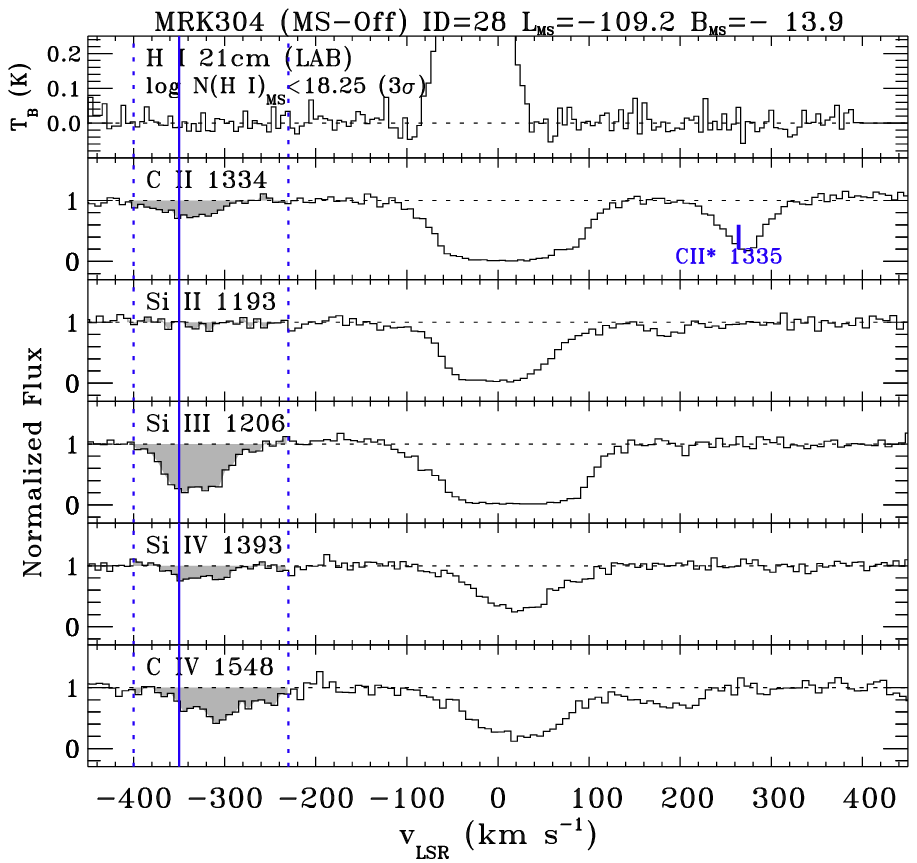}\end{figure}
\begin{figure}\epsscale{1.1}
\plottwo{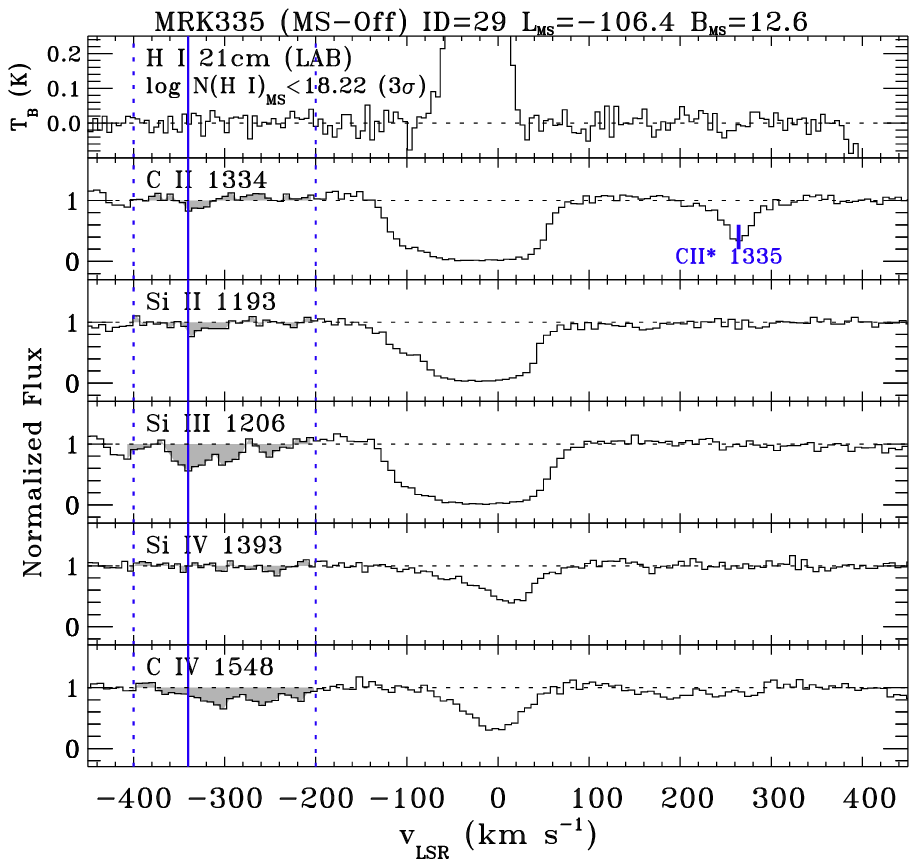}{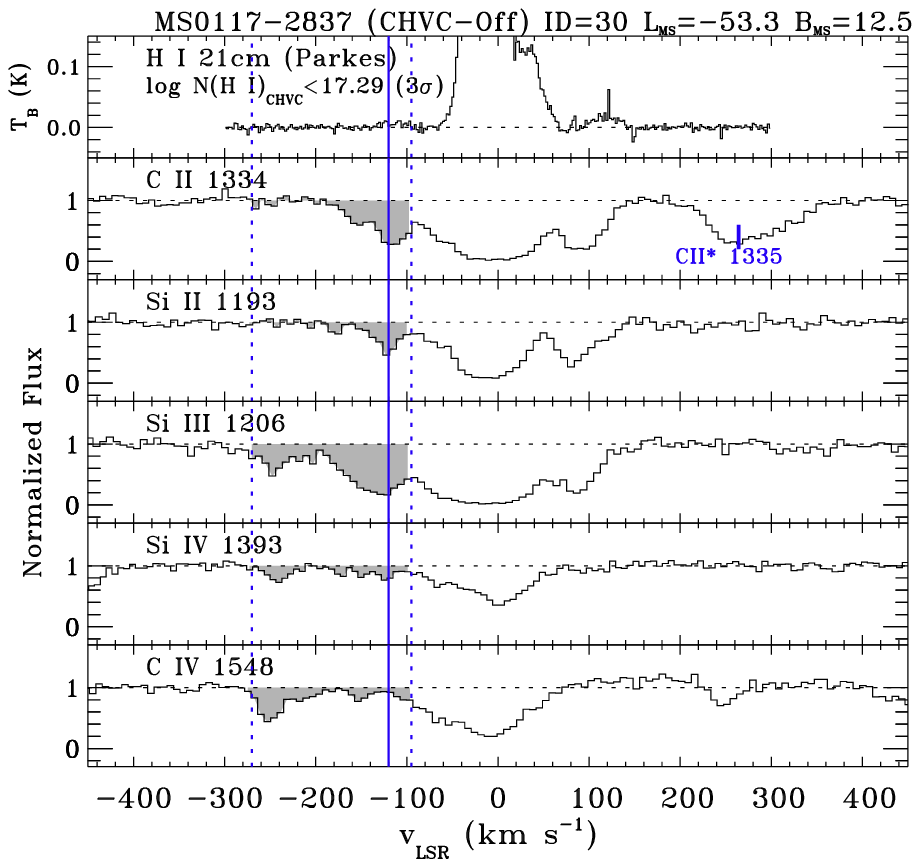}\end{figure}
\begin{figure}\epsscale{1.1}
\plottwo{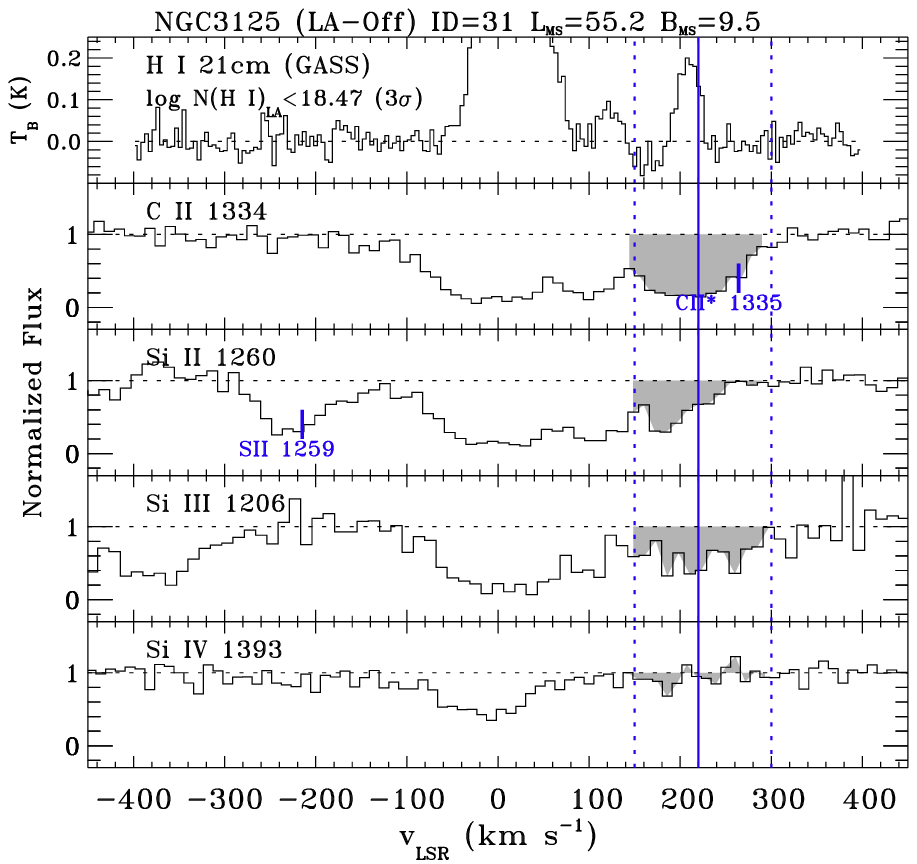}{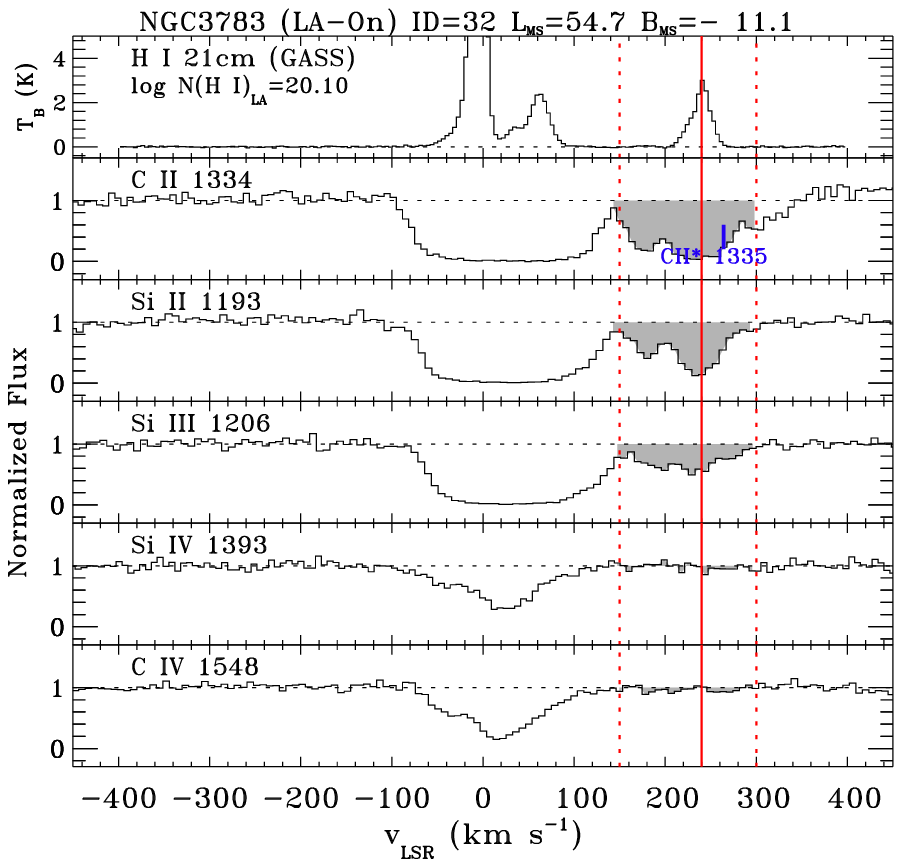}\end{figure}
\begin{figure}\epsscale{1.1}
\plottwo{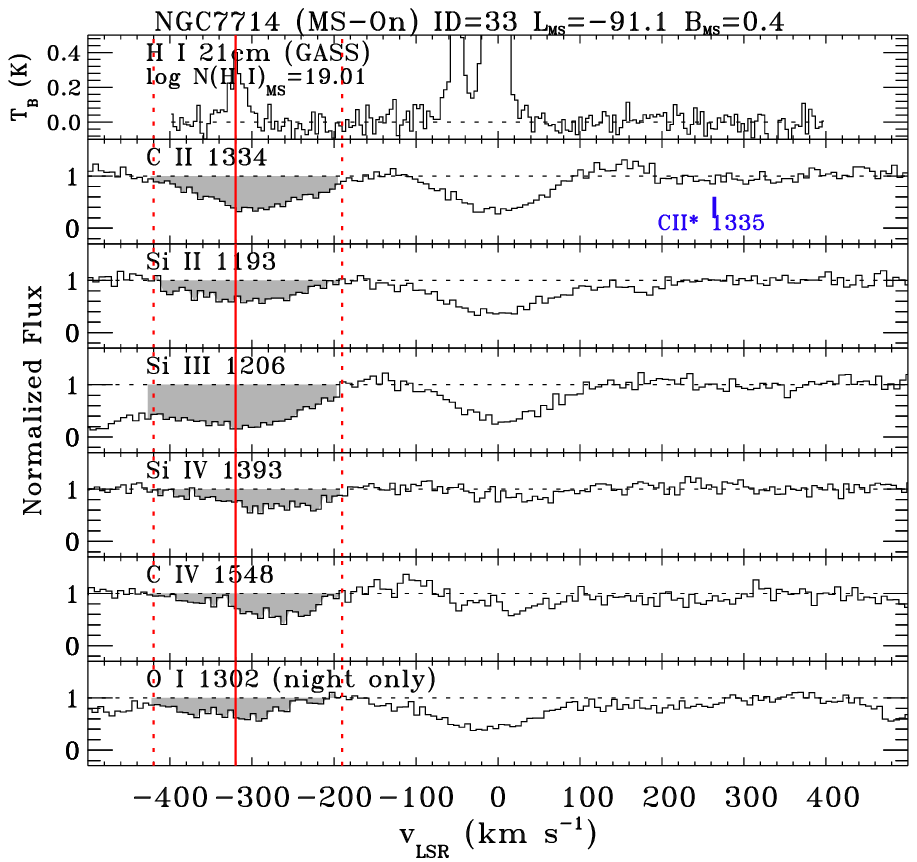}{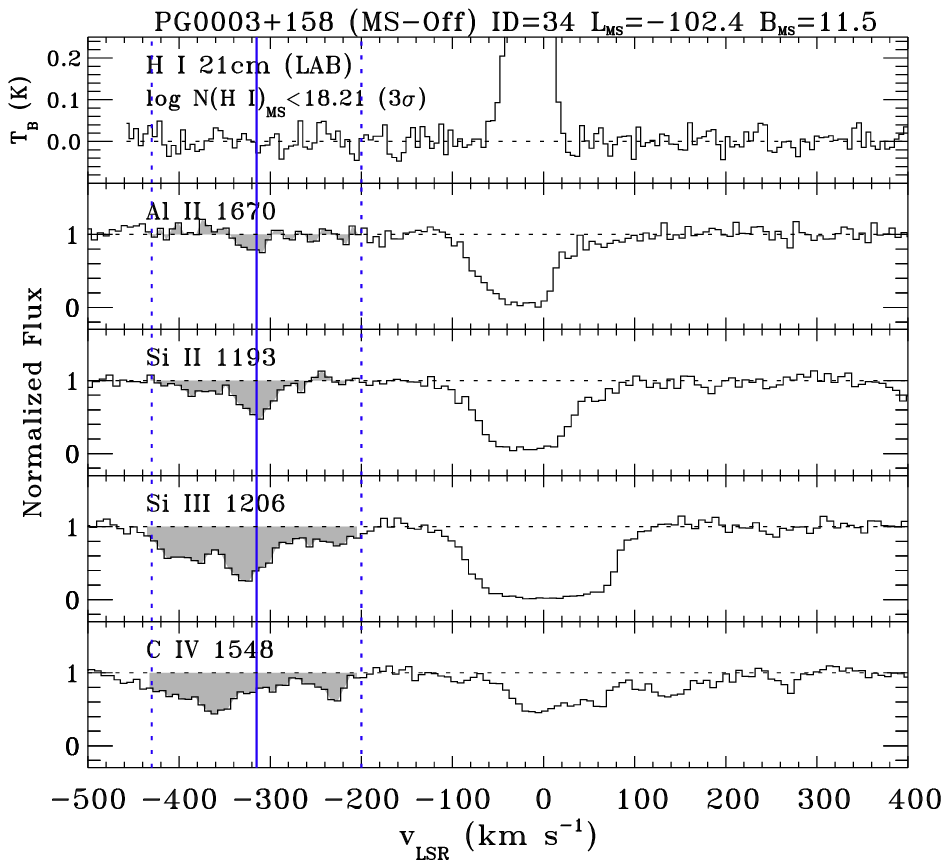}\end{figure}
\begin{figure}\epsscale{1.1}
\plottwo{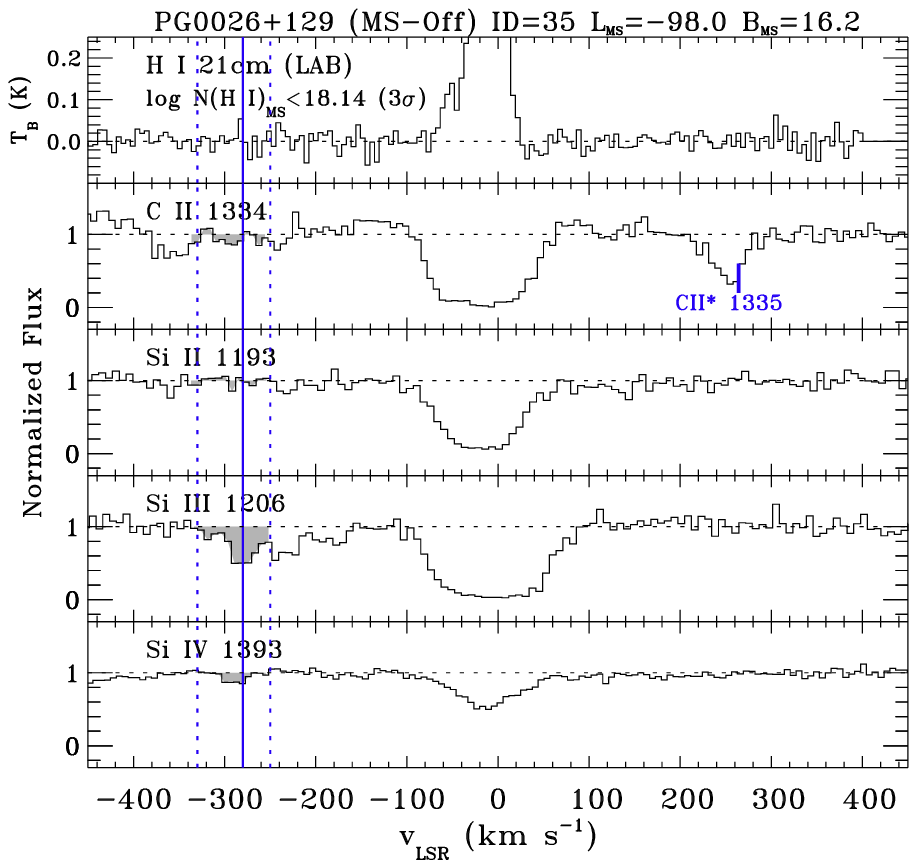}{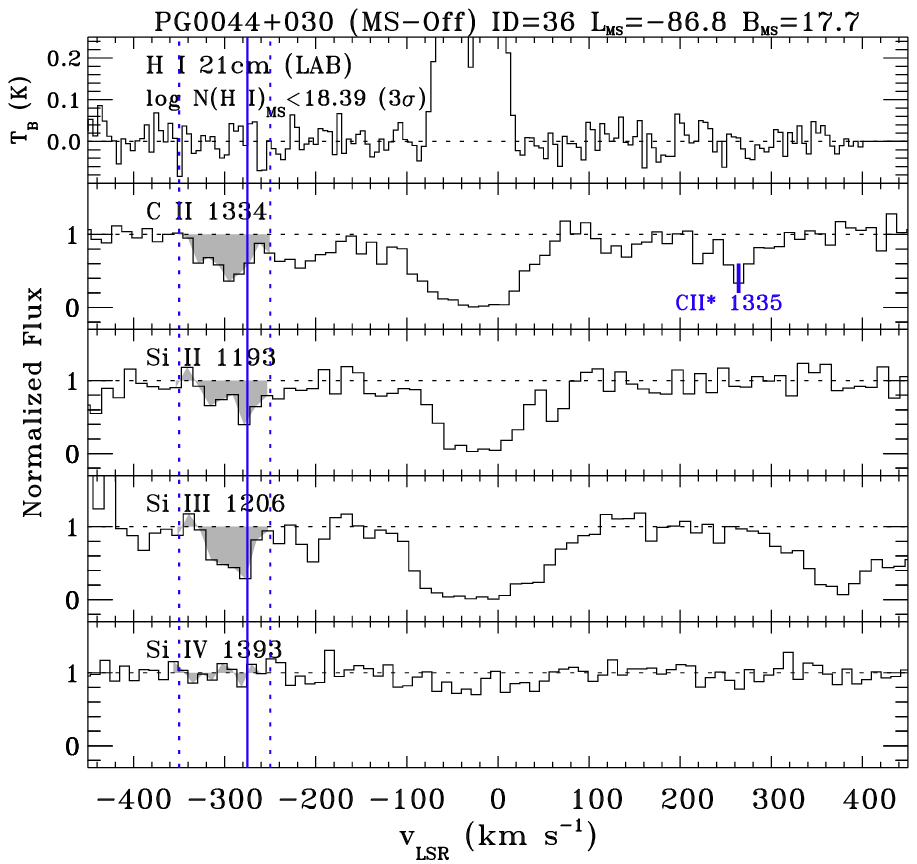}\end{figure}
\clearpage
\begin{figure}\epsscale{1.1}
\plottwo{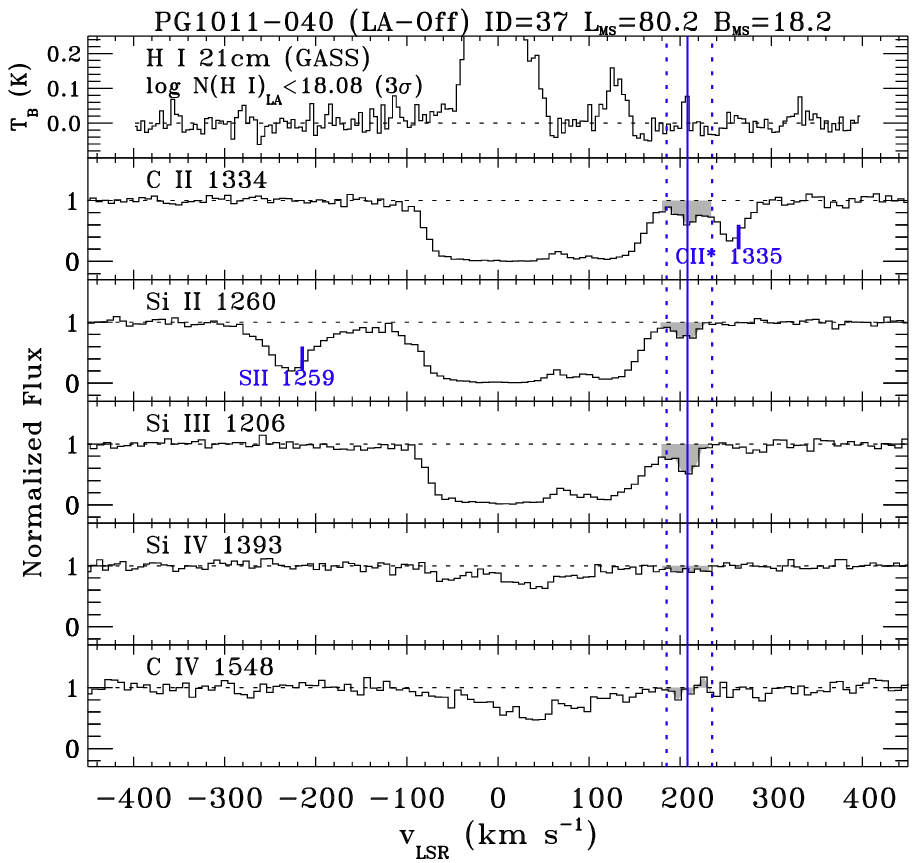}{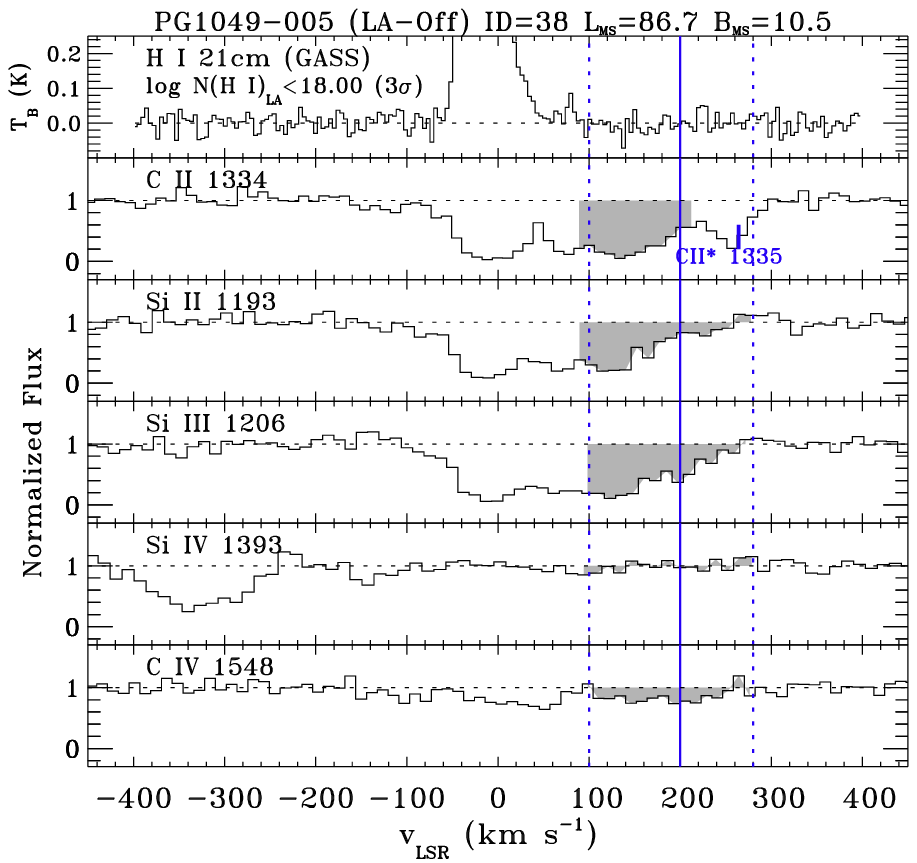}\end{figure}
\begin{figure}\epsscale{1.1}
\plottwo{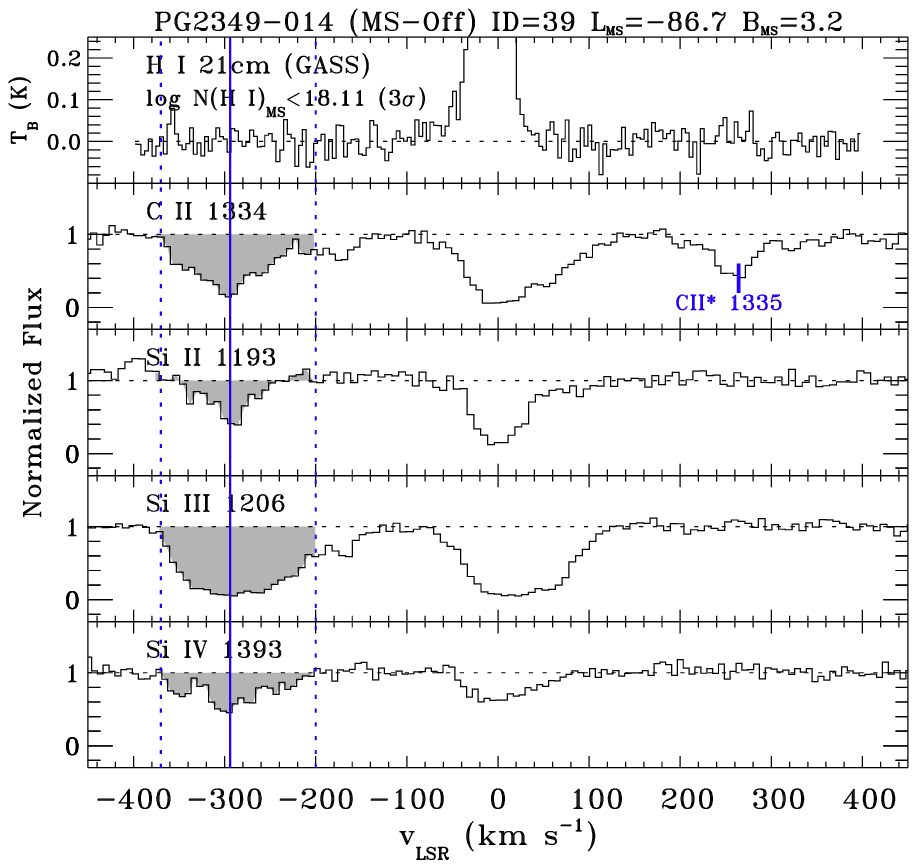}{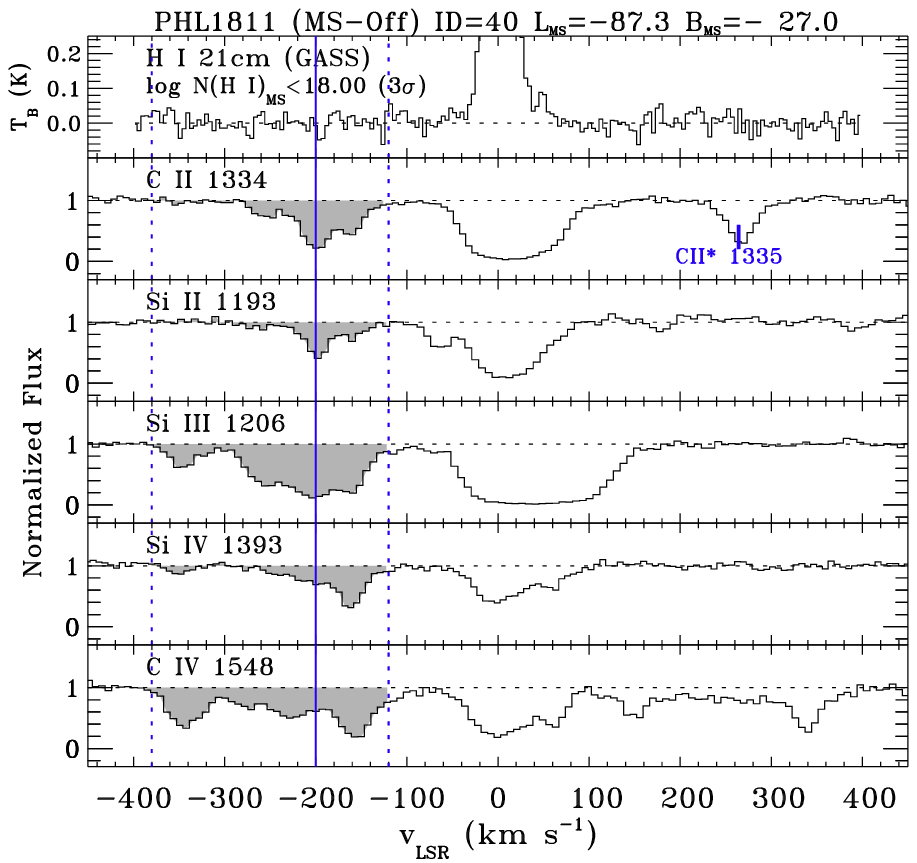}\end{figure}
\begin{figure}\epsscale{1.1}
\plottwo{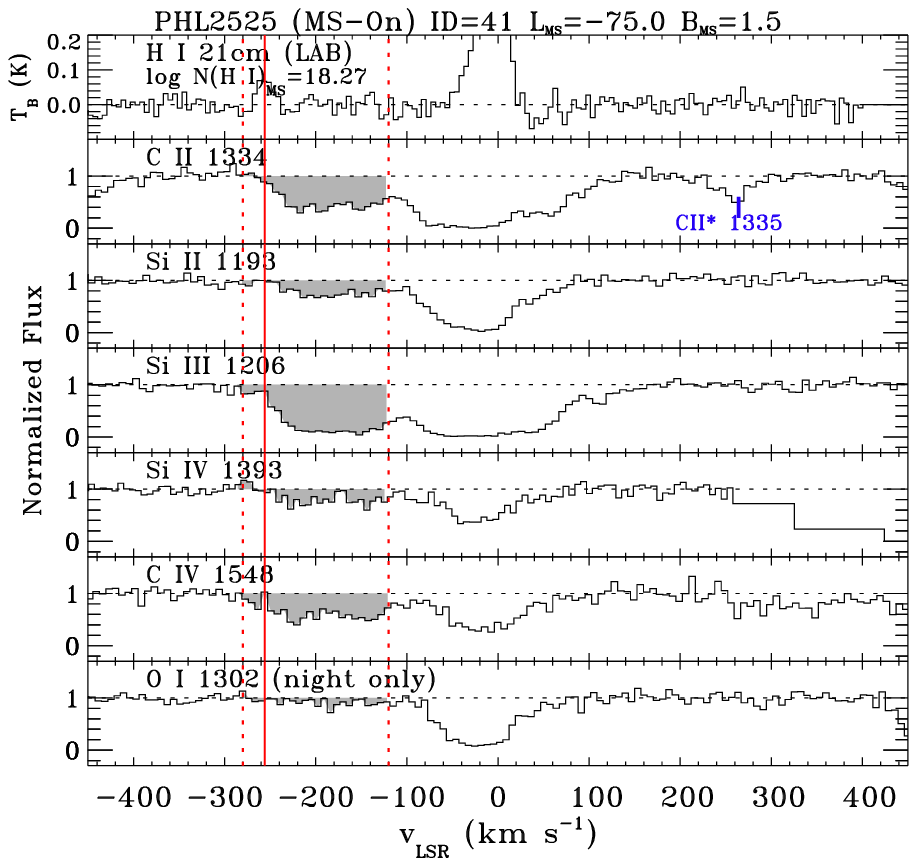}{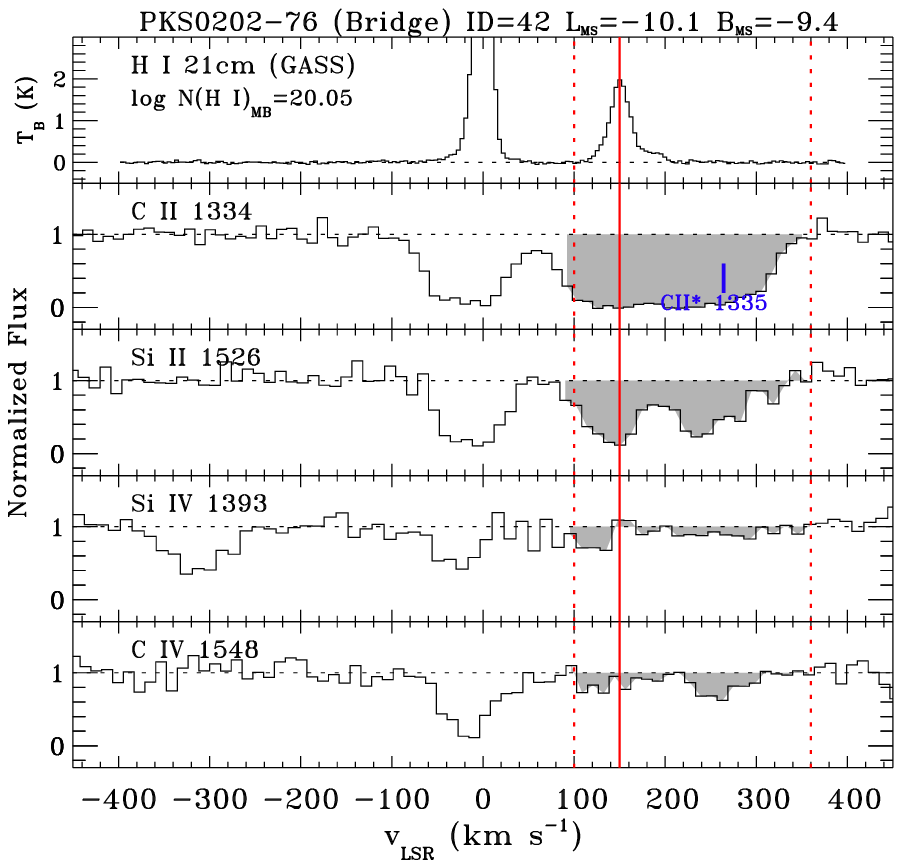}\end{figure}
\begin{figure}\epsscale{1.1}
\plottwo{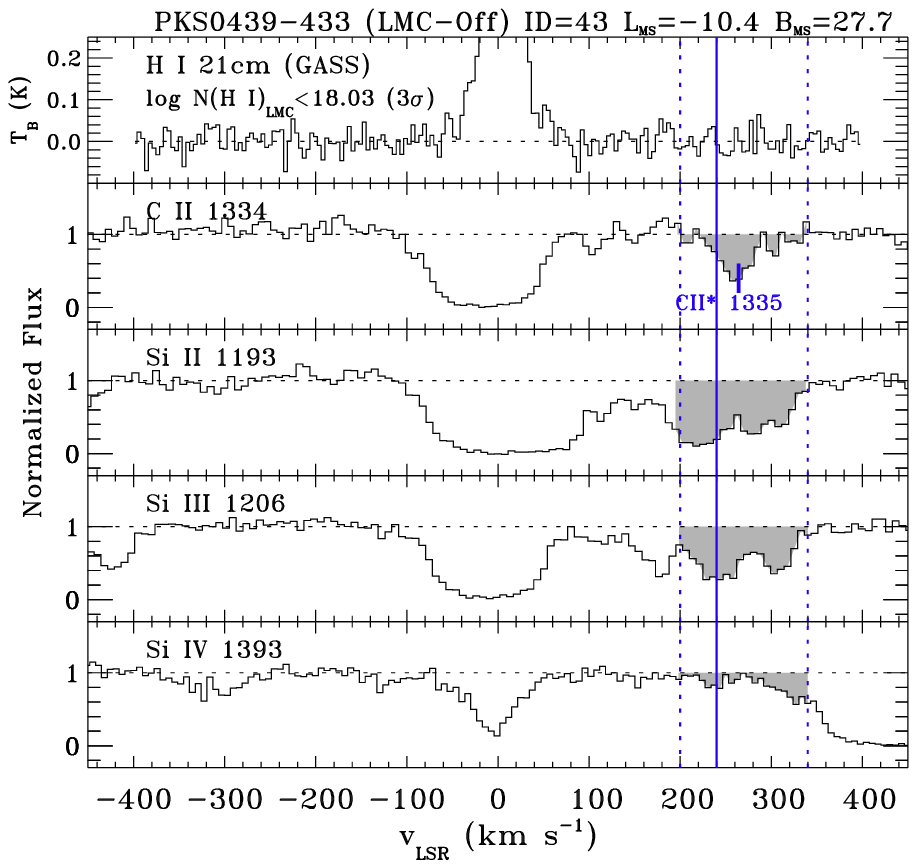}{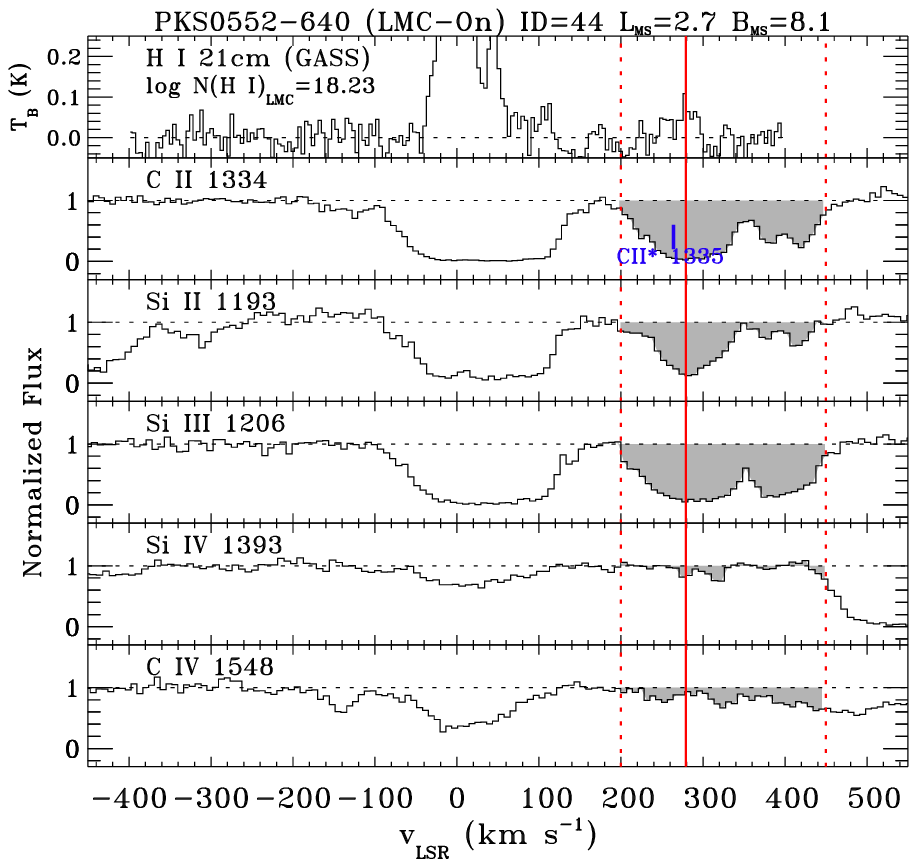}\end{figure}
\begin{figure}\epsscale{1.1}
\plottwo{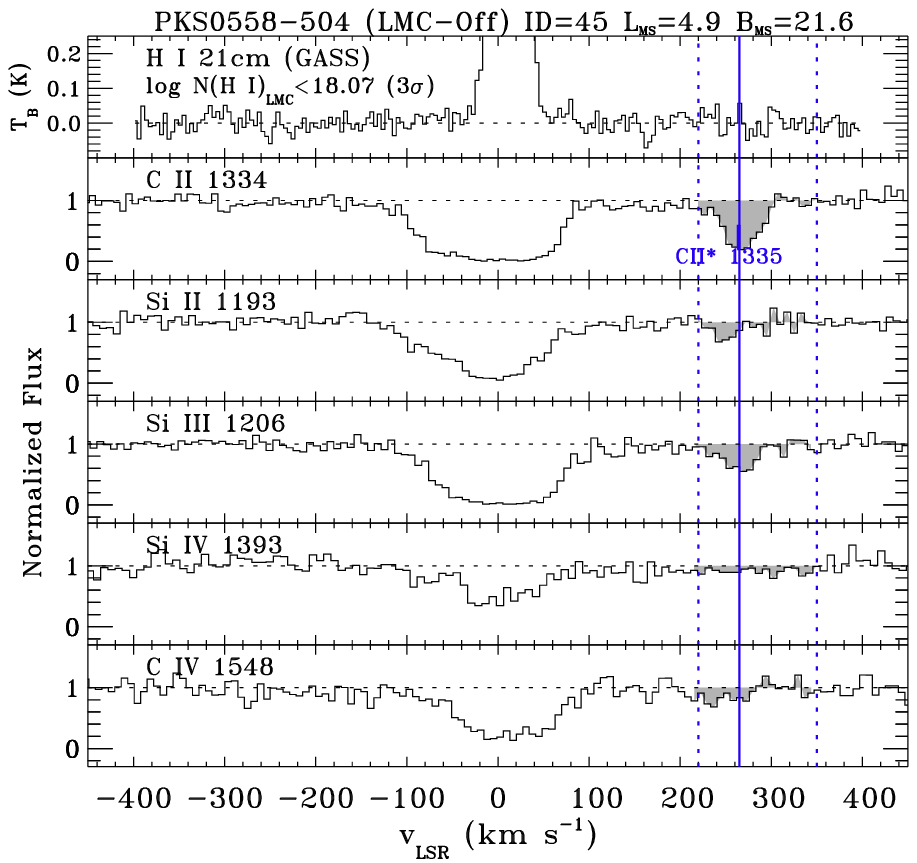}{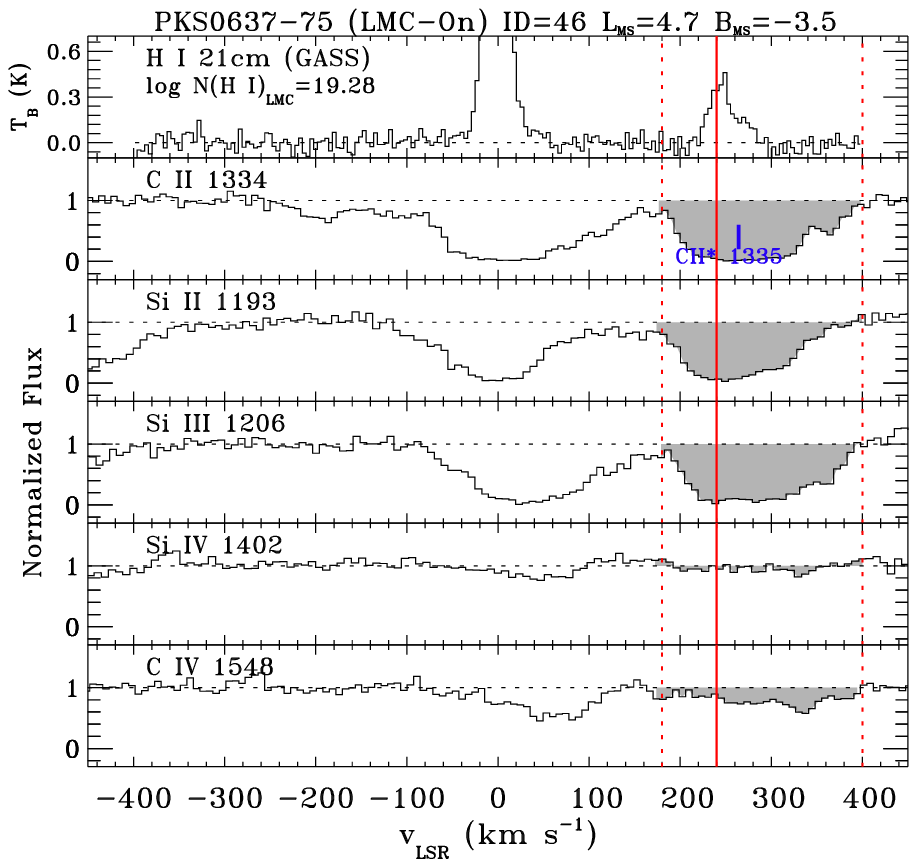}\end{figure}
\begin{figure}\epsscale{1.1}
\plottwo{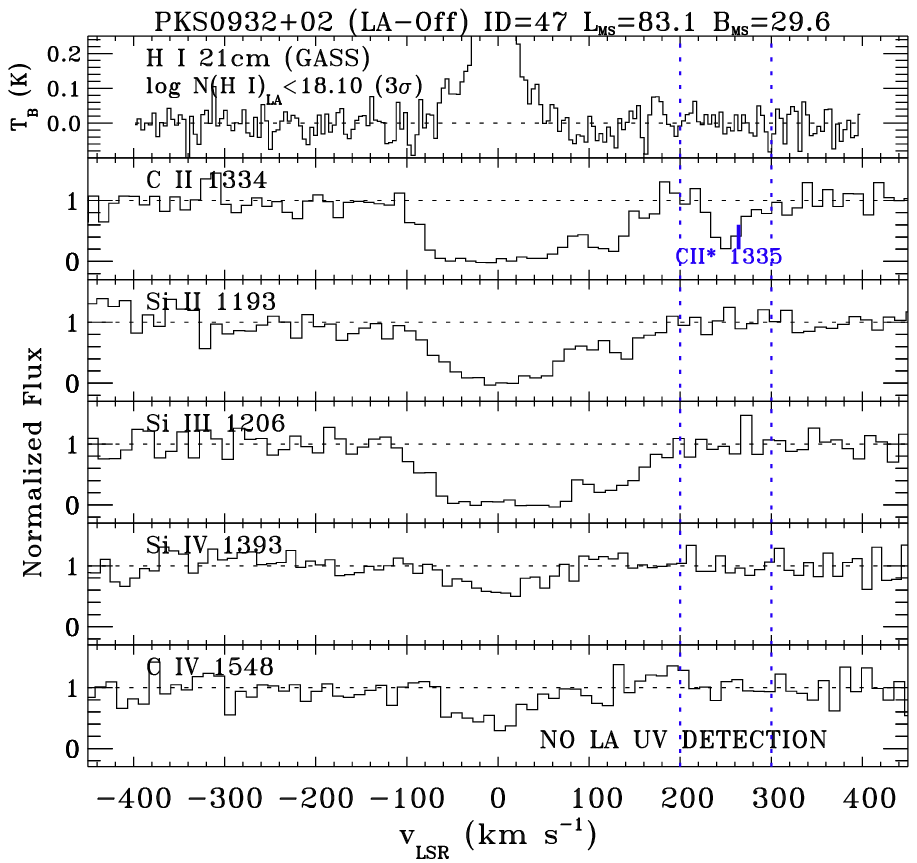}{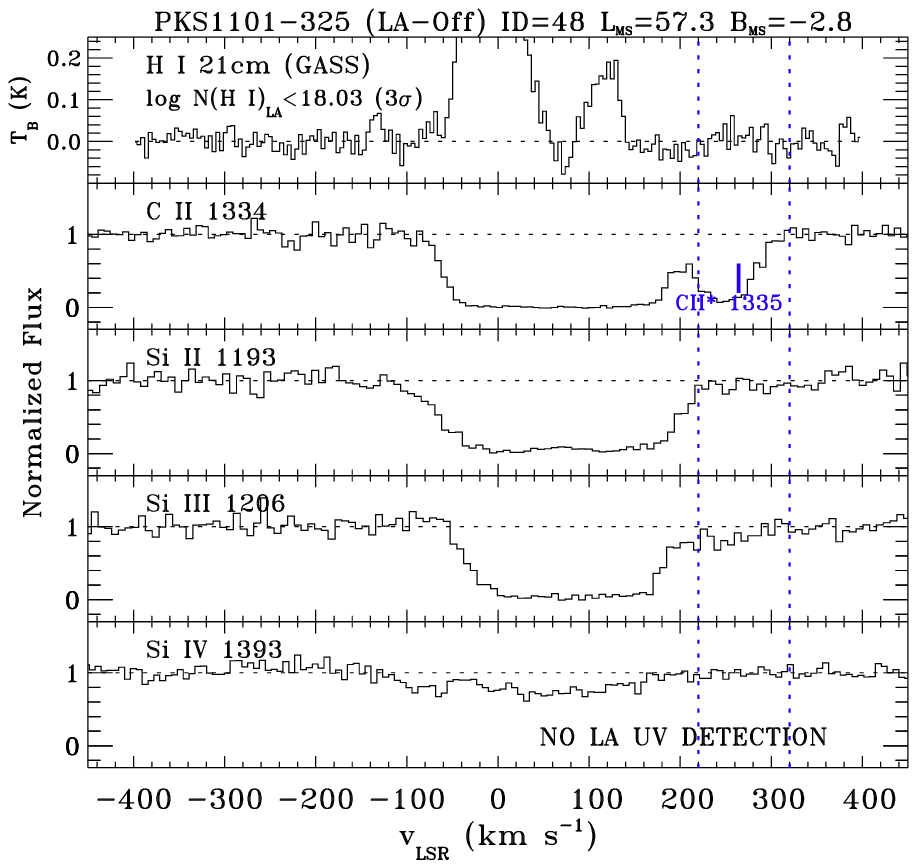}\end{figure}
\clearpage
\begin{figure}\epsscale{1.1}
\plottwo{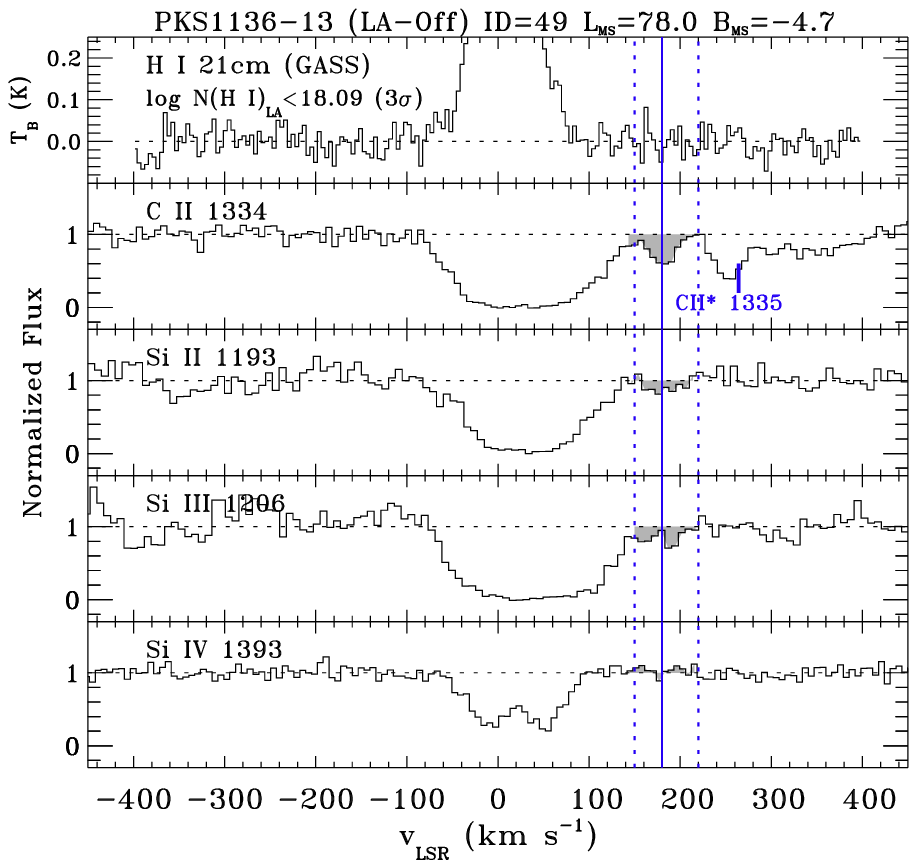}{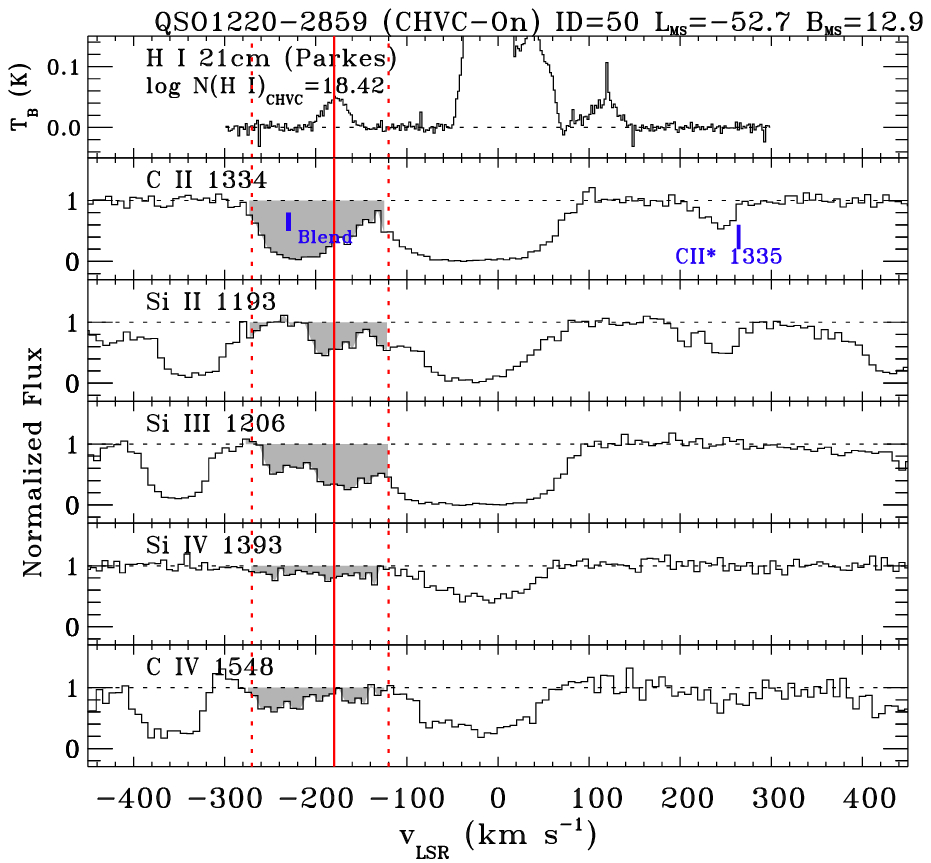}\end{figure}
\begin{figure}\epsscale{1.1}
\plottwo{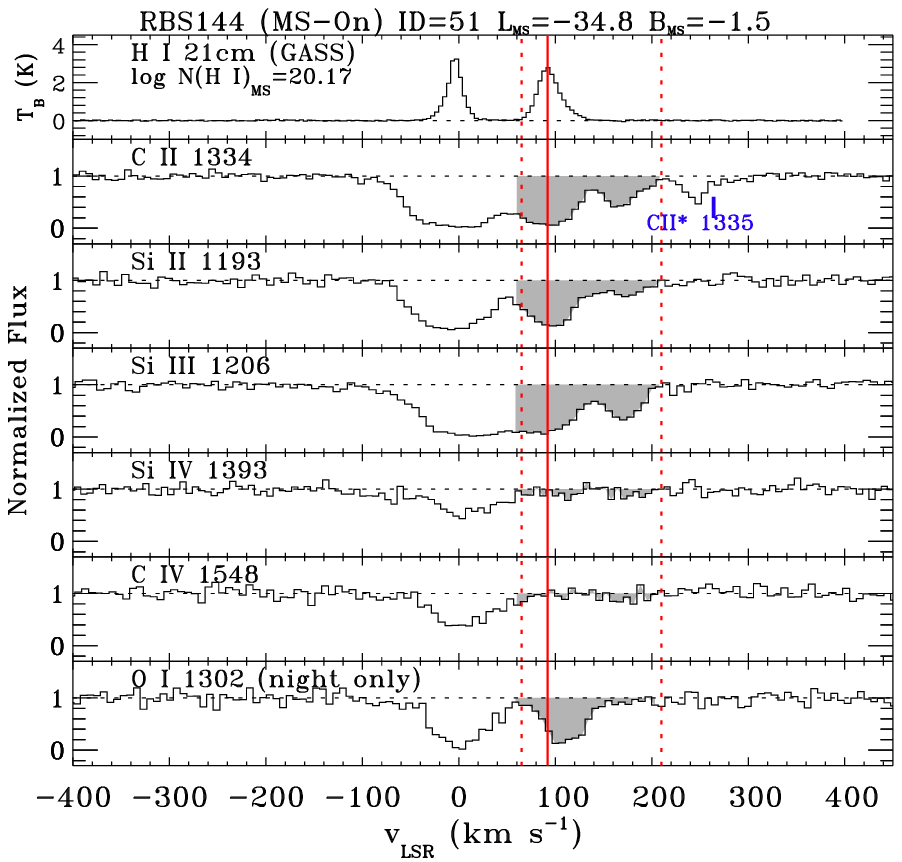}{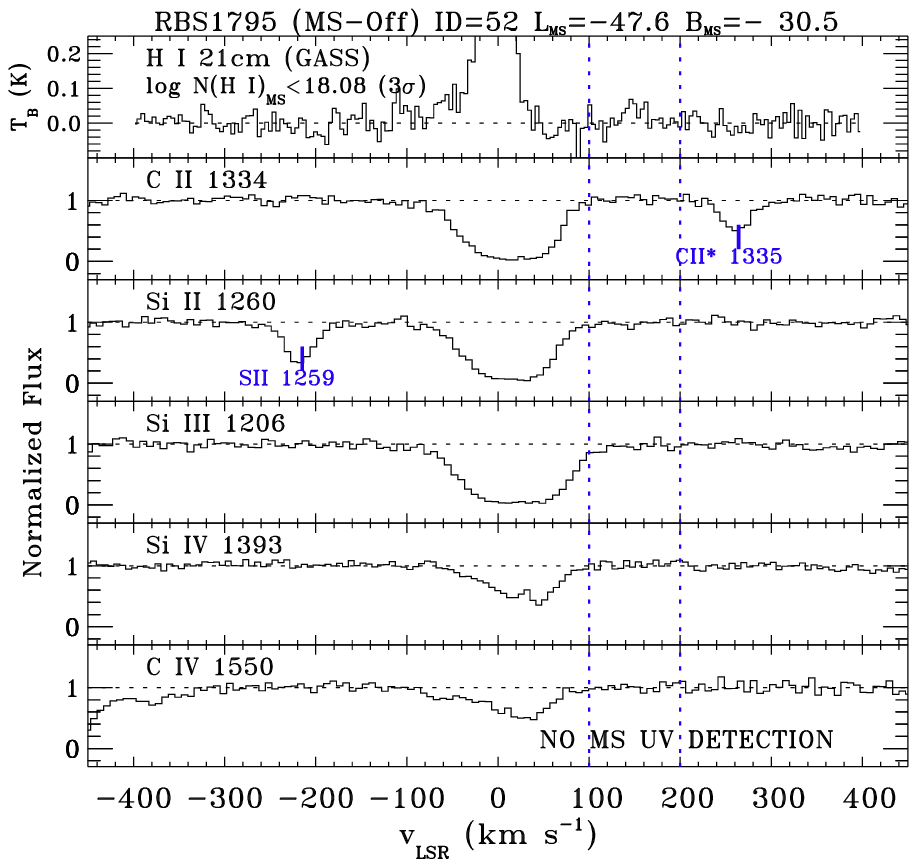}\end{figure}
\begin{figure}\epsscale{1.1}
\plottwo{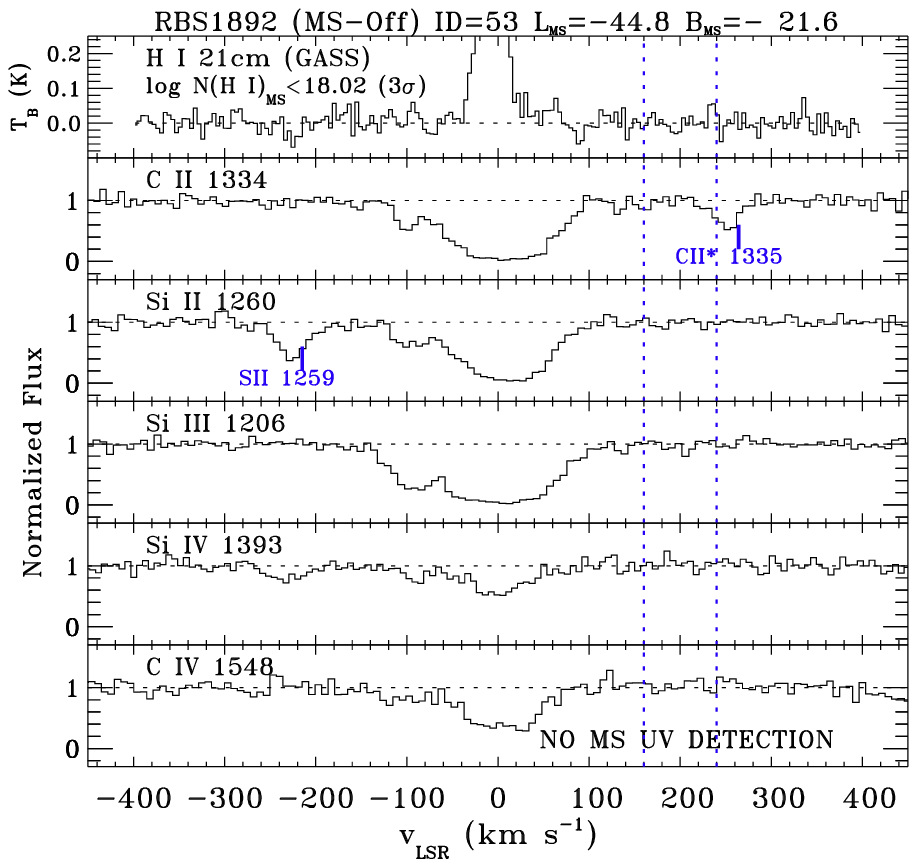}{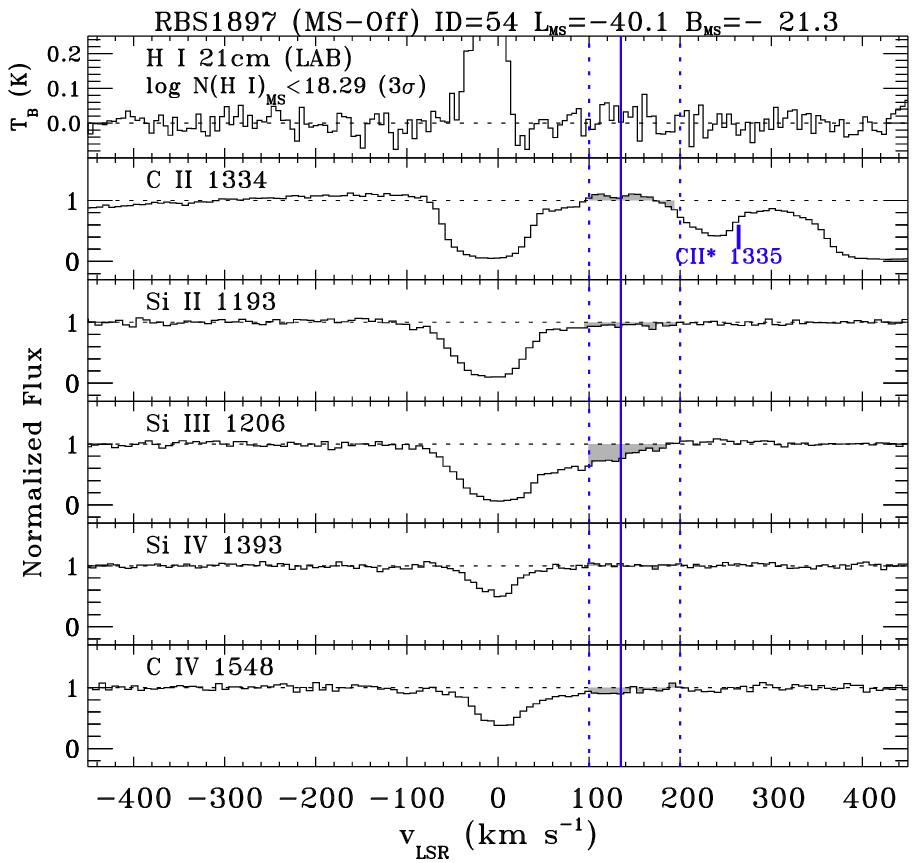}\end{figure}
\begin{figure}\epsscale{1.1}
\plottwo{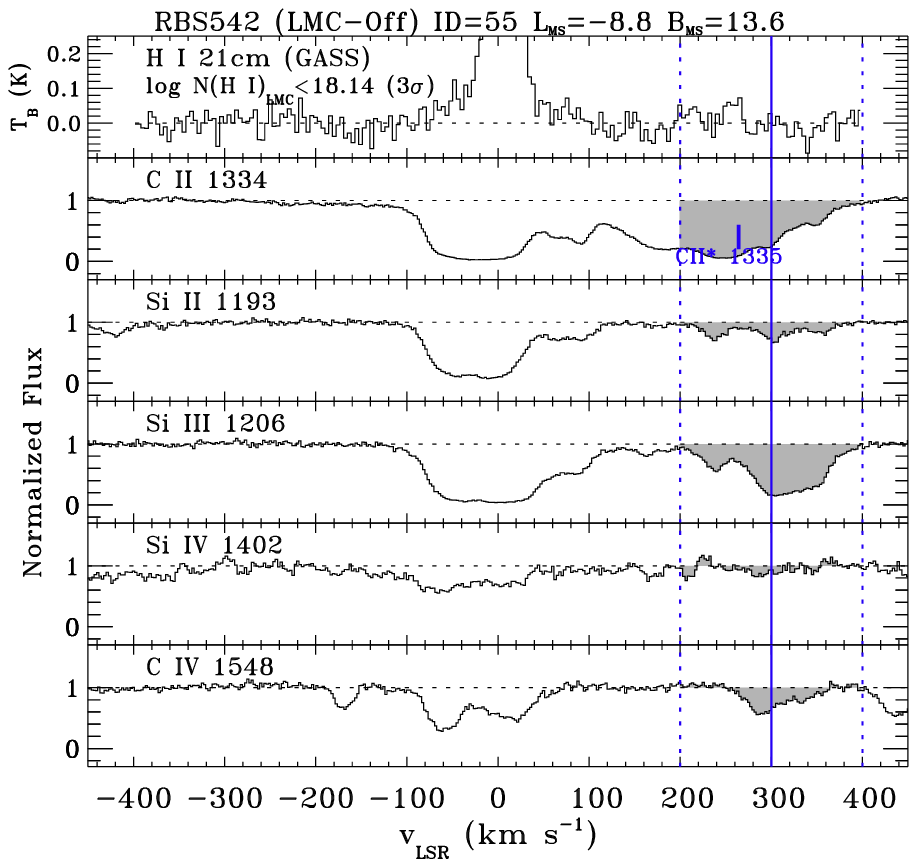}{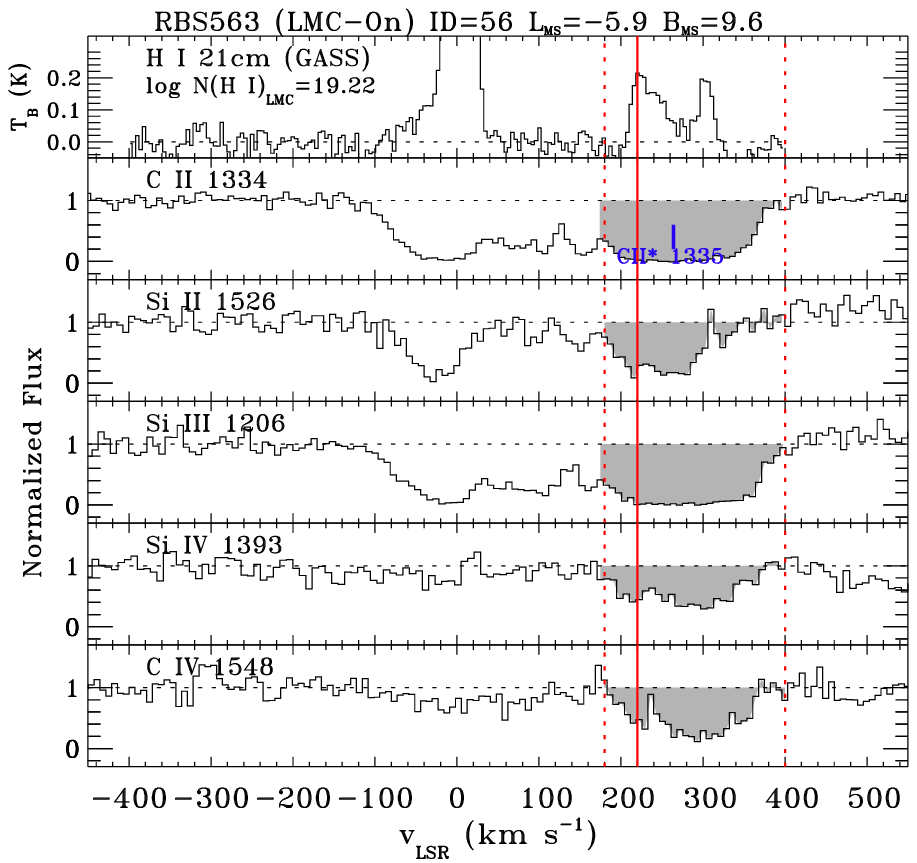}\end{figure}
\begin{figure}\epsscale{1.1}
\plottwo{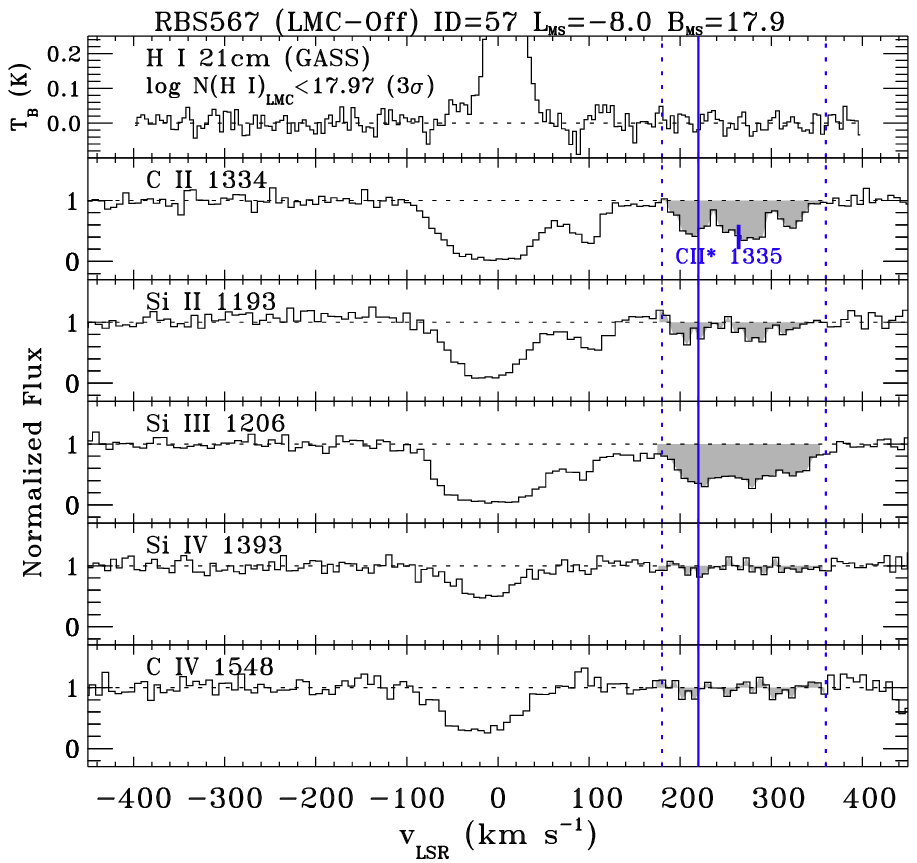}{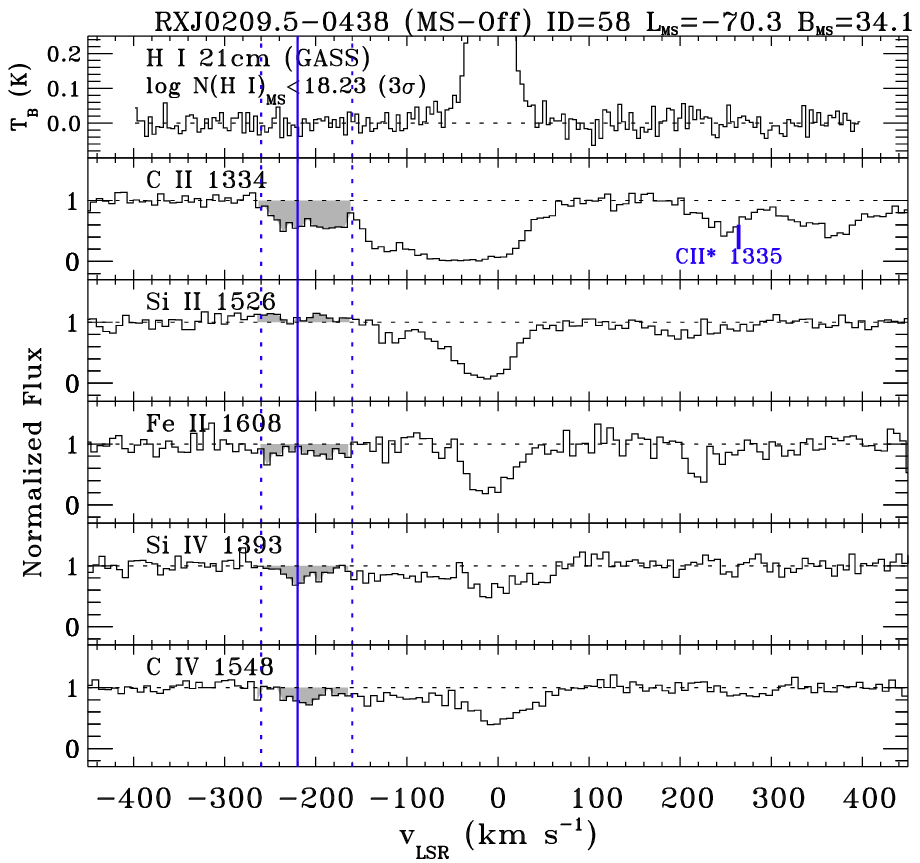}\end{figure}
\begin{figure}\epsscale{1.1}
\plottwo{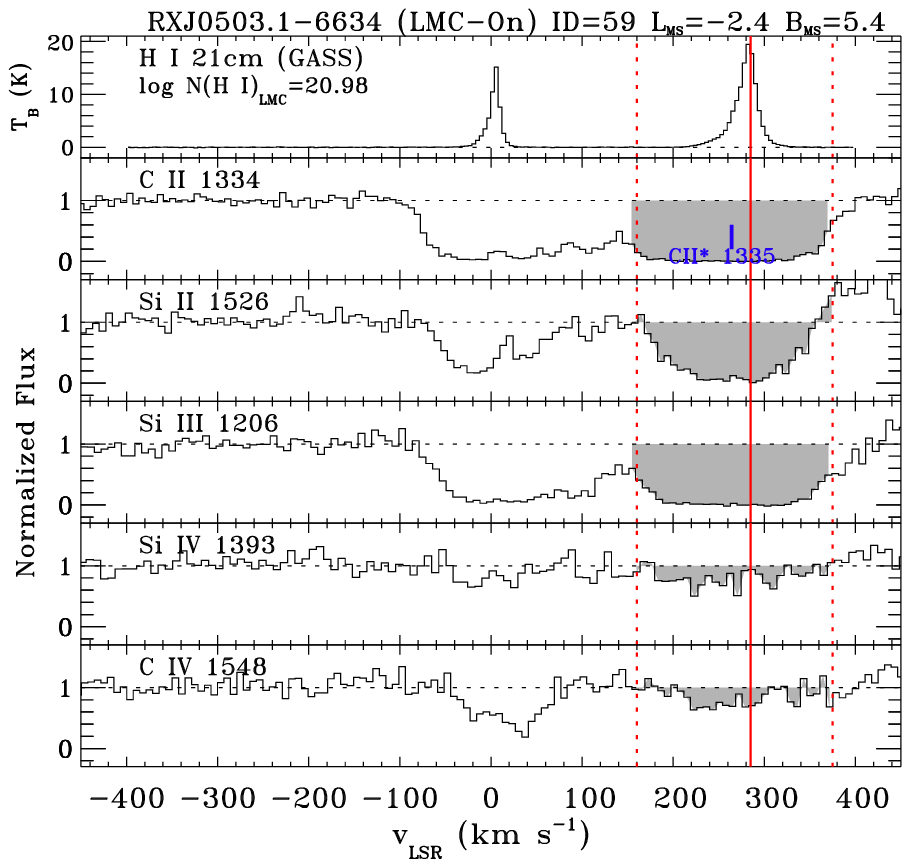}{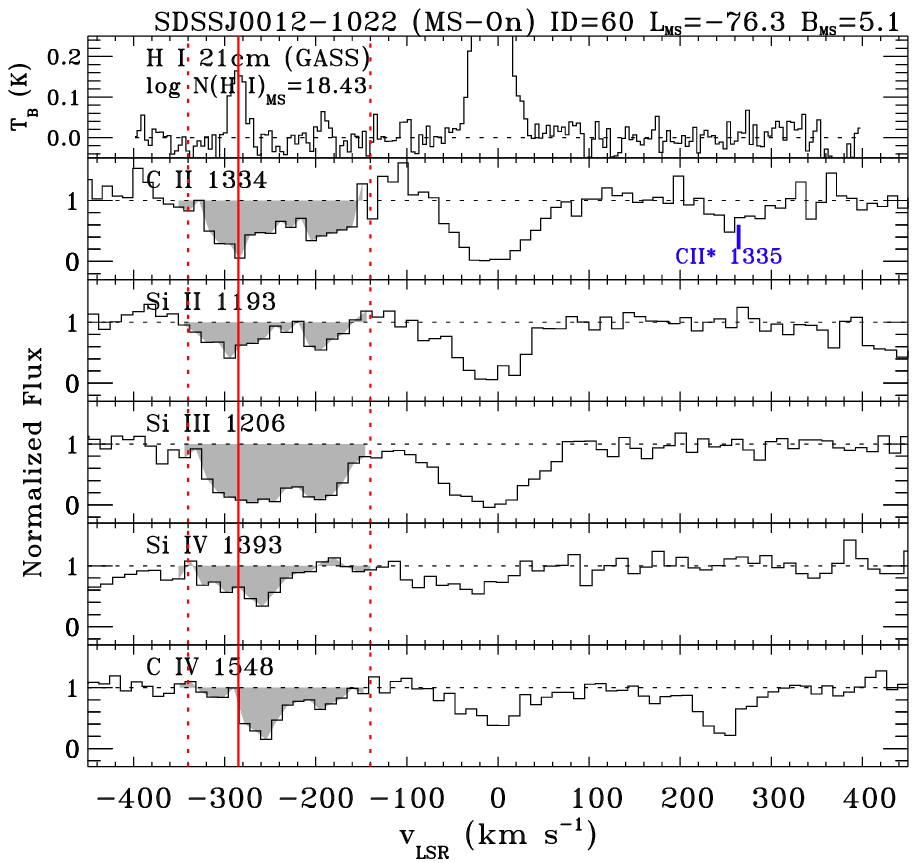}\end{figure}
\clearpage
\begin{figure}\epsscale{1.1}
\plottwo{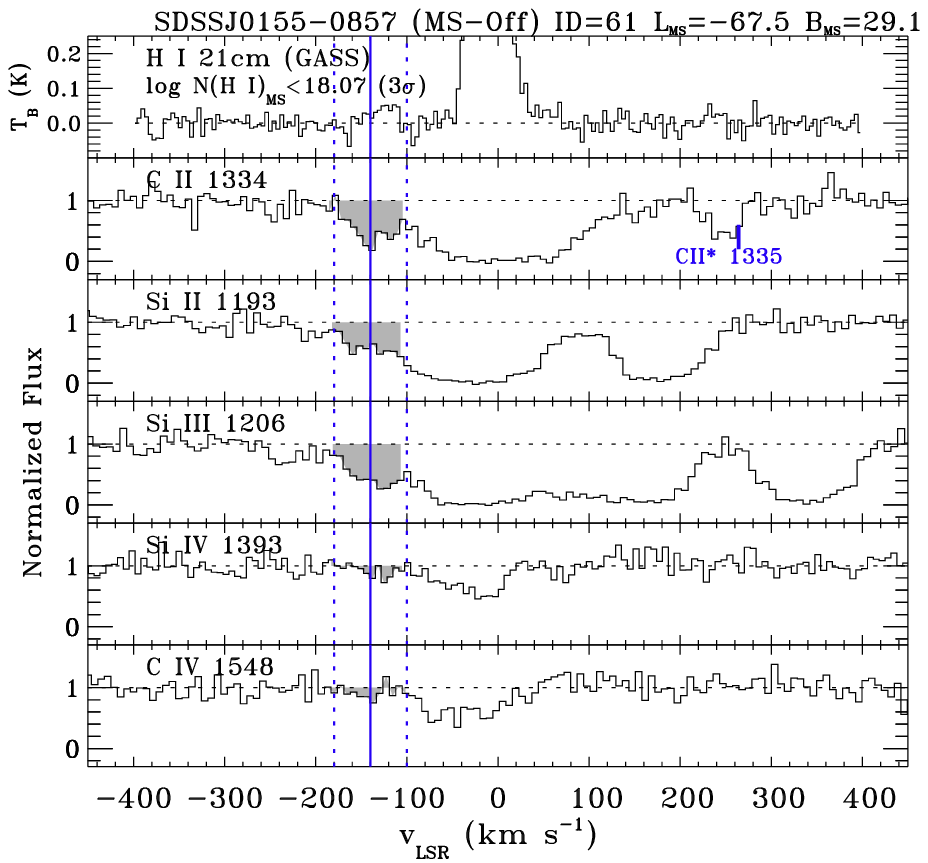}{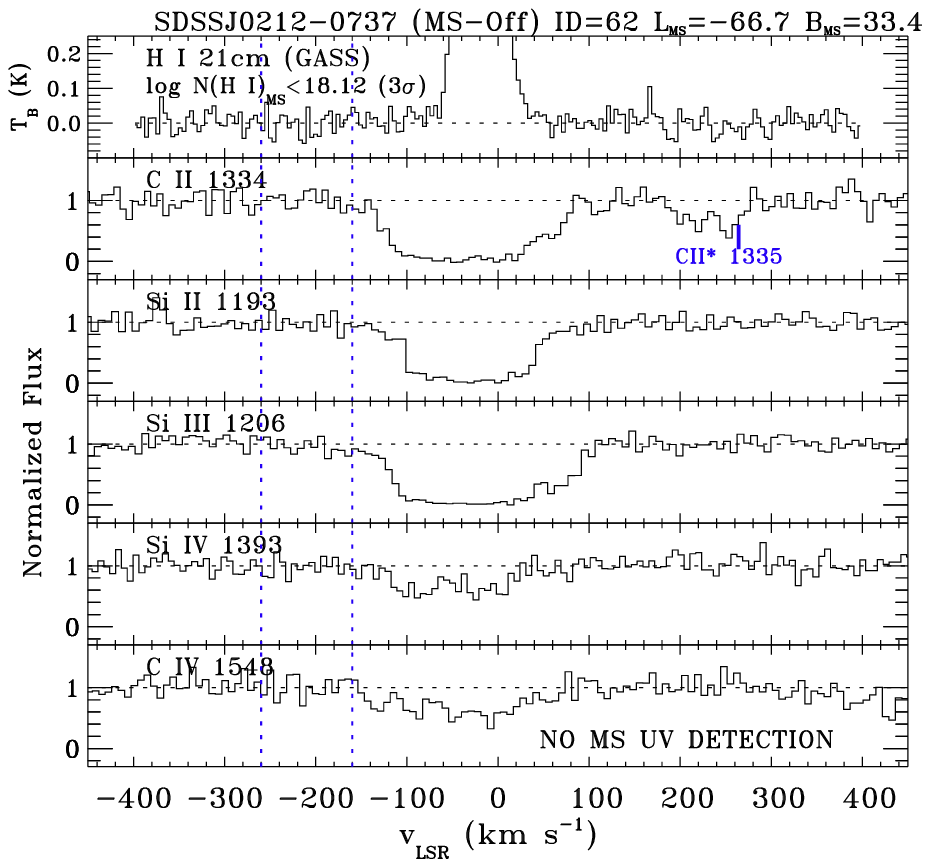}\end{figure}
\begin{figure}\epsscale{1.1}
\plottwo{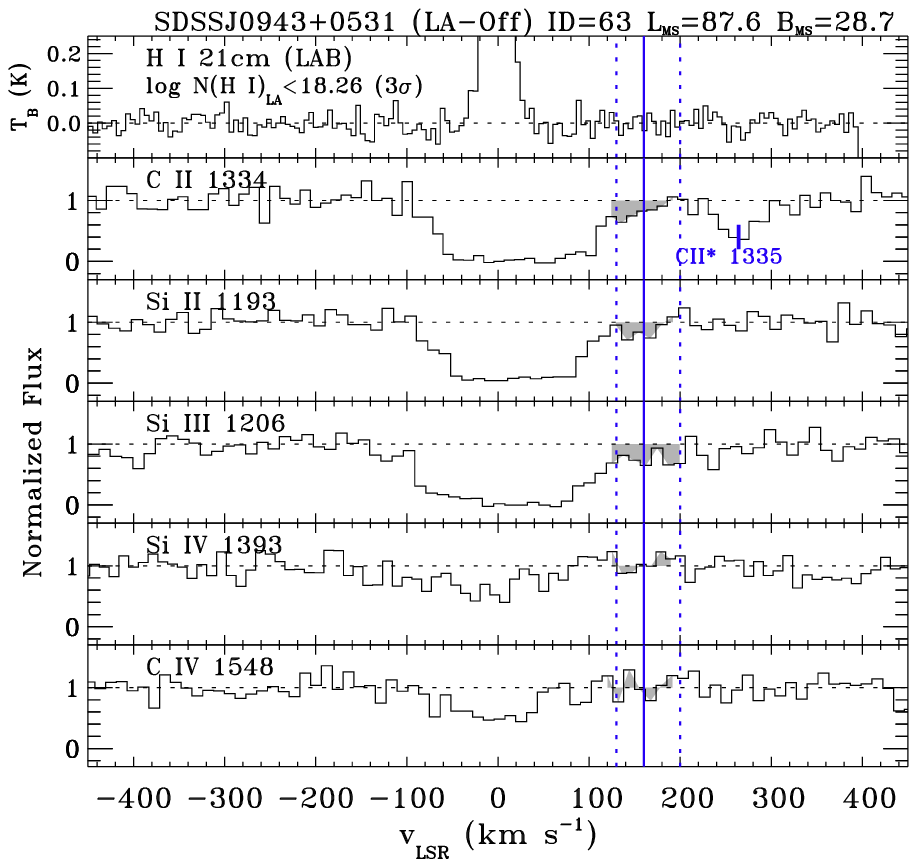}{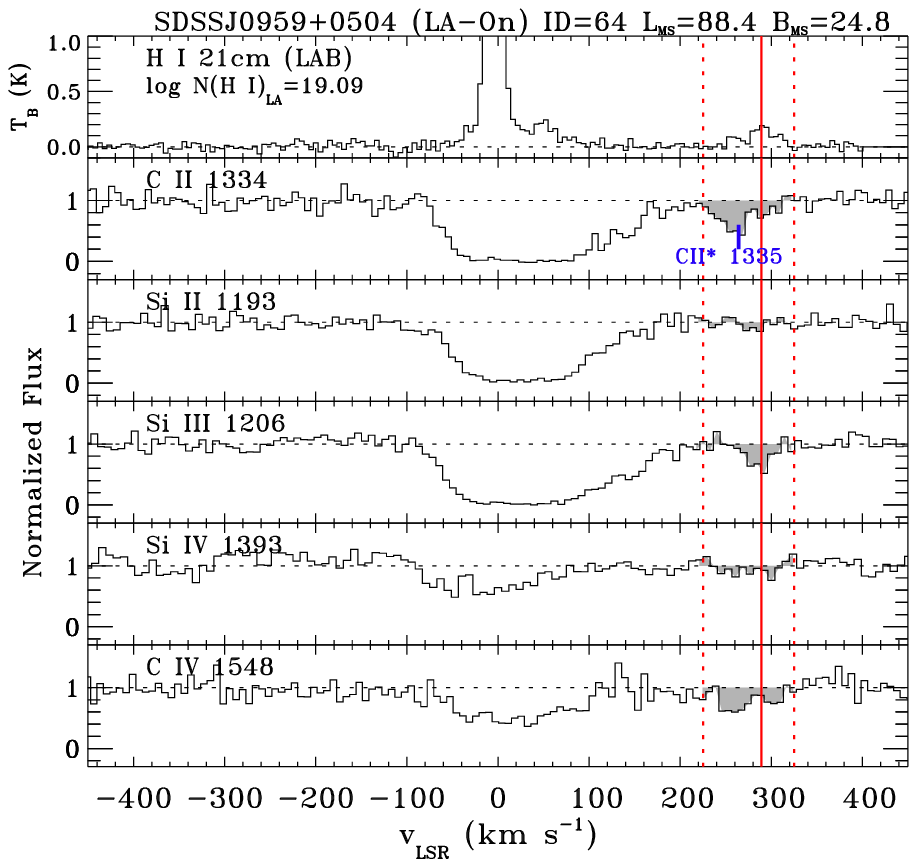}\end{figure}
\begin{figure}\epsscale{1.1}
\plottwo{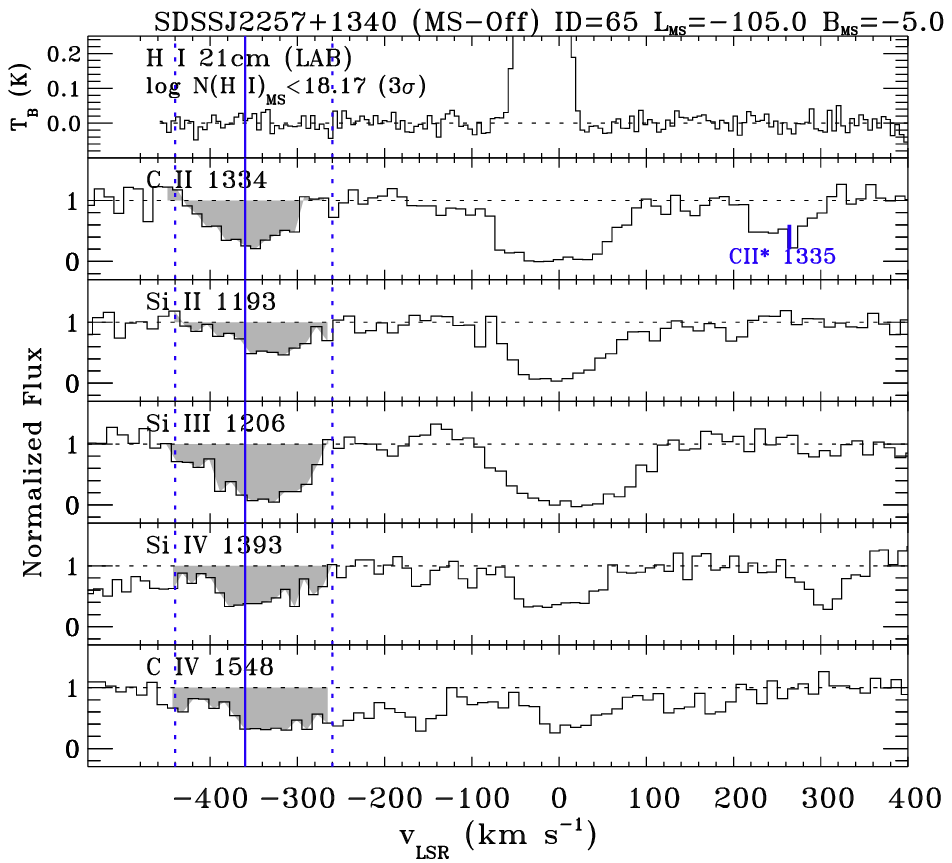}{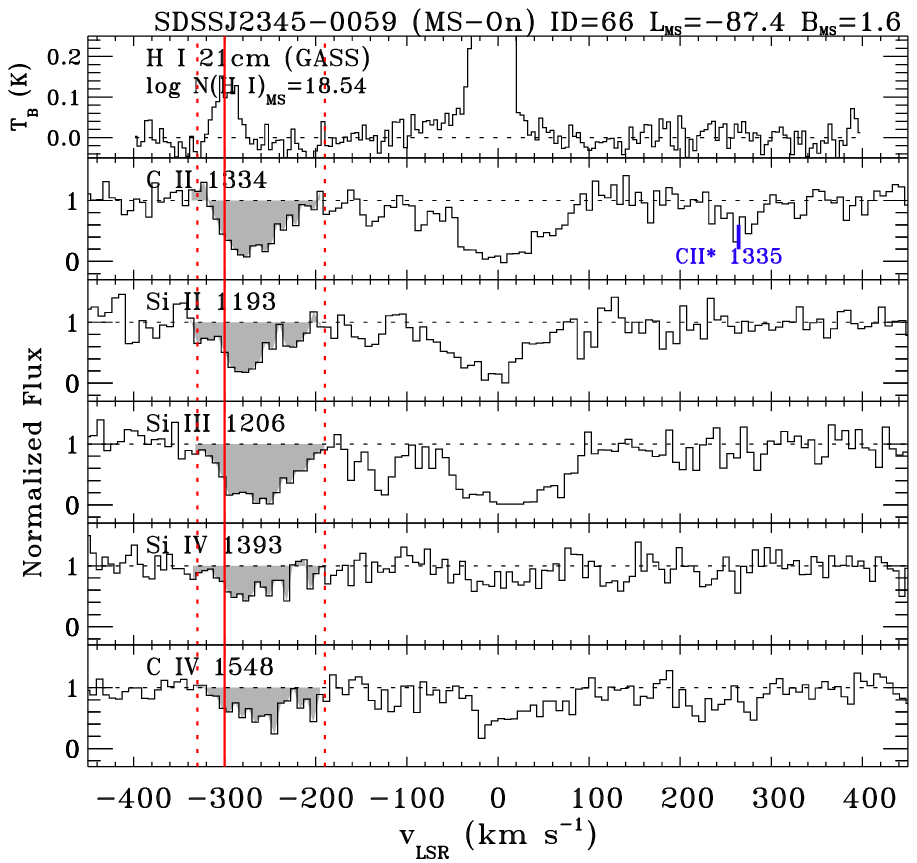}\end{figure}
\begin{figure}\epsscale{1.1}
\plottwo{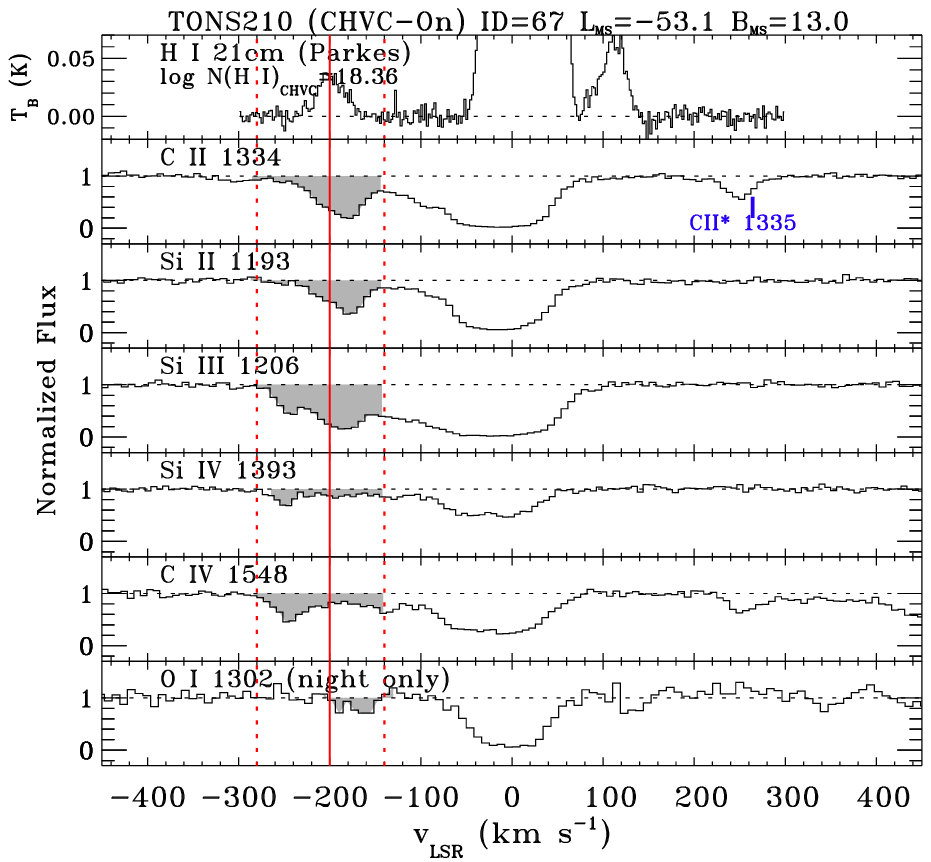}{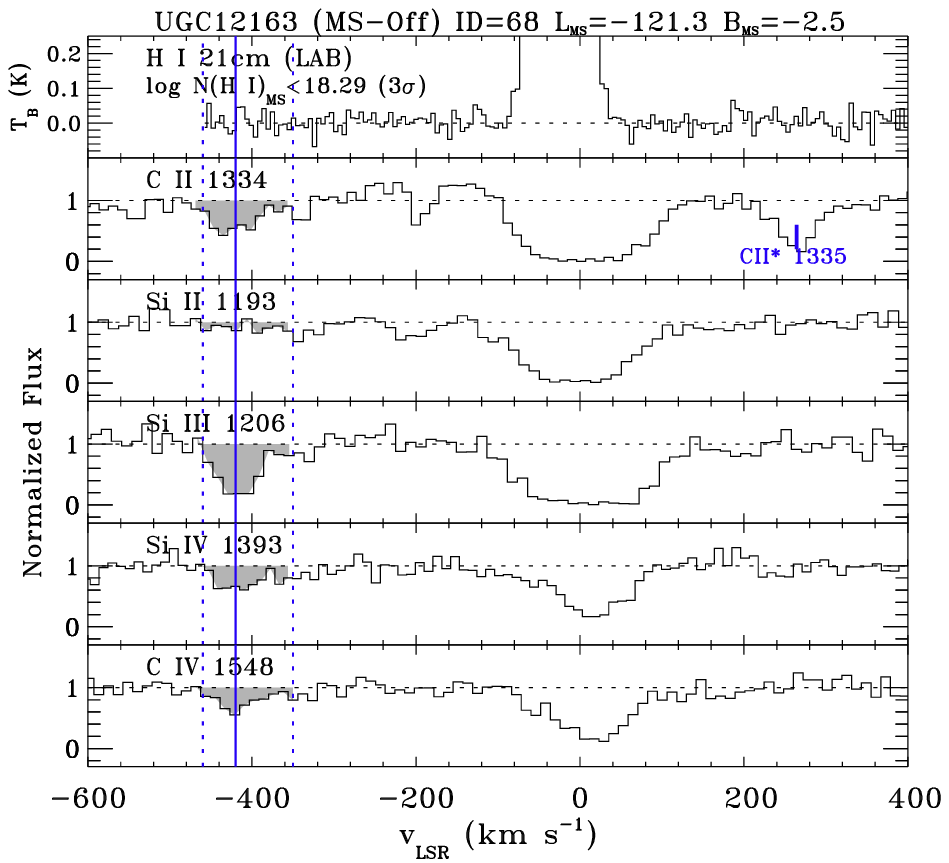}\end{figure}
\begin{figure}\epsscale{1.05}
\plottwo{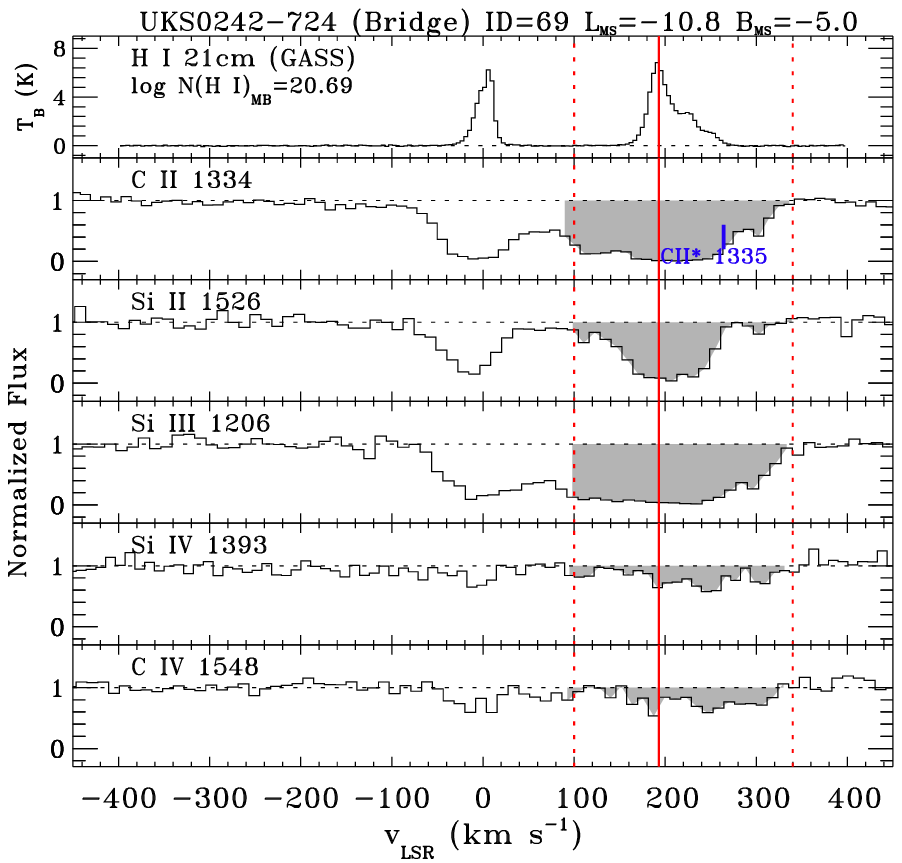}{UKS0242-724.eps}\end{figure}

\end{document}